\documentclass[11pt,english,prd,showpacs,nofootinbib,amssymb,eqsecnum,epsfig]{revtex4}
\usepackage[latin9]{inputenc}
\usepackage{graphicx}
\usepackage{amssymb}

\usepackage{babel}
\usepackage{bm}
\usepackage{amsmath}
\pdfoutput=1

\def\be{\begin{equation}}
\def\ee{\end{equation}}
\def\bea{\begin{eqnarray*}}
\def\eea{\end{eqnarray*}}
\def \erf {\rm erf}
\newcommand{\newc}{\newcommand}
\newc{\lcdm}{$\Lambda$CDM }
\newc{\ba}{\begin{eqnarray}}
\newc{\ea}{\end{eqnarray}}
\def\om{\Omega_{\rm m}}
\def\ol{\Omega_\Lambda}
\newcommand{\nn}{\nonumber}
\begin{document}

\title{Comparative analysis of model-independent methods for exploring the nature of dark energy}
\author{Savvas Nesseris}
\author{Juan Garc\'ia-Bellido}
\email{savvas.nesseris@uam.es, juan.garciabellido@uam.es}
\affiliation{Instituto de F\'isica Te\'orica UAM-CSIC, Universidad Auton\'oma de Madrid,
Cantoblanco, 28049 Madrid, Spain}

\date{\today}

\begin{abstract}
We make a comparative analysis of the various independent methods proposed in the literature for studying the nature of dark energy, using four different mocks of SnIa data. In particular, we explore a generic principal components analysis approach, the genetic algorithms, a series of approximations like Pad\'e power law approximants, and various expansions in orthogonal polynomials, as well as cosmography, and compare them with the usual fit to a model with a constant dark energy equation of state $w$. We find that, depending on the mock data, some methods are more efficient than others at distinguishing the underlying model, although there is no universally better method.
\end{abstract}


\maketitle

\section{Introduction}

Several cosmological studies point towards a cosmic dark sector that includes cold dark matter, dark energy and a spatially flat geometry, in order to explain the observed accelerating expansion of the Universe \cite{Ade:2013zuv},\cite{Suzuki:2011hu}. In this framework, the lack of a fundamental physical theory, regarding the mechanism causing the cosmic acceleration, has given rise to several alternative cosmological scenarios (see for example Ref. \cite{Copeland:2006wr} for a review).

In order to test and compare these cosmological models, so as to find the description that fits the data the best, the usual procedure involves several steps. Even though the present analysis will focus on the SnIa data, it can readily be generalized to other data as well, such as the observed baryon acoustic oscillations (BAO), the cosmic microwave background, and the observed linear growth rate of clustering, measured mainly from the PSCz, 2dF, VVDS, SDSS, 6dF, 2MASS, BOSS and {\em WiggleZ} redshift catalogs and so on.

The SnIa data are given in terms of the distance modulus $\mu_{obs}(z)\equiv m_{obs}(z)-M$; i.e., it is the difference between the absolute and the apparent magnitudes of the SnIa \cite{Suzuki:2011hu}. Then, given a specific dark energy (DE) model for which one may have a description of the equation of state $w(z)$ as
\be
w(z;p_i)=-1+\frac{1}{3}(1+z)\;\frac{d \ln\left(H^2(z;p_i)-\om(1+z)^3\right)}{dz}
\ee
where $p_i$ are the parameters of the model, the luminosity distance $d_L(z;p_i)\equiv \frac{c}{H_0}D_L(z;p_i)$ can be calculated and finally the theoretical prediction of the distance modulus itself $\mu_{th}(z;p_i)=5\log_{10}(D_L(z;p_i))+\mu_0$. If $w(z)$ is not the crucial parameter of the model, such as in $f(R)$ models, then one has to solve the modified Friedman equations numerically or semi-analytically \cite{Nesseris:2012cq} and then calculate the luminosity distance. The best-fit parameters are then found by minimizing the $\chi^2$ defined as
\be
\chi^2(p_i)=\sum_{i=1}^N\left(\frac{\mu_{th}(z;p_i)-\mu_{obs,i}}{\sigma_i}\right)^2\label{chi2snia}
\ee
The steps followed for the usual minimization of Eq.~(\ref{chi2snia}) in terms of its parameters are described in detail in  Refs.~\cite{Nesseris:2004wj,Nesseris:2005ur,Nesseris:2006er}.

Then, one can test several DE models, e.g. a model with a cosmological constant $\Lambda$ and cold dark matter ($\Lambda$CDM) that corresponds to $w(a)=-1$, a model with a constant DE equation of state $w(a)= w_0=const$ and cold dark matter (wCDM), a model with an evolving DE equation of state $w(a)= w_0+w_a(1-a)$ or even more exotic cases like the Hu-Sawicki $f(R)$ model of Ref. \cite{Hu07}. The last step is to then test the methods by implementing some sort of comparison by either ranking them with respect to their $\chi^2/dof$, for $dof=N-M$, (the degrees of freedom) where $N$ is the number of data points and $M$ the number of parameters, or by carrying out a Bayesian inference,  calculating the evidence for each model and finally using the so-called Jeffrey's scale to interpret the results, despite the problems this latter approach has been shown to have \cite{Nesseris:2012cq}. However, this methodology carries the following risks:
\begin{enumerate}
  \item It suffers from model bias, in the sense that the interpretation of the results quite obviously depends on the chosen models, e.g. \lcdm, wCDM etc, and the assumptions (priors) made in the analysis, e.g. flatness ($\Omega_K=0$).
  \item Only a limited number of models was tested, since there is only a finite number of physical theories currently in the literature and in any case, it would be impossible to test every conceivable alternative even for big collaborations or unlimited resources.
\end{enumerate}
One possibility in order to avoid these two problems is to use model-independent methods in order to extract the cosmological information from the data, while at the same time making the least possible number of assumptions on the underlying cosmology (the priors). Several methods have been proposed in the literature, and in the next sections we will briefly describe some of the more prominent ones. However, we will only consider the ones that make neither an explicit nor an implicit mention of a prior or fiducial cosmology, since in our opinion these methods suffer from the first of the two problems mentioned earlier. Our only assumption will be that the best-fit functions should be analytic, smooth and differentiable functions at all redshifts covered by our data.

In Sec. \ref{secanalysis} we will describe our methodology and some interesting theoretical results, in Sec. \ref{secmethods} we will present the model-independent methods, and in Sec. \ref{compar} we will compare the methods against each other.

\section{Analysis\label{secanalysis}}
In this section we will describe the methodology we followed in our paper. Our goal is to see which out of all the different methods works the best in reconstructing the real cosmology described by the data, so we implemented the following procedure which is based on the   following four simple and easy steps:

\begin{enumerate}
  \item Create several synthetic/mock SnIa data sets based on different cosmologies with ``real" parameters $\om{}_{,real}, w(a)_{real}, q(a)_{real}$.
  \item Apply the various model-independent reconstruction methods etc) [principal components analysis (PCA), genetic algorithm (GA), etc.] described in detail in later sections.
  \item Calculate $q_{obs}(a)$ and compare with the ``real" one.
  \item Create a test to rank each method accordingly (see Section \ref{compar}).
\end{enumerate}

Regarding the model-independent reconstruction methods we only chose those that make no explicit assumptions about an underlying fiducial cosmology since we do not want our results to be biased by our preconceptions. Such methods include the genetic algorithms, Pad\'e approximants, and the Principal Components Analysis, but  not, for example, the Gaussian processes of Ref.\cite{Shafieloo:2007cs}. Also, it is important to mention that we will explicitly focus our reconstruction methods on the deceleration parameter $q(z)$ for reasons that will be explained further in Sec. \ref{thedecel}. Finally, we describe our mock data and the model-independent methods in what follows.

\subsection{Mock SnIa data}
The mock SnIa data we used in our analysis are based on three DE models [\lcdm, $w(a)=const$ and $w(a)=w_0+w_a(1-a)$] and the Hu-Sawicki (HS) $f(R)$ model. The DE models have a Hubble parameter given by
\be
H(z)^2/H_0^2= \om (1+z)^3+(1-\om) (1+z)^{3 (1+w_0+w_a)}e^{-\frac{3 w_{a} z}{1+z}}  \label{friedman0}
\ee
The case $(w_0,w_a)=(-1,0)$ corresponds to \lcdm, $(w_0,w_a)=(w_0,0)$ corresponds to $w(a)=\textrm{const}.$ and lastly, $(w_0,w_a)= (w_0,w_a)$ corresponds to the $w(a)$ model.

The Lagrangian for the $f(R)$ model is given by \cite{Hu07}
\begin{equation}
\label{Hu}
f(R)=R-m^2 \frac{c_1 (R/m^2)^n}{1+c_2 (R/m^2)^n}
\end{equation}
where $c_1$, $c_2$ are free parameters, $m^2\simeq \om H^{2}_{0}$ is of the
order of the Ricci scalar $R_{0}$ at
the present time, $H_{0}$ is the Hubble constant, $\om$ is the
dimensionless matter density parameter at the present time,
and $m$ and $n$ are positive constants. As discussed in \cite{Basilakos:2013nfa}, the Lagrangian of Eq.~(\ref{Hu}) can also be written as
\ba
f(R)&=& R- \frac{m^2 c_1}{c_2}+\frac{m^2 c_1/c_2}{1+c_2 (R/m^2)^n} \nn\\
&=& R- 2\Lambda\left(1-\frac{1}{1+(R/(b~\Lambda)^n}\right) \nn \\
&=& R- \frac{2\Lambda }{1+\left(\frac{b \Lambda }{R}\right)^n} \label{Hu1}
\ea where $\Lambda= \frac{m^2 c_1}{2c_2}$ and $b=\frac{2 c_2^{1-1/n}}{c_1}$. In this form it is clear that the HS model can be arbitrarily close to \lcdm, depending on the parameters $b$ and $n$. Notice that the following two limits exist for $n>0$:
\bea
\lim_{b\rightarrow0}f(R)&=&R-2\Lambda \nn \\
\lim_{b\rightarrow \infty}f(R)&=&R
\eea
and therefore the HS model reduces to \lcdm for $b\rightarrow 0$. We prefer to use the HS Lagrangian in the form of Eq.~(\ref{Hu1}) as it is much easier to handle and we can also use the approximation scheme of Ref. \cite{Basilakos:2013nfa}.

Finally, the mock SnIa data we used are as follows:
\begin{itemize}
  \item Mock 1: $w=\textrm{const.}$ with $(\om, w_0, w_a)=(0.3,-0.95,0)$.
  \item Mock 2: \lcdm with $(\om, w_0, w_a)=(0.3,-1,0)$.
  \item Mock 3: $f(R)$ of Eq.~(\ref{Hu1}) with $(\om,b,n)=(0.3,0.11,1)$.
  \item Mock 4: $w=w(a)$ with $(\om, w_0, w_a)=(0.3,-1.05,0.5)$.
\end{itemize}

In all cases we used the same redshift distribution and errors as in the Union 2.1 data set \cite{Suzuki:2011hu}, but the distance modulus $\mu_i$ was calculated by the models by adding noise sampled from the normal distribution with a standard deviation equal to the error at that redshift, i.e. $(z_i,\mu_i,\sigma_{\mu_i})=\left(z_i,\mu_{th}(z_i)+\mathcal{N}(0,\sigma_{\mu_i}),\sigma_{\mu_i}\right)$.

\begin{figure*}[t!]
\centering
\vspace{0cm}\rotatebox{0}{\vspace{0cm}\hspace{0cm}\resizebox{0.60\textwidth}{!}{\includegraphics{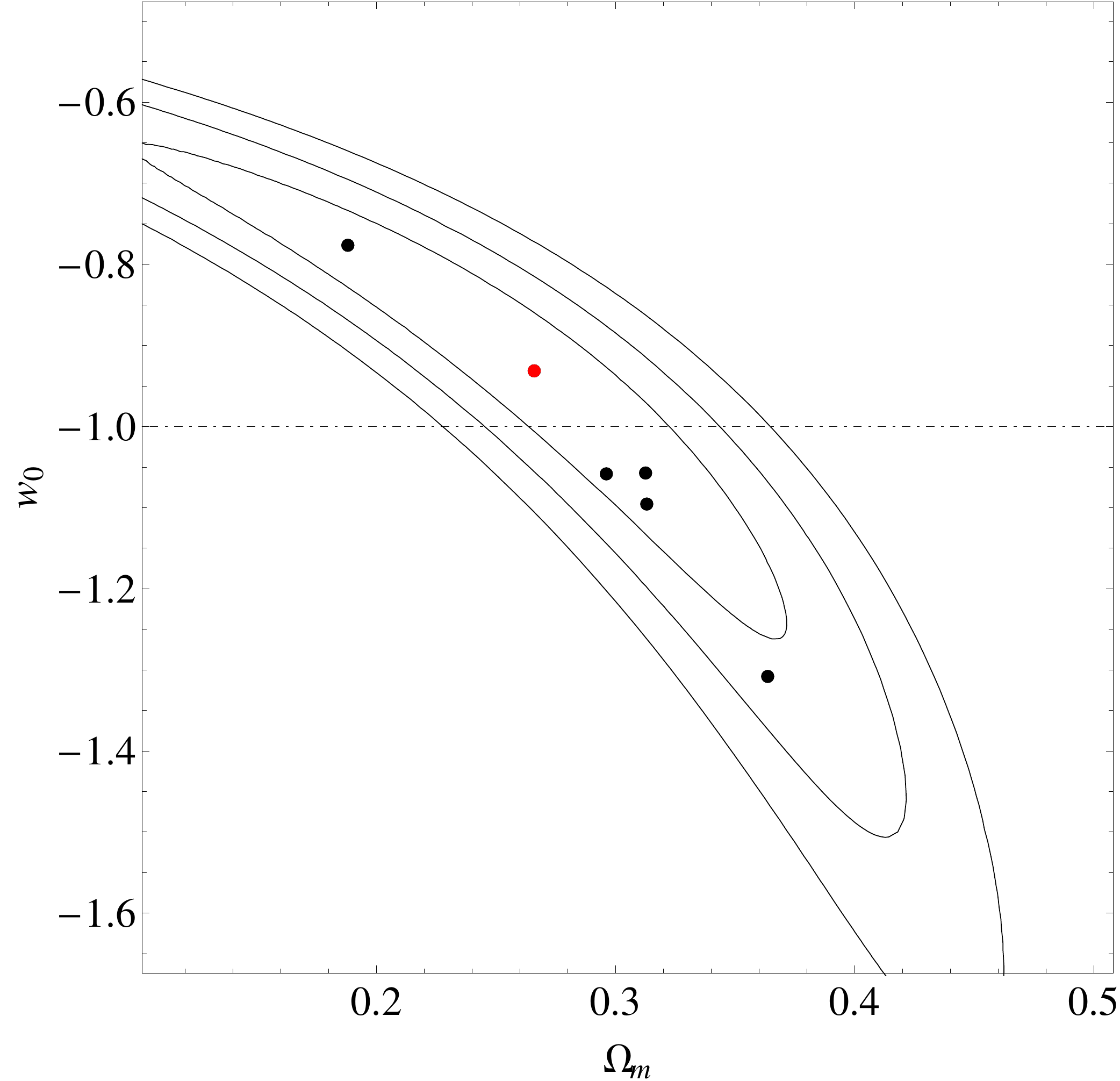}}}
\caption{The red dot corresponds to the mock we used, and the other five black dots are the different mocks.\label{contouromw0}}
\end{figure*}

In order to confirm that our results do not depend strongly on the particular mock we chose, we created several different realizations and tested them with all three DE models. As an example, in Fig. \ref{contouromw0} we show the 1, 2 and 3 $\sigma$ contours on the $(\Omega_m,w_0)$ plane. The red dot corresponds to the mock we used and the other five black dots are the other five mocks we considered in the testing of the analysis. We find that our results in all of the cases are consistent within the $\sim 1\sigma$ level, so we firmly believe that our main conclusions in the later sections are not biased by the specific choice of the mocks.

\subsection{The deceleration parameter $q(z)$ in terms of $H(z)$ and $d_L(z)$\label{thedecel}}

The deceleration parameter $q(z)$ is related to the Hubble parameter $H(z)$ through
\be
1+q(z) = \epsilon(z) = - {\dot H\over H^2} = (1+z)\frac{H'(z)}{H(z)} =\frac{d \ln H(z)}{d \ln(1+z)} \label{q(z)}\,,
\ee
or, alternatively, in terms of the luminosity density, for arbitrary curvature $\Omega_K$,
\be\label{qKz}
q(z) = \frac{1 + \Omega_K\,d_L(z)\,d_L'(z)/(1+z)}{1 + \Omega_K\,d^2_L(z)/(1+z)^2} -
\frac{(1+z)^2 d_L''(z)}{(1+z)d_L'(z) - d_L(z)}\,,
\ee
where primes denote derivatives with respect to redshift $z$. In the case of flat universes we can write it as
\be\label{qN}
q(N) =-1-\frac{H'(N)}{H(N)}= 1 + \frac{d_L''(N) + d_L'(N)}{d_L'(N) + d_L(N)}\,,
\ee
where primes here denote derivatives with respect to  $N\equiv \ln{a}=-\ln (1+z)$.

If instead of the luminosity distance we have data on the angular diameter distance, which are related in any
metric theory of gravity by $d_L(z)=(1+z)^2\,d_A(z)$, e.g. from the angular or radial BAO peak in the
matter correlation function, we can also write the deceleration parameter as
\be\label{qN}
q(N) = -\frac{\theta_{_{\rm BAO}}(N)^2 + 2 \theta_{_{\rm BAO}}(N)\theta_{_{\rm BAO}}'(N) +
2 \theta_{_{\rm BAO}}'(N)^2 - \theta_{_{\rm BAO}}(N)\theta_{_{\rm BAO}}''(N)}
{\theta_{_{\rm BAO}}(N)^2 + \theta_{_{\rm BAO}}(N)\theta_{_{\rm BAO}}'(N)}\,,
\ee
where $\theta_{_{\rm BAO}}=r_s/d_A(z)$ is the angle subtended by the sound horizon at decoupling.

In terms of $q(z)$ and $\om$ today, it is possible to recover $w(z)$ for {\rm any} background cosmology,
\be\label{wz}
3w(z) = \frac{2q(z)-1}{1- \om \, (1+z)^{(1 - 2\bar q(z))}}\,,
\ee
where
\be
\bar q(z) = \frac{1}{\ln(1+z)}\int_0^z q(s) \ d\ln(1+s)\,,
\ee
is the average deceleration parameter. As can be seen from Eq.~(\ref{wz}), the equation of state $w(z)$ depends strongly on $\om$ and clearly any measurement on $w(z)$ will be degenerate with $\om$ \cite{Kunz:2007rk} unless outside independent information is used, something which unfortunately is forgotten or simply ignored in the community. This is the main reason why in our paper we have chosen to use instead the deceleration parameter $q(z)$ given by Eq.~(\ref{q(z)}).

These expressions allow us to obtain the cosmological parameters, $w$ and $q$, from the
observed luminosity data. Alternatively, from the deceleration parameter $q(z)$, one can
obtain the rate of expansion,
\be\label{Hqz}
H(z) = H_0 \exp \int_0^z \left(\frac{1+q(z')}{1+z'}\right) dz'\,,
\ee
and from there, the luminosity distance (for $K=0$),
\be\label{dLH}
d_L(z) = (1+z)\int_0^z \frac{dz'}{H(z')}\,.
\ee

\begin{figure*}[t!]
\centering
\vspace{0cm}\rotatebox{0}{\vspace{0cm}\hspace{0cm}\resizebox{0.70\textwidth}{!}{\includegraphics{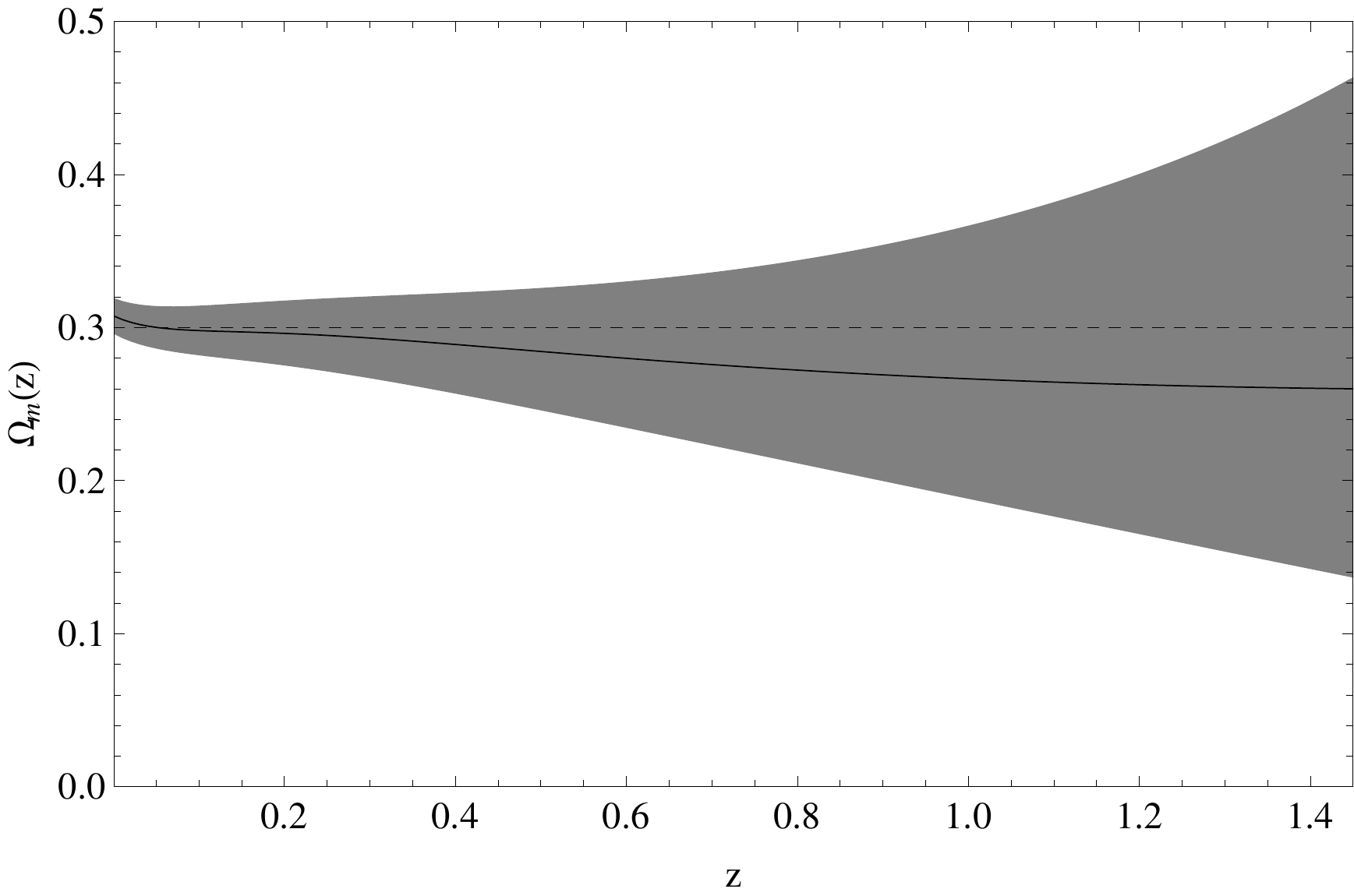}}}
\caption{$\om$ as a function of $z$, given by Eq.~(\ref{wzL}) in the case of Mock 2 with the best fit reconstructed by the genetic algorithms. Approximately, it is a constant within the errors.\label{omz}}
\end{figure*}

In any case, it is interesting to note that these expressions open up the possibility of using standard candles or standard rulers to check a consistency condition
on \lcdm models. Suppose $w=-1$ for all $z$, then the expression (\ref{wz}) can be recast into
\be\label{wzL}
3\om = 2(q(z) + 1)(1+z)^{2\bar q(z) - 1} \,.
\ee
If the rhs of Eq.~(\ref{wzL}) is not constant within the errors, then it is an indication that we may be in the
presence of a time-dependent vacuum energy. As can be seen in Fig. \ref{omz}, indeed, in the case of Mock 2 with the best fit reconstructed by the genetic algorithms, the rhs is found to be approximately constant, within the observational errors. Alternatively, the consistency condition for \lcdm can be recast as a very simple differential equation for the acceleration parameter,
\be\label{conscond}
q'(N) = (1+q)(2q-1)\,,
\ee
which can be easily checked with present and future data. However, since Eq.~(\ref{conscond}) contains one more derivative we expect it not to give as good constraints as the earlier integral equation of (\ref{wzL}), as differentiation of noisy data makes deviations more prominent.

\subsection{The deceleration parameter in f(R) theories}

In order to calculate the deceleration parameter $q(z)$ in the case of the HS Lagrangian of Eq.~(\ref{Hu1}), it is much easier to use the approximation scheme of \cite{Basilakos:2013nfa} and the series expansion of $H(z)$ in terms of the parameter $b$.

In Ref. \cite{Basilakos:2013nfa} it was found that for $n=1$, the Hubble parameter can be written as
\be
H_{HS}^2(N,b)=H_{\Lambda}^2(N)+b~\delta H_1^2(N)+ ...\label{expansionHS1}
\ee
where
\be
\frac{H_{\Lambda}^2(N)}{H_0^2}=\om e^{-3N}+(1-\om)\label{LCDM1}
\ee
is the Hubble parameter for \lcdm and $\delta H_1^2(N)$ is the first-order correction, given in the Appendix of Ref. \cite{Basilakos:2013nfa}.

Then it is easy to see that the deceleration parameter can also be written as a series expansion in terms of $b$ around $b=0$, i.e. the \lcdm model,
\be
q_{HS}(N)=q_{\Lambda}(N)+\delta q^{(1)}(N)\;b +O(b^2) \label{qfrhs}
\ee
where
\be
\delta q^{(1)}(N)\equiv-\frac{\left(\left(1-2 q_{\Lambda}(N)\right){}^2 \left(1+q_{\Lambda}(N)\right) \left(-2+q_{\Lambda}(N) \left(13+q_{\Lambda}(N) \left(-7+2 q_{\Lambda}(N)\right)\right)\right)\right)}{18 \left(-1+q_{\Lambda}(N)\right){}^4}
\ee
and $q_{\Lambda}(N)$ is the deceleration parameter for the \lcdm given by

\be
q_{\Lambda}(N)=-1+\frac{3 e^{-3 N} \om}{2 \left(1-\om+e^{-3 N} \om\right)}
\ee
Also, we can do something similar for the dark energy equation of state $w(z)$,
\ba
w(z)&=&-1+\frac{1}{3}(1+z)\;\frac{d \ln\left(H_{HS}^2(z)-\om(1+z)^3\right)}{dz} \nn \\
&=&-1+\frac{\left(q_{\Lambda }(z) \left(6 q_{\Lambda }(z){}^2+q_{\Lambda }(z)-4\right)+1\right)}{3 \left(q_{\Lambda }(z)-1\right){}^4}b+O\left(b^2\right)
\ea which clearly shows us the correction picked up by the equation of state of the \lcdm model $(w=-1)$ due to the $f(R)$ theory.

It is interesting to note that the value of the deceleration parameter today, i.e. $q_0\equiv q(N=0)=q(z=0)$, picks up a correction with respect to its \lcdm value $q_{\Lambda ,0}=-1+\frac{3 }{2}\om$, which obviously depends on the parameter $b$,

\be
q_0=q_{\Lambda ,0}-\frac{\left(1-2 q_{\Lambda ,0}\right){}^2 \left(1+q_{\Lambda ,0}\right) \left(-2+q_{\Lambda ,0} \left(13+q_{\Lambda ,0} \left(-7+2 q_{\Lambda ,0}\right)\right)\right)}{18 \left(-1+q_{\Lambda ,0}\right){}^4} \; b +... \label{qfrhs0}
\ee

We have tested numerically the expressions of Eqs.~(\ref{qfrhs}), (\ref{qfrhs0}) and found them to be in excellent agreement with the numerical solutions of the $f(R)$ differential equations. Finally, it should be noted that similar expressions can be found for other values of $n$, but also for different $f(R)$ models.

\section{Methods\label{secmethods}}
\subsection{Principal components analysis}
\subsubsection{Constant $q(z)$ in redshift bins}

We now present a way of parametrizing the deceleration parameter by assuming it constant or at least it does not vary much in each redshift bin. If we write
\be
q(z)=\sum_{i=1}^{n}q_i\,\theta(z_i)
\ee
where $q_i$ are constant in each redshift bin $z_i$, i.e. $\theta(z_i)=1$ for $z_{i-1}\leq z<z_i$ and 0 elsewhere, then we can solve Eq.(\ref{q(z)}) and write the Hubble parameter in terms of the deceleration parameter. Assuming that $z$ is in the nth bin:
\be
H_n(z)/H_0=c_n~(1+z)^{1+q_n} \label{H(z)}
\ee
where

\be
c_n=\prod_{j=1}^{n-1}(1+z_j)^{q_j-q_{j+1}}
\ee
For example, we have:
\bea
n=1&,& H_1(z)/H_0=(1+z)^{1+q_1} \\
n=2&,& H_2(z)/H_0=(1+z)^{1+q_2} (1+z_1)^{q_1-q_2} \\
n=3&,& H_3(z)/H_0=(1+z)^{1+q_3} (1+z_1)^{q_1-q_2}(1+z_2)^{q_2-q_3}
\eea

\subsubsection{The luminosity distance $d_{L}$ in terms of $q(z)$}

The luminosity distance is defined
\be
d_{L}(z)= \frac{c}{H_0}\left(1+z\right) \int_{0}^{z}{\frac{{\rm d}x}{H(x)/H_0}}
\ee
with $H$ being the Hubble parameter. Using Eq.~(\ref{H(z)}) we can evaluate the luminosity distance in terms of the deceleration parameters $q$'s,

\be
d_{L,n}(z)=\frac{c}{H_0}\left(1+z\right)\left(f_n-\frac{(1+z)^{-q_n}}{c_{n} q_n}\right)
\label{lumPCA}\ee
where

\be
f_n\equiv\frac{(1+z_{n-1})^{-q_n}}{c_{n} q_n}+\sum_{j=1}^{n-1}\frac{(1+z_{j-1})^{-q_j}-(1+z_j)^{-q_j}}{c_{j} q_j}
\ee

The advantages of Eq.~(\ref{lumPCA}) are twofold. First, they are simple analytic expressions that allow for fast and efficient evaluation of the best-fit parameters and second, the parameter $\om$ does not appear at all, thus lifting the problem of the standard PCA analysis with $w(z)$ where one always has to either fix $\om$ to some value based on some prior knowledge or allow it to vary as a free parameter. At this point we should note that the PCA approach for the deceleration parameter has also been considered in Ref. \cite{Shapiro:2005nz}, but as far as we know our analytic expression of Eq.~(\ref{lumPCA}) is new in the literature.

\subsubsection{The PCA}
\begin{figure*}[t!]
\centering
\vspace{0cm}\rotatebox{0}{\vspace{0cm}\hspace{0cm}\resizebox{0.425\textwidth}{!}{\includegraphics{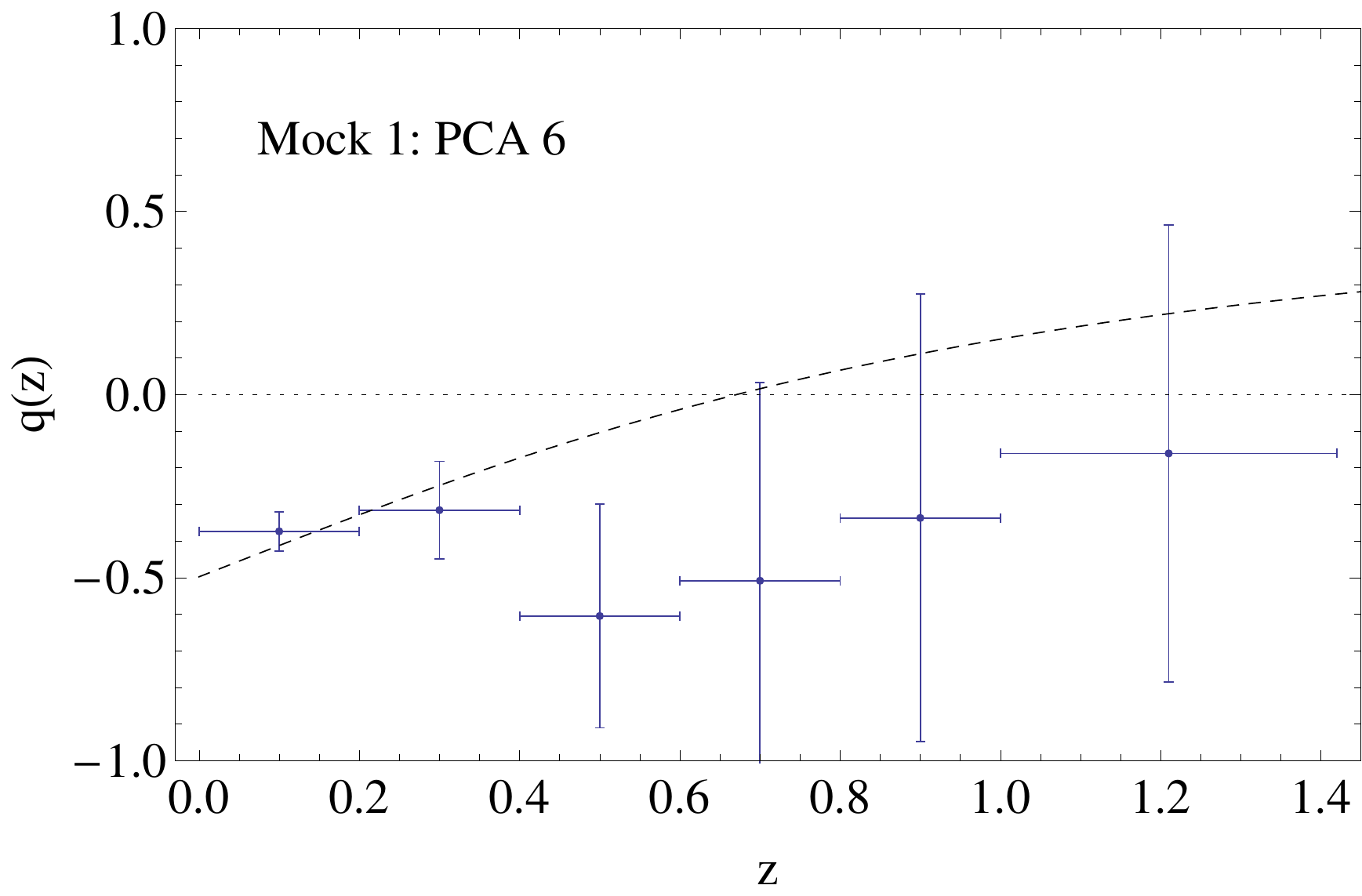}}}
\vspace{0cm}\rotatebox{0}{\vspace{0cm}\hspace{0cm}\resizebox{0.425\textwidth}{!}{\includegraphics{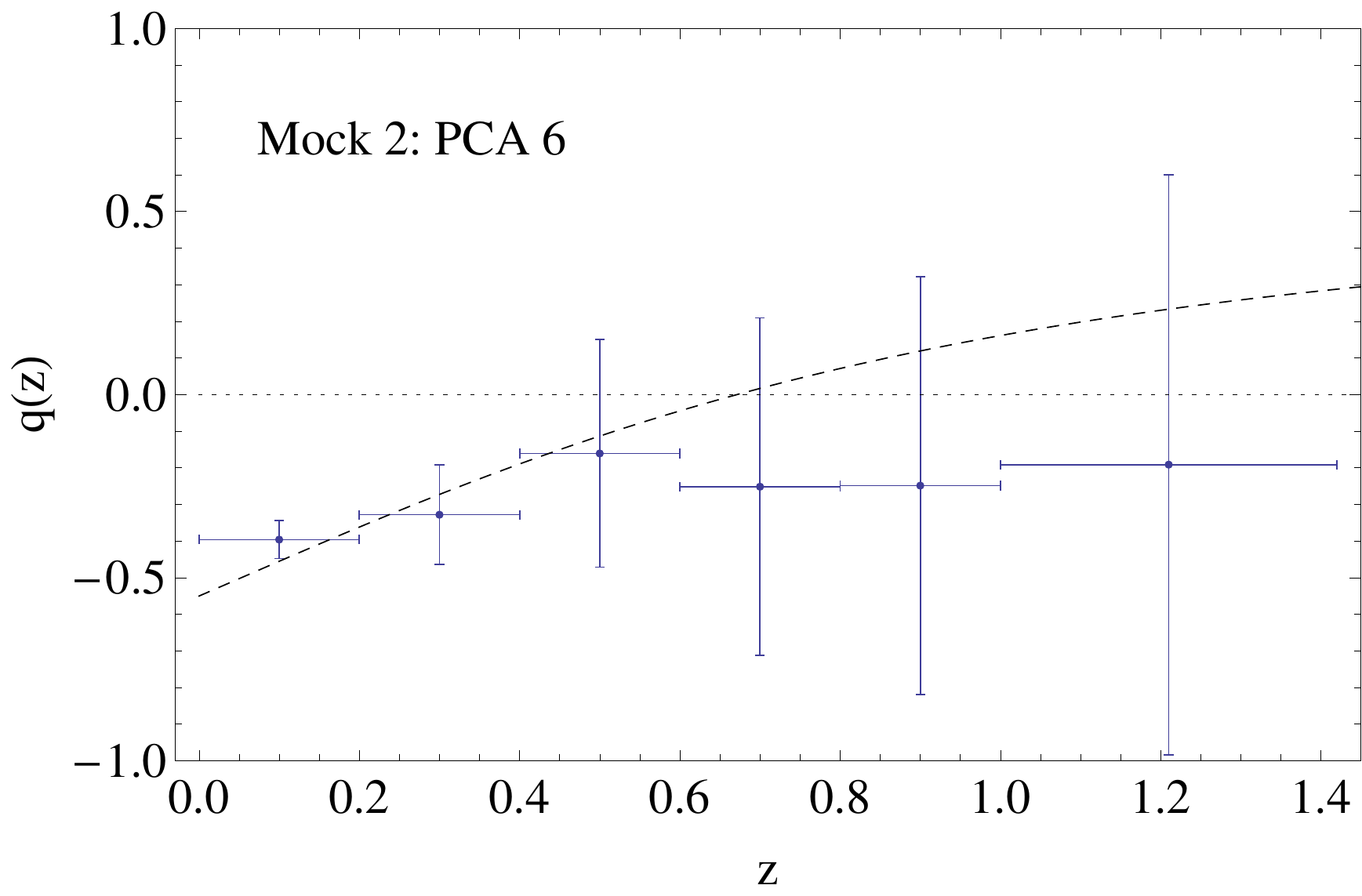}}}
\vspace{0cm}\rotatebox{0}{\vspace{0cm}\hspace{0cm}\resizebox{0.425\textwidth}{!}{\includegraphics{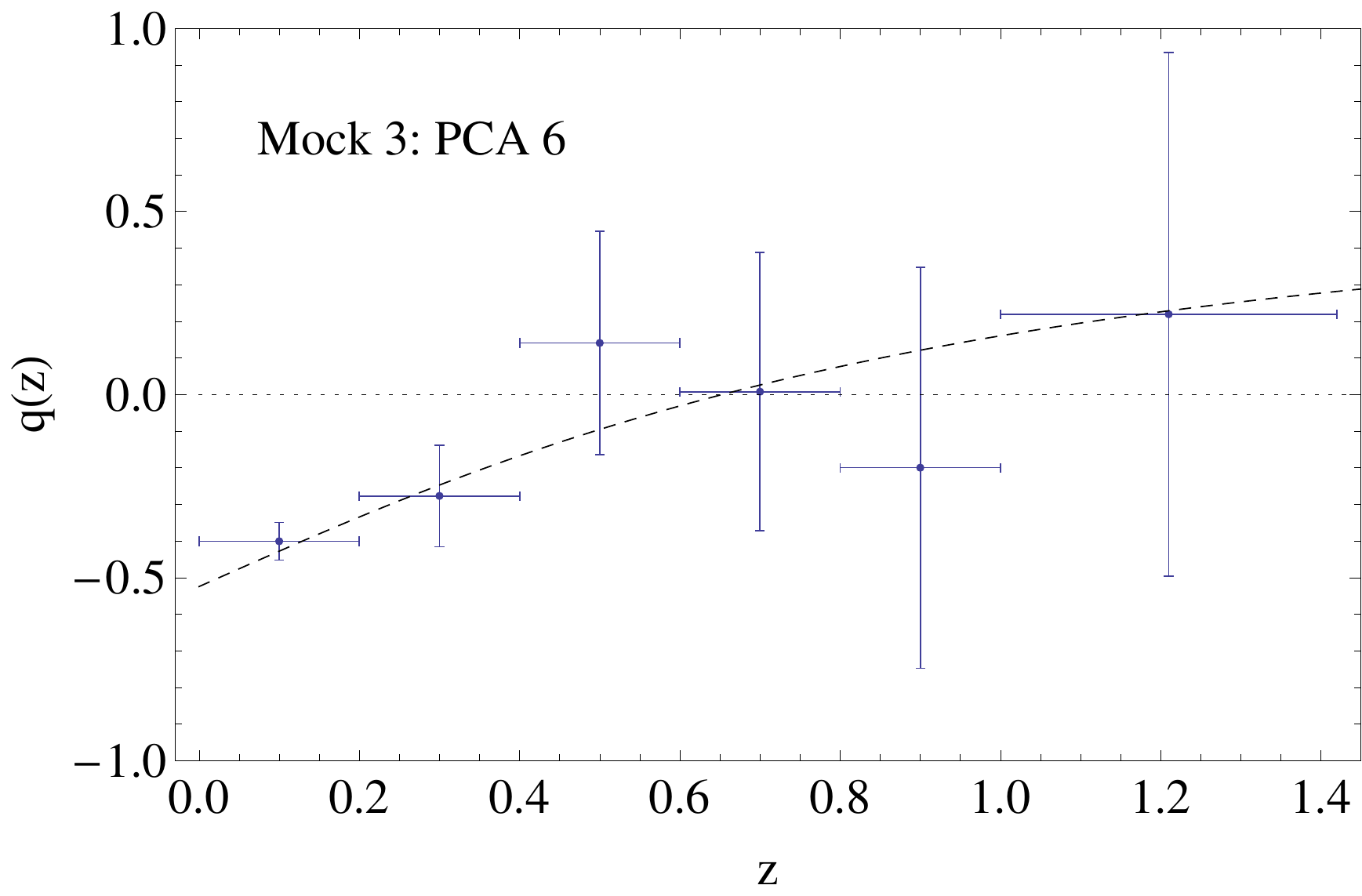}}}
\vspace{0cm}\rotatebox{0}{\vspace{0cm}\hspace{0cm}\resizebox{0.425\textwidth}{!}{\includegraphics{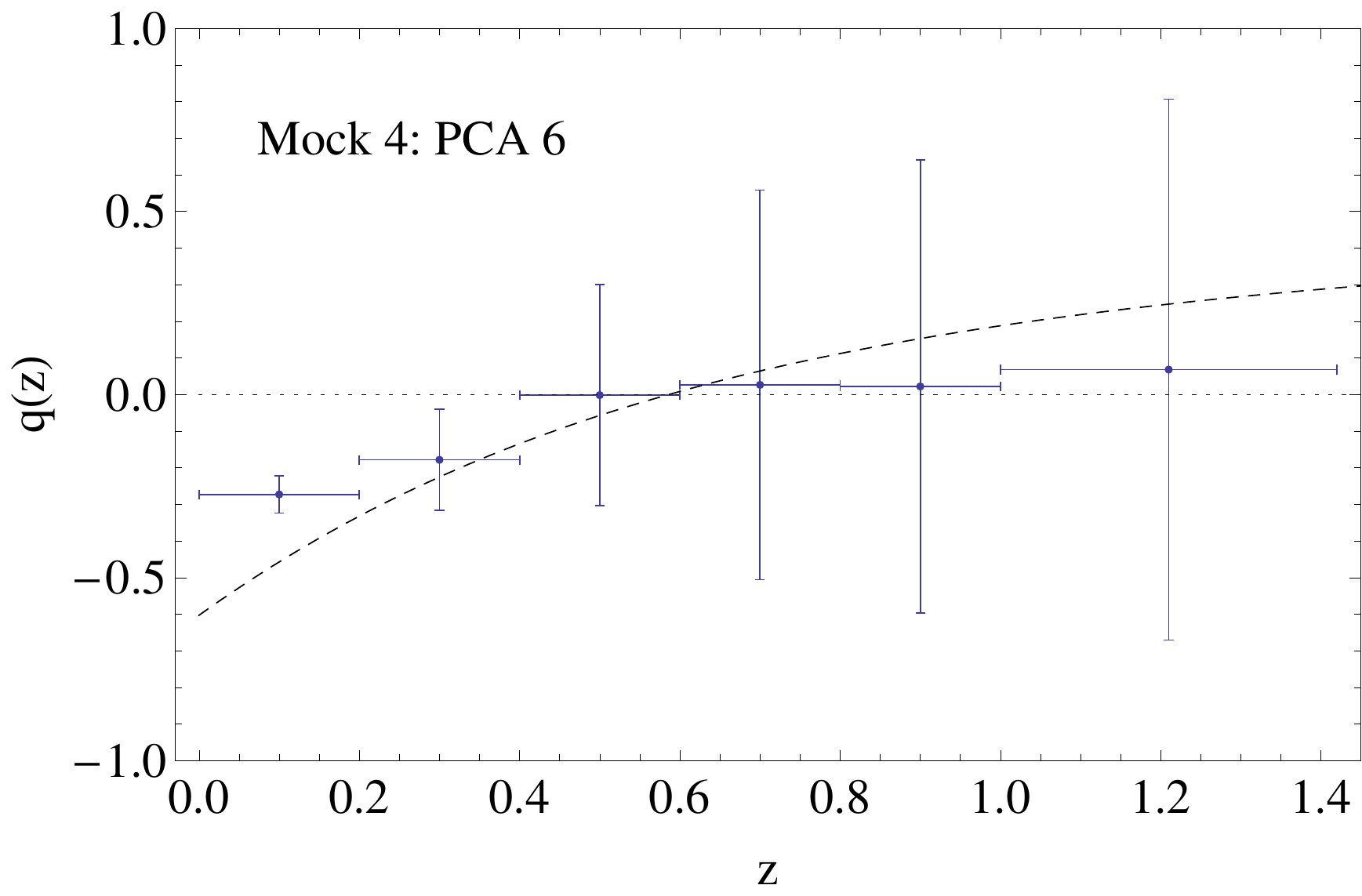}}}
\caption{The deceleration parameter $q(z)$ for all four mocks for six bins. The dashed line corresponds to the real model.\label{qzPCA6}}
\end{figure*}

\begin{figure*}[t!]
\centering
\vspace{0cm}\rotatebox{0}{\vspace{0cm}\hspace{0cm}\resizebox{0.425\textwidth}{!}{\includegraphics{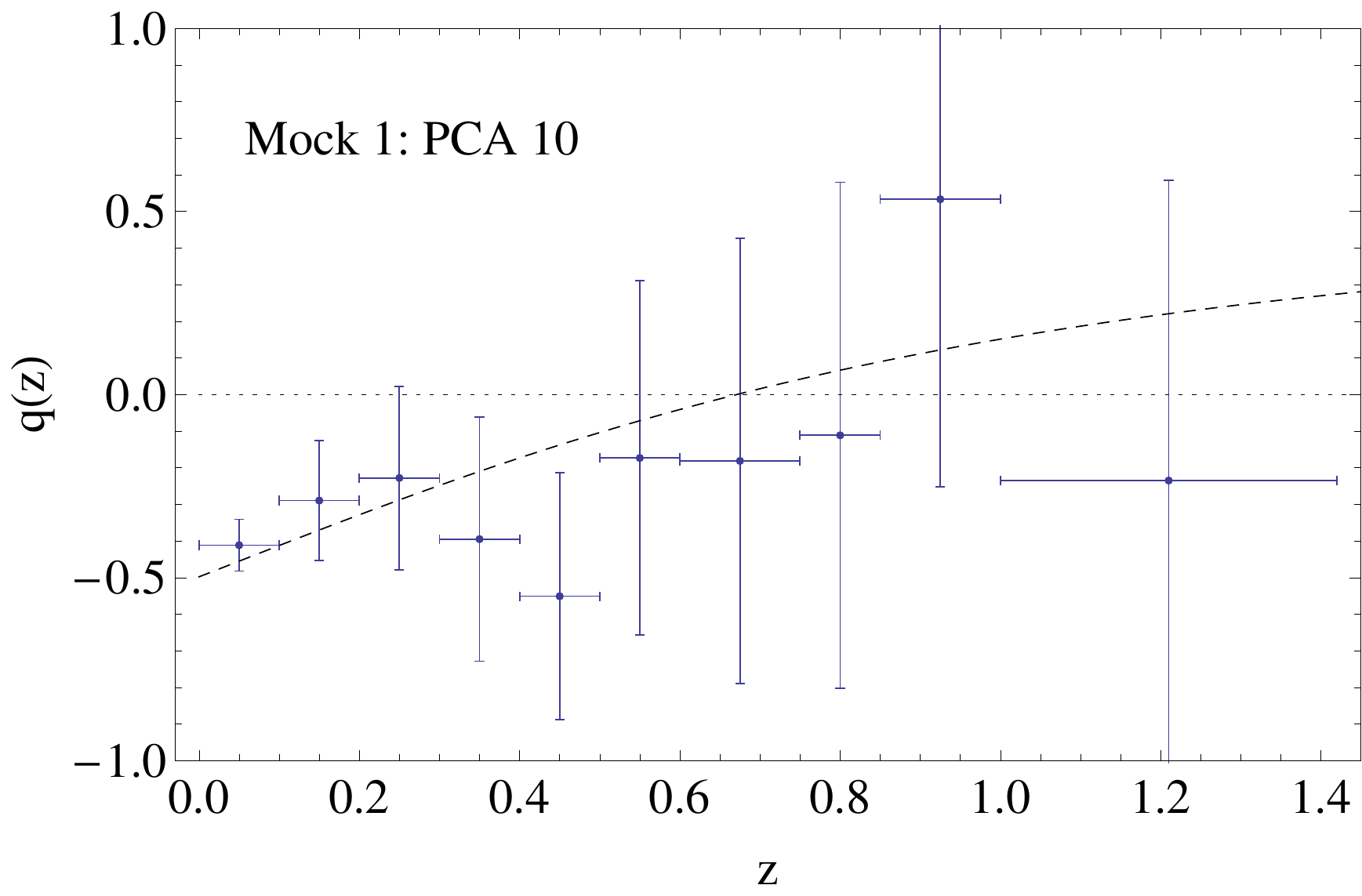}}}
\vspace{0cm}\rotatebox{0}{\vspace{0cm}\hspace{0cm}\resizebox{0.425\textwidth}{!}{\includegraphics{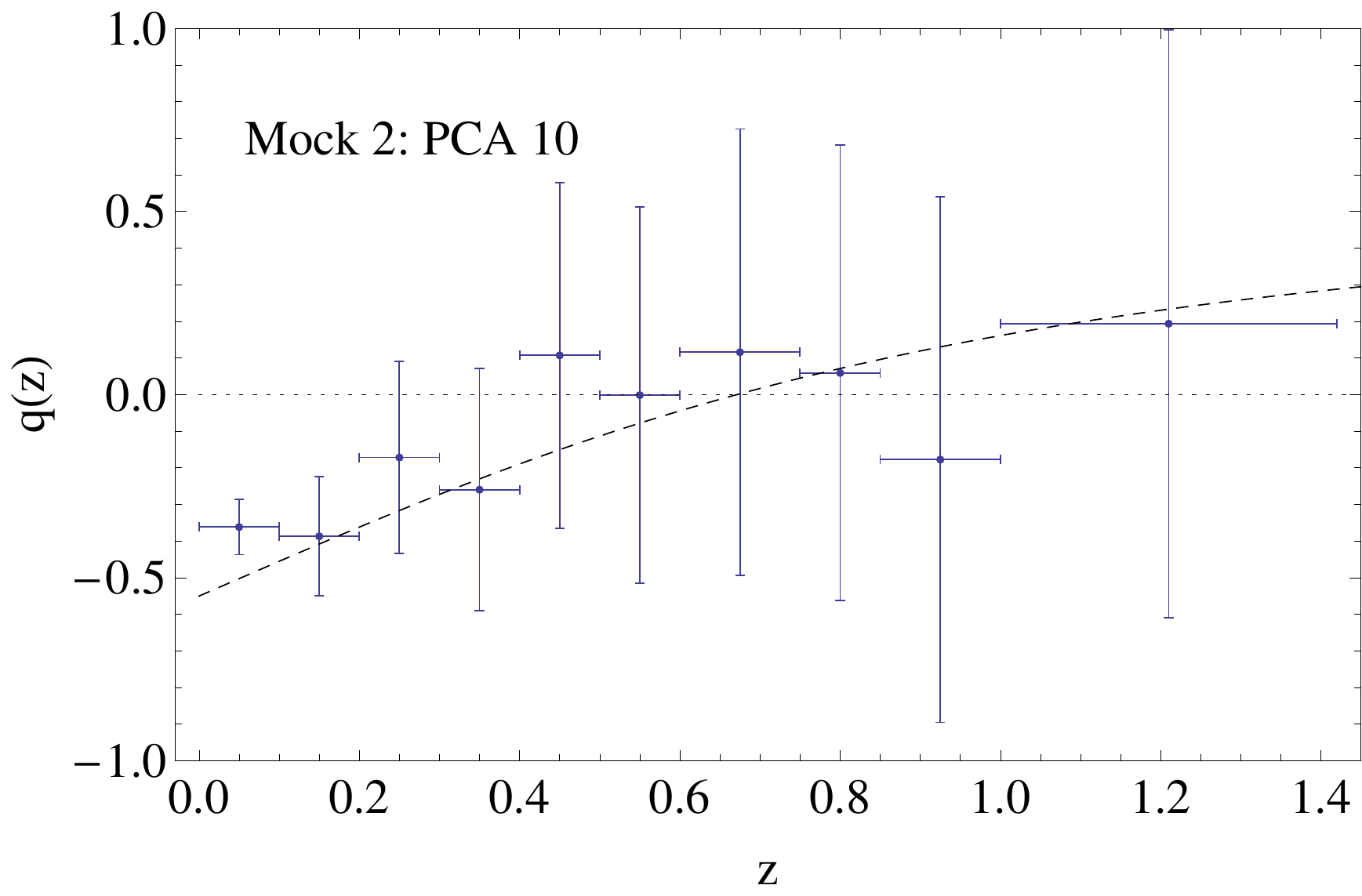}}}
\vspace{0cm}\rotatebox{0}{\vspace{0cm}\hspace{0cm}\resizebox{0.425\textwidth}{!}{\includegraphics{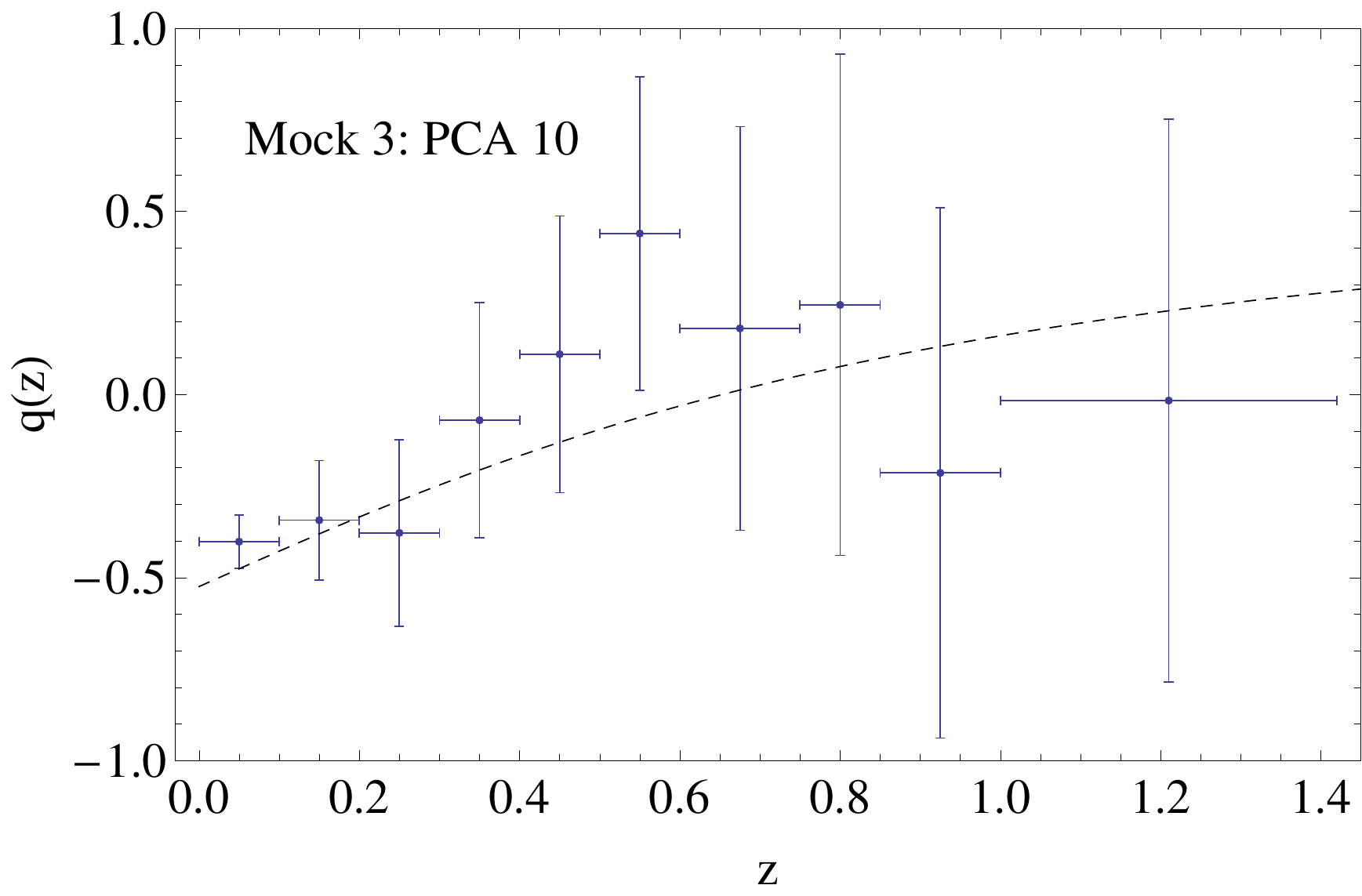}}}
\vspace{0cm}\rotatebox{0}{\vspace{0cm}\hspace{0cm}\resizebox{0.425\textwidth}{!}{\includegraphics{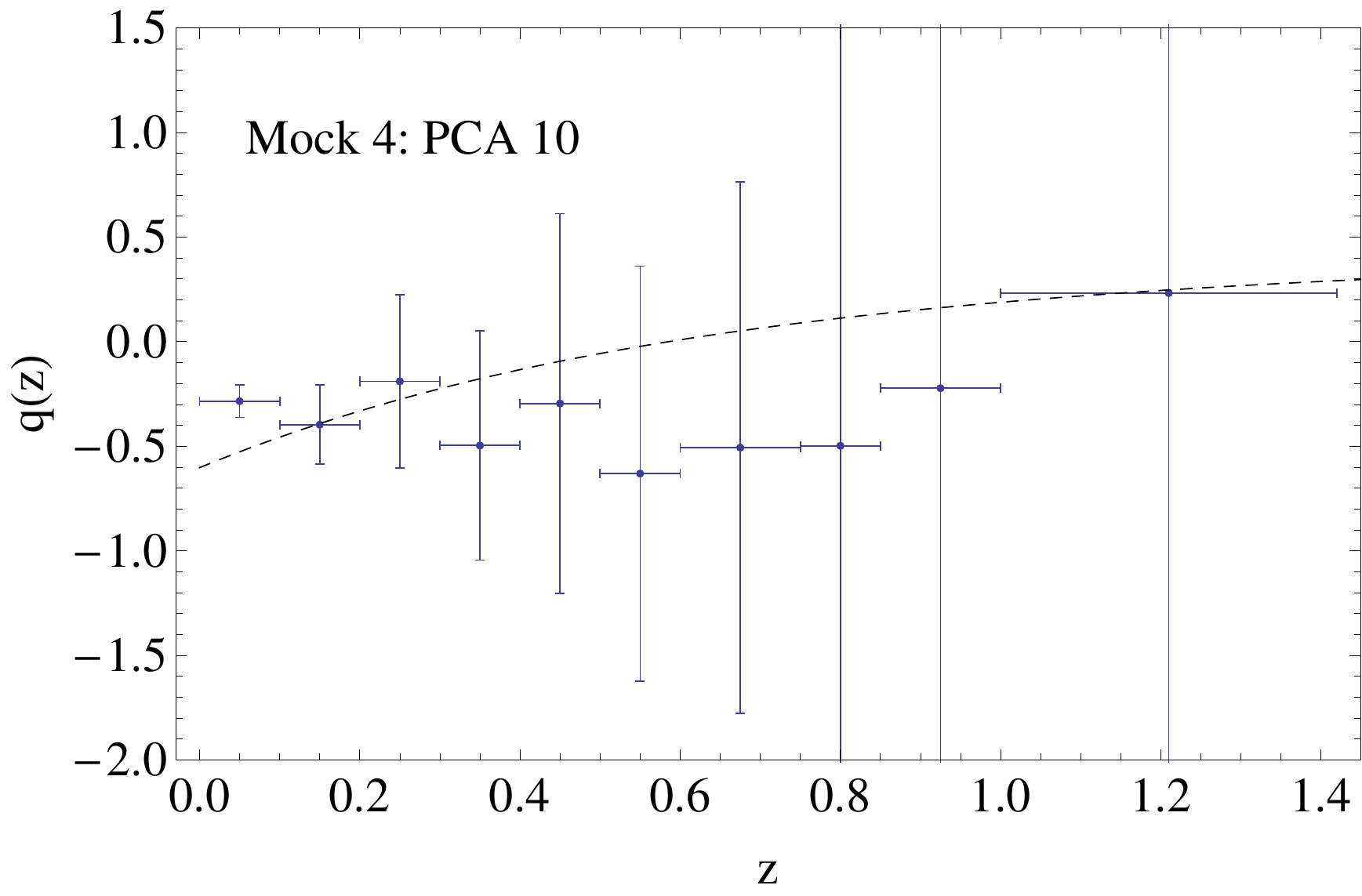}}}
\caption{The deceleration parameter $q(z)$ for all four mocks for ten bins. The dashed line corresponds to the real model.\label{qzPCA10}}
\end{figure*}

We use the expression for the luminosity distance of Eq.~(\ref{lumPCA}) in order to fit the four mock SnIa data for two different cases: for six and ten redshift bins. The bins we used were $z_6=(0,0.2,0.4,0.6,0.8,1,1.42)$ and $z_{10}=(0,0.1,0.2,0.3,0.4,0.5,0.6,0.75,0.85,1,1.42)$.

First, we determine the best fit parameters $q_n$ by implementing a Markov Chain Monte Carlo (MCMC) method, which has the added advantage of providing the best-fit and the covariance matrix at the same time, and then we use the PCA approach in order to uncorrelate the coefficients $q_n$. For the actual process to uncorrelate the parameters we follow Ref. \cite{Huterer:2004ch}. We diagonalize the Fisher matrix $F$, i.e. the inverse of the covariance matrix $C_{ij}=\langle (q_i-\langle q_i\rangle)(q_j-\langle q_j\rangle)\rangle$, by using an orthogonal matrix $W$ such that $F=W^T\Lambda W$, where $\Lambda$ is diagonal and contains the eigenvalues $\lambda_i$ of $F$. Then, we define $\widetilde{W}\equiv W^T \Lambda^{1/2} W=F^{1/2}$ and finally, we normalize $\widetilde{W}$ so that its rows sum to unity. With these definitions, the uncorrelated parameters $p_i$ are given by $p=\widetilde{W} q$, which can also be written as $p_i=\sum_{j=1}^M \widetilde{W}_{ij} q_j$ for $M=6$ or $M=10$ for the two different bins, and they each have a variance of $\sigma^2(p_i)=1/\lambda_i$ \cite{Huterer:2004ch}. In order to choose how many coefficients we will keep for each, we follow Ref. \cite{Amendola:2011qp} and we only keep these $N$ coefficients for which $\sigma_i^2\leq1$ or in other words we truncate the sum at the term $N\leq M$ or $p_i=\sum_{j=1}^M\widetilde{W}_{ij} q_j \hookrightarrow \sum_{j=1}^N \widetilde{W}_{ij} q_j$. Then we normalize the error such that $\sigma^2=1$ for the worst determined mode and $\sigma_i^2 \rightarrow\frac{\sigma_i^2}{1+\sigma_i^2}$ for the rest.

In Figs. \ref{qzPCA6} and \ref{qzPCA10} we show the deceleration parameter $q(z)$ for all four mocks for six and ten bins, respectively. The dashed line corresponds to the real models. As can be seen, in all cases the PCA prediction is relatively close to the real models; however, the errors become unacceptably large at $z>0.6$, signifying a failure of the method to give solid predictions at high redshifts by using the SnIa alone, in agreement with Ref. \cite{Amendola:2011qp}.

\subsection{Genetic algorithms}
\subsubsection{Brief introduction}
In what follows for the sake of completeness we will briefly introduce the Genetic algorithms (GA). For a more detailed description and the application of GAs to cosmology we refer the interested reader to Refs. \cite{Bogdanos:2009ib},\cite{Nesseris:2010ep} and \cite{Nesseris:2012tt}. The GAs are algorithms that are loosely based on the principles of biological evolution via natural selection, where a population of individuals evolves over a time period under the combined influence of two operators: the mutation (a random change in an individual) and the crossover (the combination of two or more different individuals). The probability or  ``reproductive success'' that an individual will produce offspring is proportional to the fitness of the individual. The fitness function in our case is taken to be a $\chi^2$, and it measures how accurately each individual describes the data.

The algorithm initializes with a population of individuals, which in our case are functions, randomly generated based on a predefined grammar of allowed basis, e.g. $\exp,\sin, \log$ etc., and the standard set of operations $+,-,\times,\div$. In each consecutive generation, the fitness for each individual of the population is evaluated and the genetic operations of mutation and crossover are applied. This process is iterated until certain termination criteria are reached, e.g. the maximum number of generations. To make the whole process more clear we will also summarize the various steps of the algorithm as follows:
\begin{enumerate}
    \item Start by generating an initial random population of functions $P(0)$ based on a predefined grammar.
    \item Calculate the fitness for all individuals in the population
    $P(t)$.
    \item Create the next generation $P(t+1)$ by choosing individuals from $P(t)$ to produce offsprings via crossover and mutation, but possibly also keeping a part of the previous generation $P(t)$.
    \item Repeat step 2 until a termination goal has been achieved, e.g. the maximum number of generations.
\end{enumerate}

We should point out that the initial population $P(0)$ depends solely on the choice of the grammar and the available operations and therefore it only affects how fast the GA converges to the best fit. Using a not optimal grammar may result in the algorithm not converging fast enough or being trapped in a local minimum. Also, two important factors that affect the convergence speed of the GA are the mutation rate and the selection rate. The selection rate is typically of the order of $10\% \sim 20\%$, and it determines the number of individuals that will be allowed to produce offspring. The mutation rate is usually much smaller, of the order of $5\% \sim 10\%$, and it expresses the probability that an arbitrary number of individuals will be changed. If either of the two rates is much larger than these values then the GA may not converge at all, while if the two rates are much smaller the GA will converge very slowly and will usually get stuck at some local minimum.

The difference between the GAs and the standard analysis of observational data, i.e. having an \textit{a priori} defined model with a number of free parameters, is that the later method introduces model-choice bias and in general models with more parameters
tend to give better fits to the data. Also, GAs have a definite advantage to the usual methods when the parameter space is too large, quite complex or not well enough understood, as is the case with DE. Finally, our goal is to minimize a function, in our case the $\chi^2$, not using some \textit{a priori} defined model, but through a stochastic process based on a GA evolution. In this way, no prior knowledge of a theoretical model is needed and our result will be completely parameter free.

In other words, the GA does not require us to choose some arbitrary DE model, but uses the data themselves to find this model. Also, it is parameter free as the end result does not have any free parameters, like $\om$ in the case of the usual DE models, that can be changed in order to fit the data. So, in this sense this method has far less bias than any of the other standard methods for the reconstruction of the expansion history of the Universe that we will mention later on. This is one of the main reasons for the use of the GAs in this paper. For more details on the genetic algorithms and their application in the analysis of cosmological data, see Refs. \cite{Bogdanos:2009ib}, \cite{Nesseris:2010ep} and \cite{Nesseris:2012tt}.

\subsubsection{Results and error estimates}
The error estimates of the best fit are calculated by implementing the ``path integral" approach first developed by the authors of Ref. \cite{Nesseris:2012tt}. Our likelihood functional is 
\be 
\mathcal{L}=\mathcal{N}\exp\left(-\chi^2(f)/2\right) \ee where $f(x)$ is the function to be determined by the genetic algorithm, $\chi^2(f)$ is the corresponding chi-squared for $N$ data points $(x_i, y_i, \sigma_i)$, defined as
\be \chi^2(f)\equiv \sum_{i=1}^N \left(\frac{y_i-f(x_i)}{\sigma_i}\right)^2 \label{chi2def}
\ee
Determining the normalization constant $\mathcal{N}$ is much more complicated than that of a normal distribution, as we have to integrate over all possible functions $f(x)$ or in other words perform a ``path integral"
\be 
\int \mathfrak{D}f~\mathcal{L}= \int \mathfrak{D}f~\mathcal{N}\exp\left(-\chi^2(f)/2\right)=1, \label{path1}
\ee 
where $\mathfrak{D}f$ indicates integration over all possible values of $f(x)$. The reason for this is that as the GA is running, it may consider any possible function no matter how bad a fit it represents, due to the mutation and crossover operators. Of course, even though these ``bad" fits will in the end be discarded, they definitely contribute in the total likelihood and have to be included in the calculation of the error estimates of the best fit.

The infinitesimal quantity $\mathfrak{D}f$ can be written as $\mathfrak{D}f=\prod_{i=1}^{N} df_i$, where $df_i$ and $f_i$ are assumed to mean $df(x_i)$ and $f(x_i)$ respectively, and we will for the time being assume that the function $f$ evaluated at a point $x_i$ is uncorrelated (independent) from that at a point $x_j$. Therefore, Eq. (\ref{path1}) can be recast as 
\ba 
\int \mathfrak{D}f~\mathcal{L}&=&\int_{-\infty}^{+\infty} \prod_{i=1}^{N}df_i~ \mathcal{N} \exp\left(-\frac{1}{2}\sum_{i=1}^N \left(\frac{y_i-f_i}{\sigma_i}\right)^2\right) \nn\\&=& \prod_{i=1}^{N}\int_{-\infty}^{+\infty}df_i~ \mathcal{N} \exp\left(-\frac{1}{2}\left(\frac{y_i-f_i}{\sigma_i}\right)^2\right) \nn\\ &=& \mathcal{N}\cdot\left(2 \pi \right)^{N/2} \prod_{i=1}^{N}\sigma_i \nn\\ &=&1 \nn\label{path2} \ea 
which means that $\mathcal{N}=\left(\left(2 \pi \right)^{N/2} \prod_{i=1}^{N}\sigma_i\right)^{-1}$. Therefore, the likelihood becomes 
\be 
\mathcal{L}=\frac{1}{\left(2 \pi \right)^{N/2} \prod_{i=1}^{N}\sigma_i}\exp\left(-\chi^2(f)/2\right) \ee or, if we take into account our assumption that the function $f$ evaluated at each point $x_i$ is independent, 
\be 
\mathcal{L}= \prod_{i=1}^{N}\mathcal{L}_i=\prod_{i=1}^{N} \frac{1}{\left(2 \pi \right)^{1/2}\sigma_i}\exp\left(-\frac{1}{2}\left(\frac{y_i-f_i}{\sigma_i}\right)^2 \right),\ee where  \be \mathcal{L}_i\equiv\frac{1}{\left(2 \pi \right)^{1/2}\sigma_i}\exp\left(-\frac{1}{2}\left(\frac{y_i-f_i}{\sigma_i}\right)^2 \right). \ee

We can calculate the $1\sigma$ error $\delta f_i$ around the best-fit $f_{bf}(x)$ at a point $x_i$ as 
\bea 
CI(x_i,\delta f_i)&=&\int_{f_{bf}(x_i)-\delta f_i}^{f_{bf}(x_i)+\delta f_i} df_i \frac{1}{\left(2 \pi \right)^{1/2}\sigma_i}\exp\left(-\frac{1}{2}\left(\frac{y_i-f_i}{\sigma_i}\right)^2 \right)\nn\\&=&\frac{1}{2}\left( \erf \left(\frac{\delta f_i+f_{bf}(x_i)-y_i}{\sqrt{2}\sigma_i}\right)+\erf \left(\frac{\delta f_i-f_{bf}(x_i)+y_i}{\sqrt{2}\sigma_i}\right)\right).\label{conf1}
\eea 
If we demand that the errors $\delta f_i$ correspond to the $1\sigma$ error of a normal distribution,  then from Eq.~(\ref{conf1}) we can solve the following equation for $\delta f_i$ numerically, 
\be 
CI(x_i,\delta f_i)=\erf\left(1/\sqrt{2}\right), 
\ee 
and therefore determine the $1\sigma$ error $\delta f_i$ of the best-fit function $f_{bf}(x)$ at each point $x_i$. However, this will lead to knowledge of the error in specific points $x_i$, which is not ideal for our purpose, which is to have a smooth, continuous and differentiable function. Therefore, we will create a new chi-square defined as 
\be 
\chi^2_{CI}(\delta f_i)=\sum_{i=1}^N \left(CI(x_i,\delta f_i)-\erf\left(1/\sqrt{2}\right)\right)^2 \label{chi2ci}
\ee 
and we will also parametrize $\delta f$ with a second-order polynomial $\delta f(x)=a+b x+c x^2$. Finally, we minimize the combined chi-squared $\chi^2(f_{bf}+\delta f)+ \chi^2_{CI}(\delta f)$ for the parameters $(a,b,c)$, where $\chi^2$ is given by Eq.~(\ref{chi2def}) and $\chi^2_{CI}$ is given by Eq.~(\ref{chi2ci}). Then, the $1\sigma$ region for the best-fit function $f_{bf}(x)$ will be contained within the region $[f_{bf}(x)-\delta f(x),f_{bf}(x)+\delta f(x)]$. For more details on the path integral approach to error estimation and the case of correlated data, see Ref. \cite{Nesseris:2012tt}.

\begin{figure*}[t!]
\centering
\vspace{0cm}\rotatebox{0}{\vspace{0cm}\hspace{0cm}\resizebox{0.43\textwidth}{!}{\includegraphics{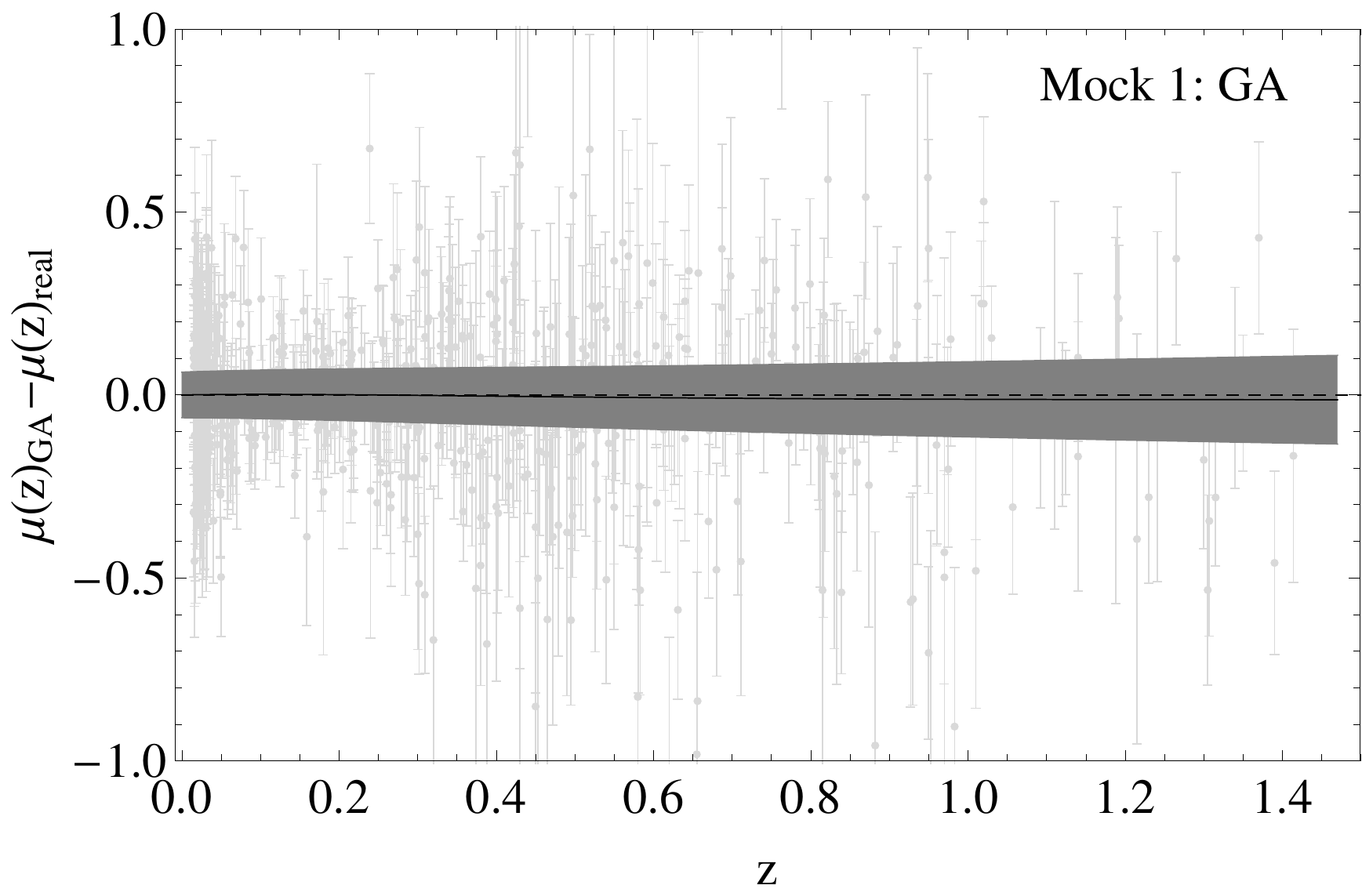}}}
\vspace{0cm}\rotatebox{0}{\vspace{0cm}\hspace{0cm}\resizebox{0.43\textwidth}{!}{\includegraphics{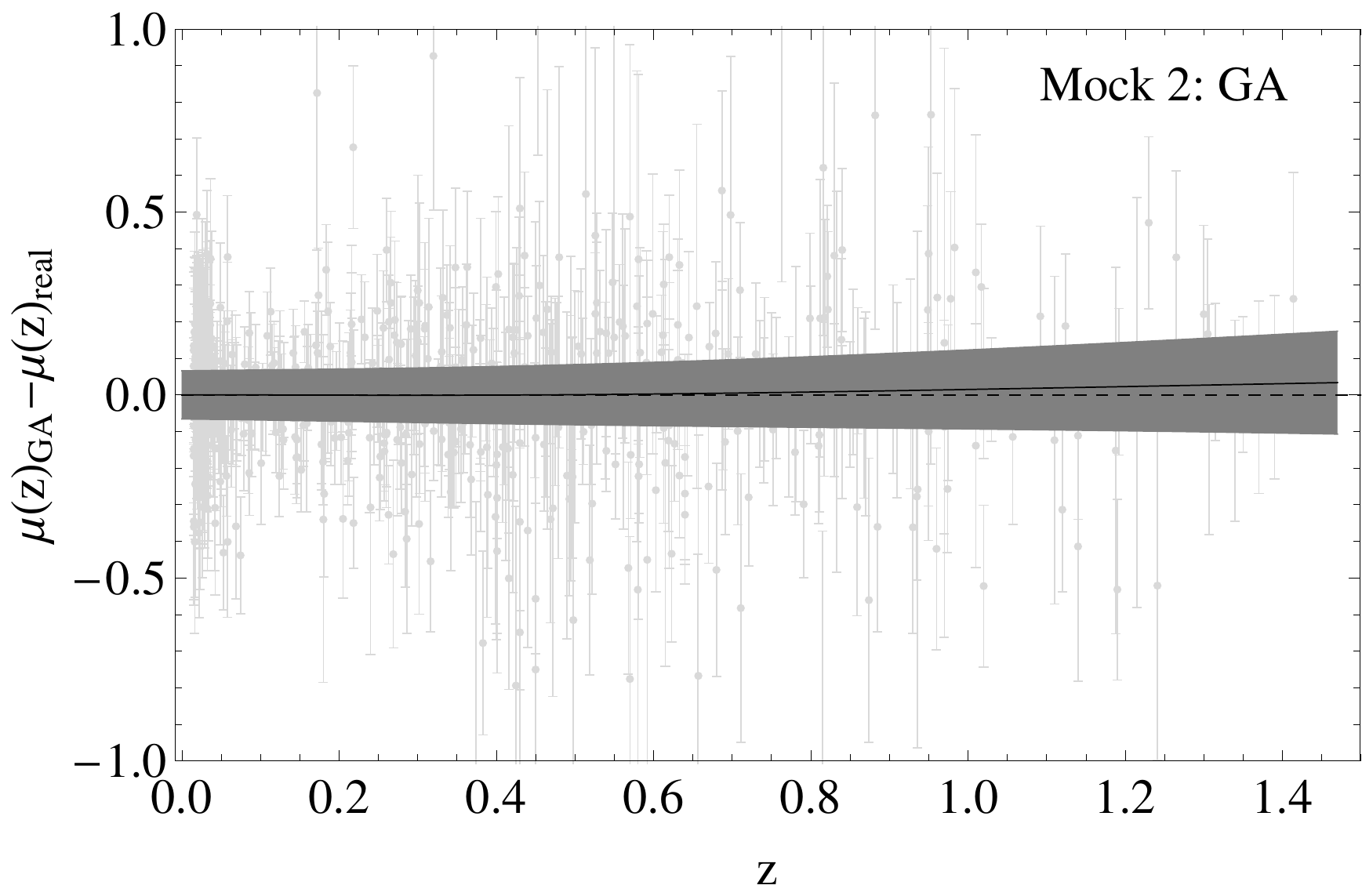}}}
\vspace{0cm}\rotatebox{0}{\vspace{0cm}\hspace{0cm}\resizebox{0.43\textwidth}{!}{\includegraphics{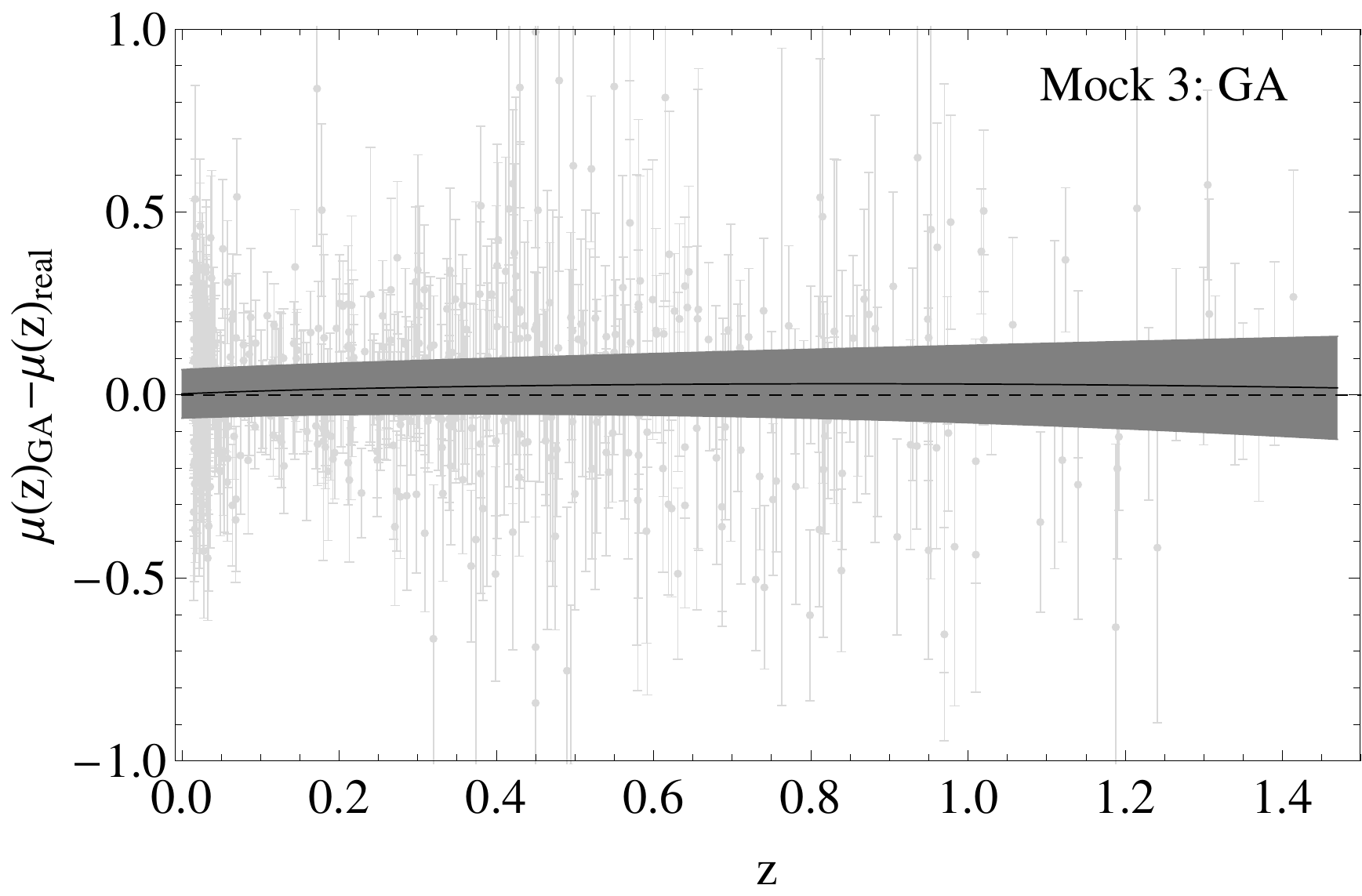}}}
\vspace{0cm}\rotatebox{0}{\vspace{0cm}\hspace{0cm}\resizebox{0.43\textwidth}{!}{\includegraphics{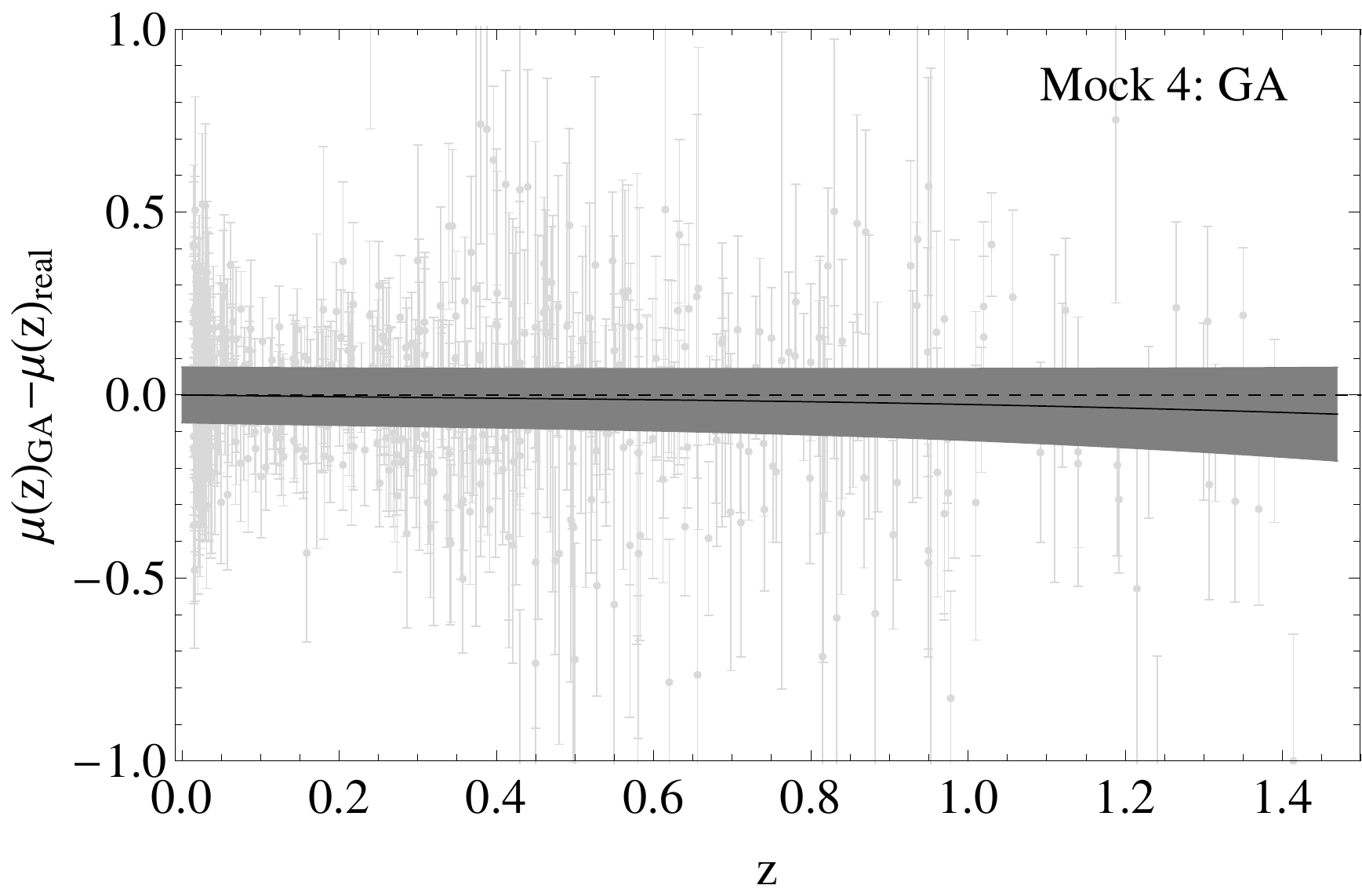}}}
\caption{The residues $\mu_{GA}(z)-\mu_{real}(z)$ for all four mocks.\label{dmuGA}}
\vspace{1.5cm}
\centering
\vspace{0cm}\rotatebox{0}{\vspace{0cm}\hspace{0cm}\resizebox{0.43\textwidth}{!}{\includegraphics{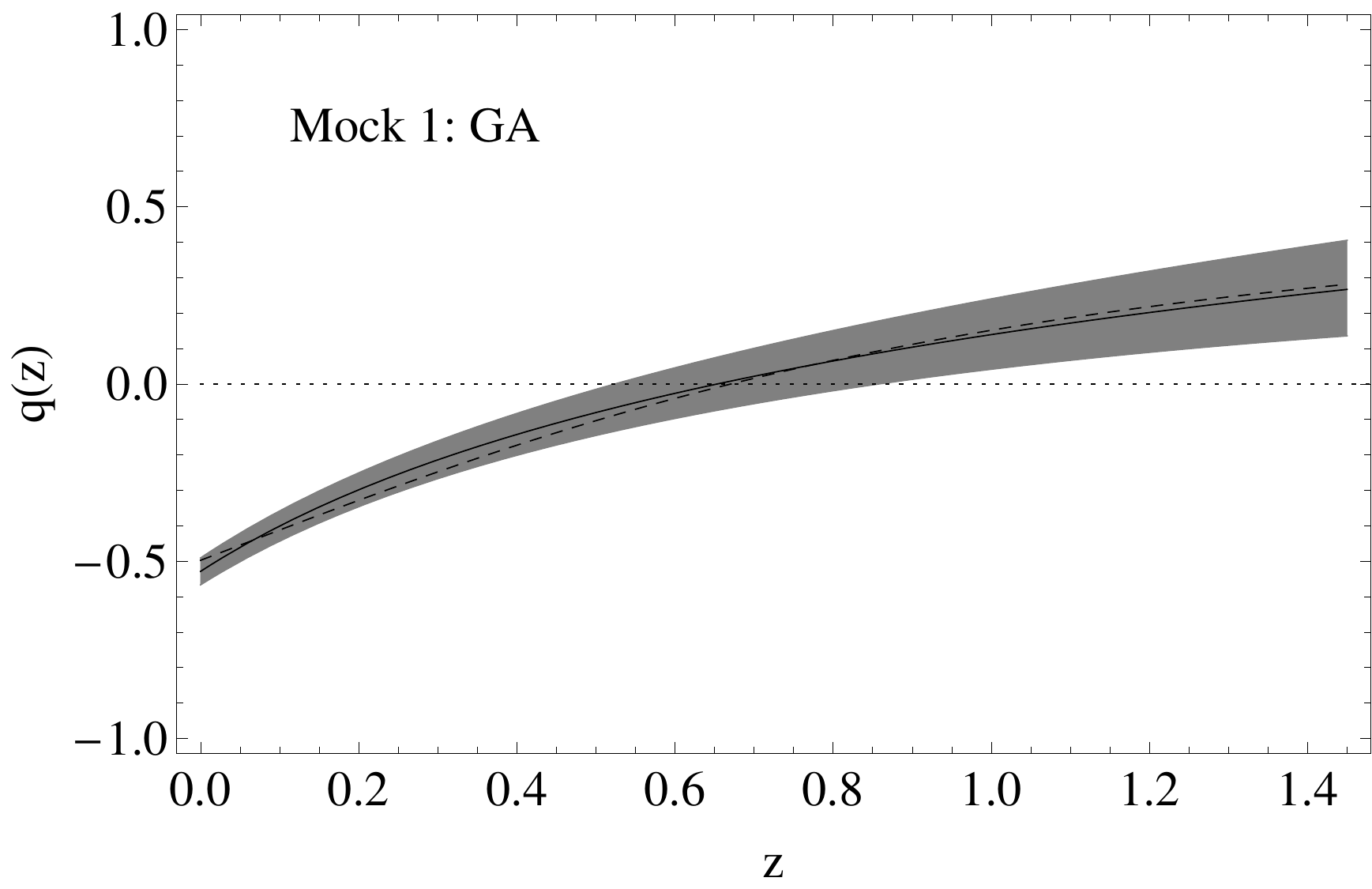}}}
\vspace{0cm}\rotatebox{0}{\vspace{0cm}\hspace{0cm}\resizebox{0.43\textwidth}{!}{\includegraphics{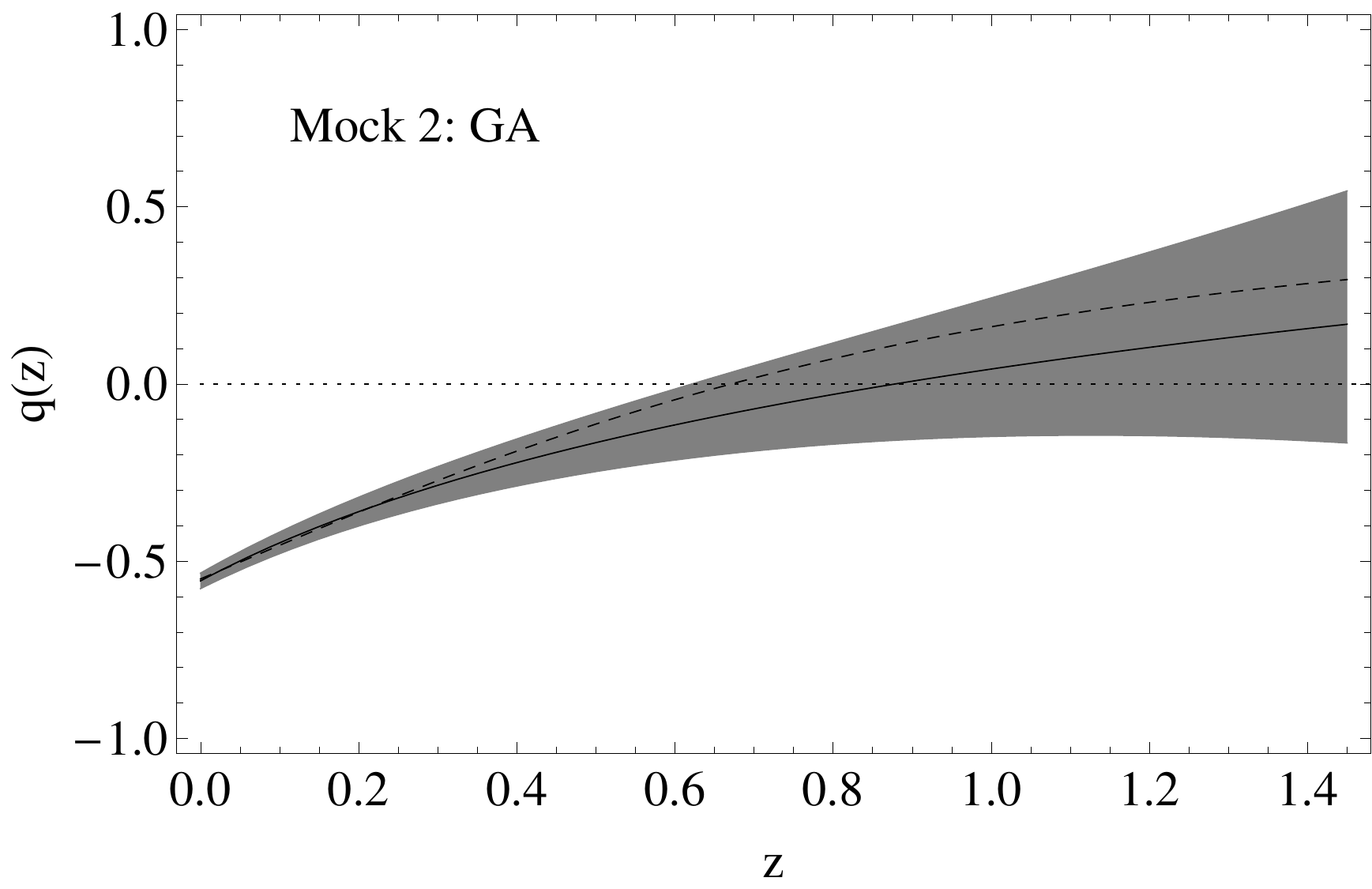}}}
\vspace{0cm}\rotatebox{0}{\vspace{0cm}\hspace{0cm}\resizebox{0.43\textwidth}{!}{\includegraphics{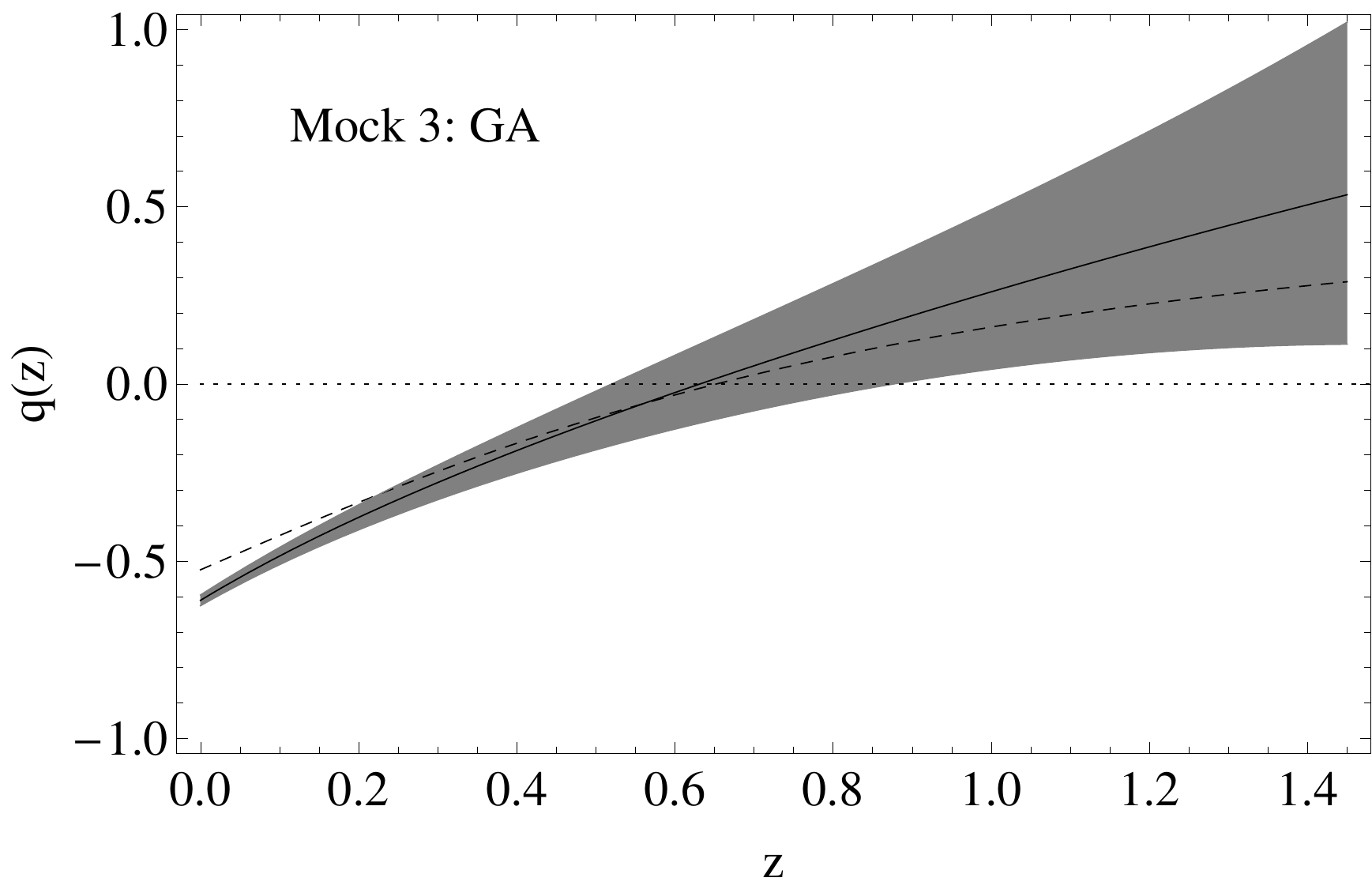}}}
\vspace{0cm}\rotatebox{0}{\vspace{0cm}\hspace{0cm}\resizebox{0.43\textwidth}{!}{\includegraphics{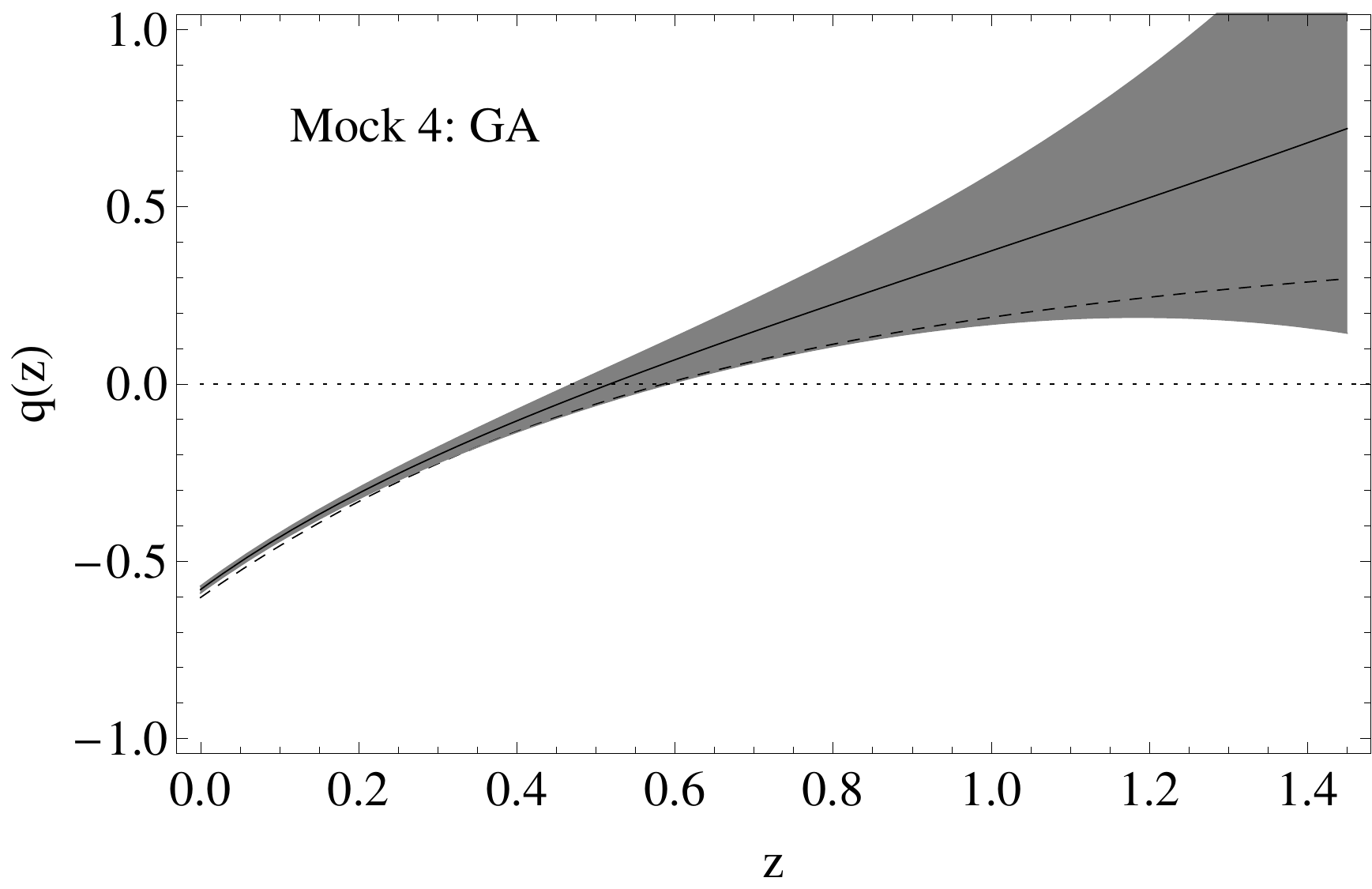}}}
\caption{The deceleration parameter $q(z)$ for all four mocks. The dashed line corresponds to the real model.\label{qzGA}}
\end{figure*}

In the present analysis, the actual fitting of the data was done with a modified version of the GDF v2.0 C++ program\footnote{Freely available at http://cpc.cs.qub.ac.uk/summaries/ADXC} developed by I. Tsoulos et al. \cite{Tsoulos} and a \textsc{Mathematica} code written by one of the authors.\footnote{Freely available at http://www.uam.es/savvas.nesseris/codes.html} The residues $\mu_{GA}(z)-\mu_{real}(z)$ for all four mocks can be seen in Fig. \ref{dmuGA}, while the deceleration parameter $q(z)$ for all four mocks can be seen in Fig. \ref{qzGA}. Clearly in all cases the GAs achieve very good agreement with the real models.

\subsection{Pad\'e approximants for $q(z)$}
\subsubsection{Linear}
If we expand $q(z)$ in Taylor series around $z=0$ for a $(w_0,w_a)$ flat \lcdm model, we find
$1+q(z) = \sum_{n=0}^\infty q_n z^n$, with
\bea\label{q0123}
q_0  &\!=&\!   \frac{3}{2}\Big(1+w_0(1-\om)\Big)\,,\nonumber\\
q_1  &\!=&\!   \frac{3}{2}(1-\om)\Big(3w_0^2\om + w_a\Big)\,,\nonumber\\
q_2  &\!=&\!   \frac{3}{4}(1-\om)\Big(18w_0^3\om^2-3w_0\om(3w_0^2+w_0-3w_a)-2w_a\Big)\,,\\
q_3  &\!=&\!   \frac{3}{4}(1-\om)\Big(54w_0^4\om^3-18w_0^2\om^2(3w_0^2+w_0-2w_a)+\nonumber\\
&&\hspace{2cm} \om(9w_0^4+9w_0^3+2w_0^2(1-9w_a)-13w_0w_a+3w_a^2) + 2w_a\Big)\,.\nonumber\\
q_4  &\!=&\!  \dots \nonumber
\eea

The problem is that this series converges very slowly, and the fourth-order expansion is not enough for describing the deceleration parameter in the whole range of observations, $z\in[0,1.5]$. A possibility worth exploring is to produce such a series and then find the Pad\'e approximant of order $(m,n)$ that better fits the data in the whole range. Then we can write
\be
1+q(z) = \frac{\prod_{i=1}^m (c_i + z)}{\prod_{j=1}^n (d_j + z)}\,,
\ee
for certain $\{c_i,\,d_j\}$. The rate of expansion can then be integrated via Eq.~(\ref{Hqz}), and from there the luminosity distance can be obtained with Eq.~(\ref{dLH}).

An alternative (and equivalent) way is to assume the deceleration parameter can be written as
\be\label{qPade}
1+q(z) = \sum_{i=1}^n \left(\frac{a_i + z}{b_j + z}\right)\,,
\ee
which can be integrated to give
\be\label{HPade}
H(z) = \frac{H_0}{\alpha}\,(1+z)^\beta\,\prod_{i=1}^n(b_i + z)^{\alpha_i}\,,
\ee
where
\be\label{abi}
\alpha_i = \frac{a_i - b_i}{1-b_i}\,,\hspace{1cm}
\beta_i = \frac{1- a_i}{1-b_i} = 1 - \alpha_i\,,\hspace{1cm}
\alpha = \prod_{i=1}^n b_i^{\alpha_i} \,,\hspace{1cm}
\beta = \sum_{i=1}^n \beta_i\,,
\ee
which itself can be integrated to give $d_L(z)$ via Eq.~(\ref{dLH}),
\be
\frac{H_0 d_L(z)}{1+z} =  \int_0^z \frac{\alpha\,ds\,(1+s)^{-\beta}}{\prod_{i=1}^n(b_i + s)^{\alpha_i}}\,.
\ee
For example, if we use just four parameters (n=2), we have the exact solution
\ba\label{HdL}
\frac{H_0 d_L(z)}{1+z} &\!=&\!   \frac{b_1^{\alpha_1}b_2^{\alpha_2}}
{(\alpha_1+\alpha_2-1)(b_1-1)^{\alpha_1}(b_2-1)^{\alpha_2}}\times \nn \\
&&  \left[ (1+z)^{\alpha_1+\alpha_2-1}\,
AF_1\left(\alpha_1+\alpha_2-1,\,\alpha_1,\,\alpha_2,\,\alpha_1+\alpha_2;
\,\frac{1+z}{1-b_1},\,\frac{1+z}{1-b_2}\right) - \right.\,, \nn \\
&& \hspace{2cm} \left.  AF_1\left(\alpha_1+\alpha_2-1,\,\alpha_1,\,\alpha_2,\,\alpha_1+\alpha_2;
\,\frac{1}{1-b_1},\,\frac{1}{1-b_2}\right)\right]\,,
\ea
where we have used the Abell hypergeometric function of two variables,
\be
AF_1\Big(a,\,b_1,\,b_2,\,c\,;\,x,\,y\Big) = \sum_{m,n=0}^\infty
\frac{(a)_{m+n}(b_1)_m(b_2)_n}{(c)_{m+n}\,m!\,n!}\ x^m\, y^n\,,
\ee
with $(q)_n\equiv\Gamma(q+n)/\Gamma(q)=q(q+1)\dots(q+n-1)$.

Then, we could fit the observations of the luminosity distances of SnIa to function (\ref{HdL}) and obtain, from the fit, the parameters $\{\alpha_i,\,b_i\}$, deduce the $a_i = \alpha_i + b_i(1-\alpha_i)$, and then write directly the deceleration parameter (\ref{qPade}). We have checked with explicit examples that this procedure is convergent and gives quite good fits to cosmological parameters. For example, from $\{a_i,\,b_i\}$ we can obtain $\{\om,\,w_0,\,w_a\}$ using (\ref{q0123}), and from there we deduce $w(z)$ and $H(z)$.

Inverting (\ref{q0123}) we find
\bea
\om &\!=&\! 1 - \frac{2q_0-3}{3w_0}\,, \nonumber\\
\frac{w_a}{w_0} &\!=&\! \frac{2q_1}{2q_0-3} +2q_0- 3(1+w_0)\,, \nonumber\\
w_0 &\!=&\! \frac{1}{6(2q_0-3)}\Big(6q_0^2-17q_0+3q_1+12 \pm \nonumber\\
&& \hspace{-3mm}
\sqrt{4q_0^4-28q_0^3+q_0^2(73-12q_1)+2q_0(13q_1-8q_2-42)+9q_1^2-12q_1+24q_2+36}\Big)\,,
\eea
where we can then use the expressions for $q_i$ from the fit to (\ref{HdL})
\be
q_0 = \frac{a_1}{b_1} + \frac{a_2}{b_2}\,, \hspace{1cm}
q_n = (-1)^n\left(\frac{a_1-b_1}{b_1^{n+1}} + \frac{a_2-b_2}{b_2^{n+1}}\right)\,,
 \hspace{5mm}  {\rm for}  \hspace{5mm} n\geq 1\,.
\ee

\subsubsection{Power law with fixed exponent}
\begin{figure*}[t!]
\centering
\vspace{0cm}\rotatebox{0}{\vspace{0cm}\hspace{0cm}\resizebox{0.425\textwidth}{!}{\includegraphics{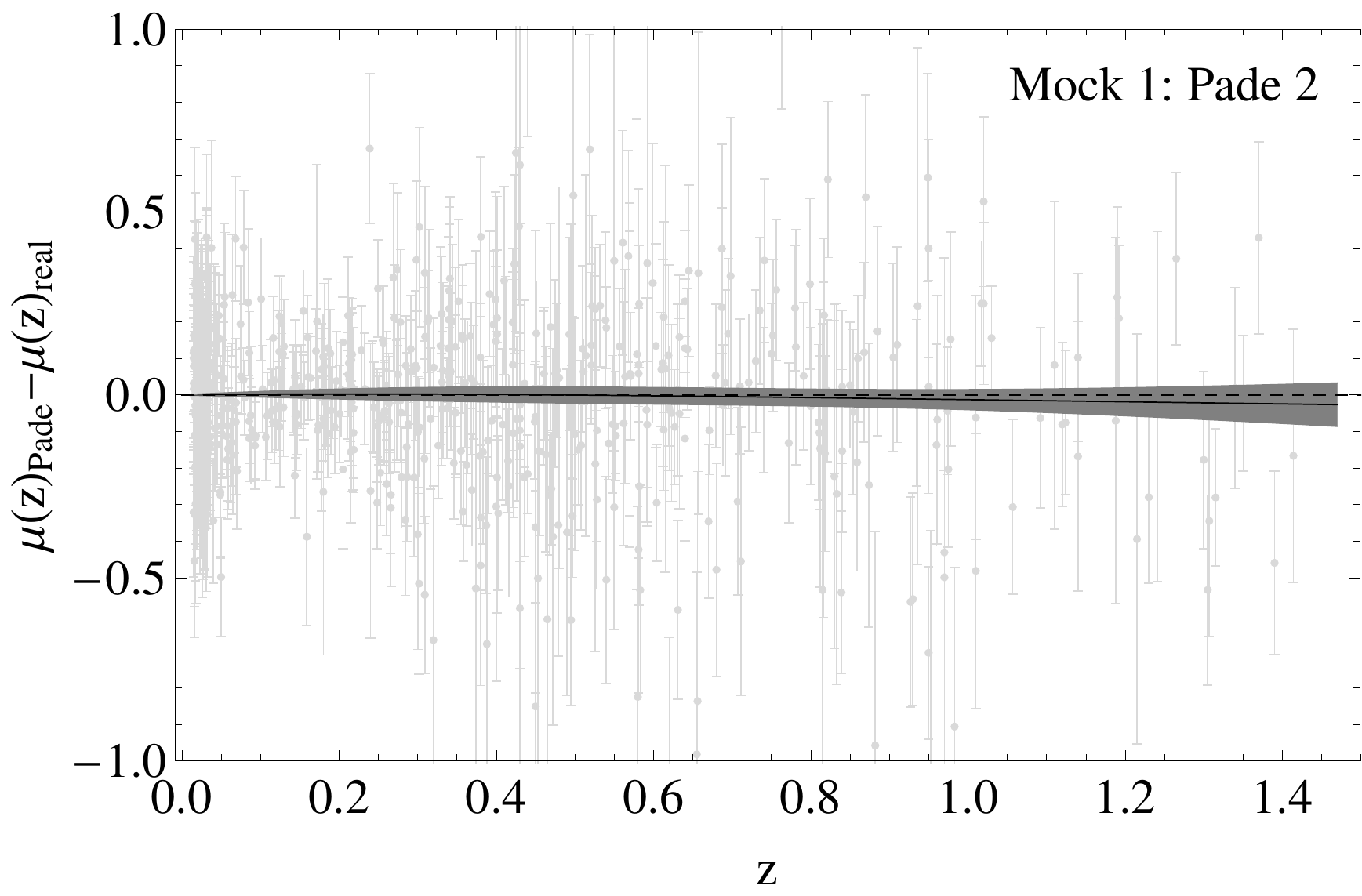}}}
\vspace{0cm}\rotatebox{0}{\vspace{0cm}\hspace{0cm}\resizebox{0.425\textwidth}{!}{\includegraphics{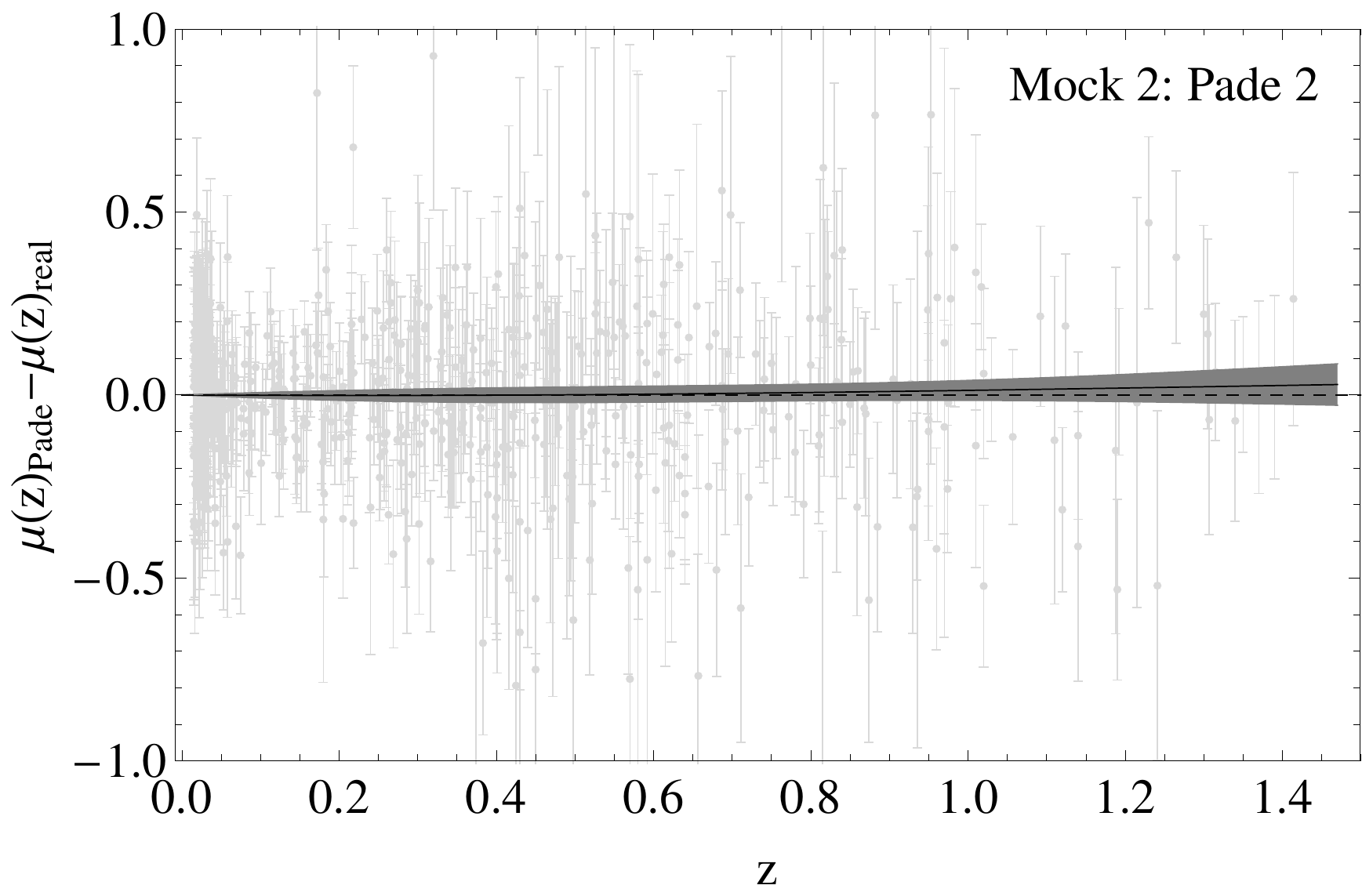}}}
\vspace{0cm}\rotatebox{0}{\vspace{0cm}\hspace{0cm}\resizebox{0.425\textwidth}{!}{\includegraphics{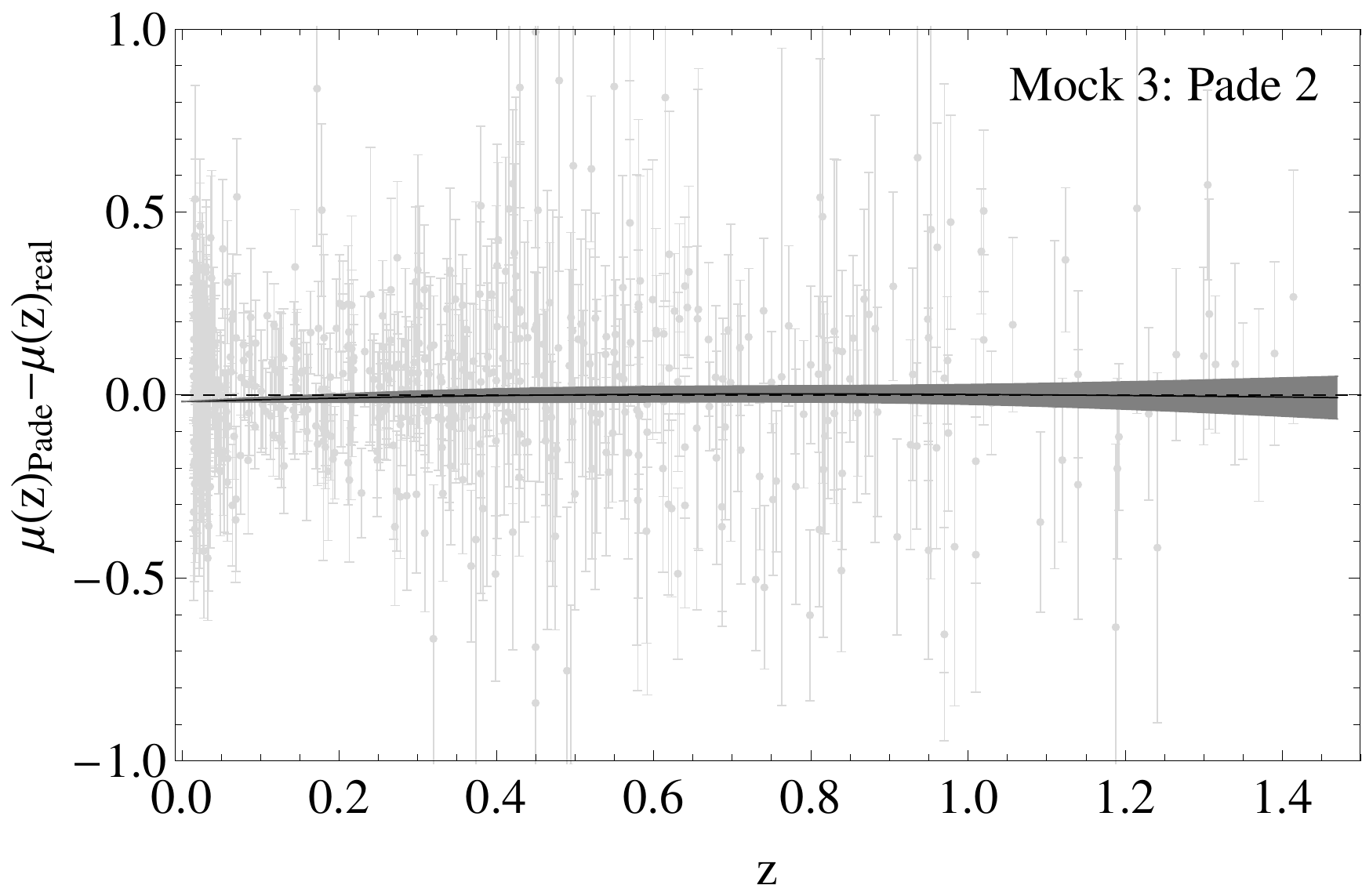}}}
\vspace{0cm}\rotatebox{0}{\vspace{0cm}\hspace{0cm}\resizebox{0.425\textwidth}{!}{\includegraphics{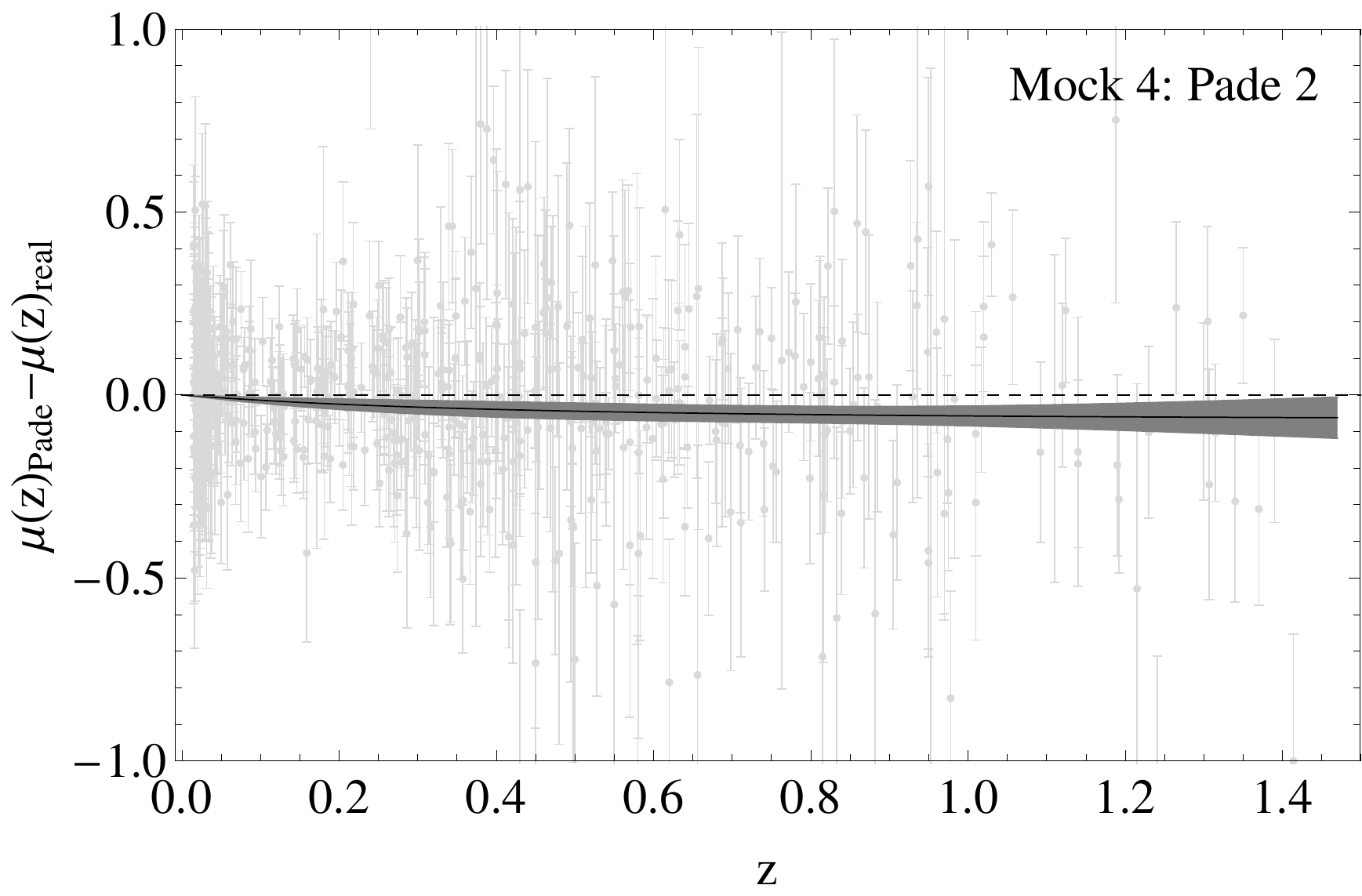}}}
\caption{The residues $\mu_{Pade}(z)-\mu_{real}(z)$ for all four mocks.\label{dmuPade2}}
\vspace{1.5cm}
\centering
\vspace{0cm}\rotatebox{0}{\vspace{0cm}\hspace{0cm}\resizebox{0.425\textwidth}{!}{\includegraphics{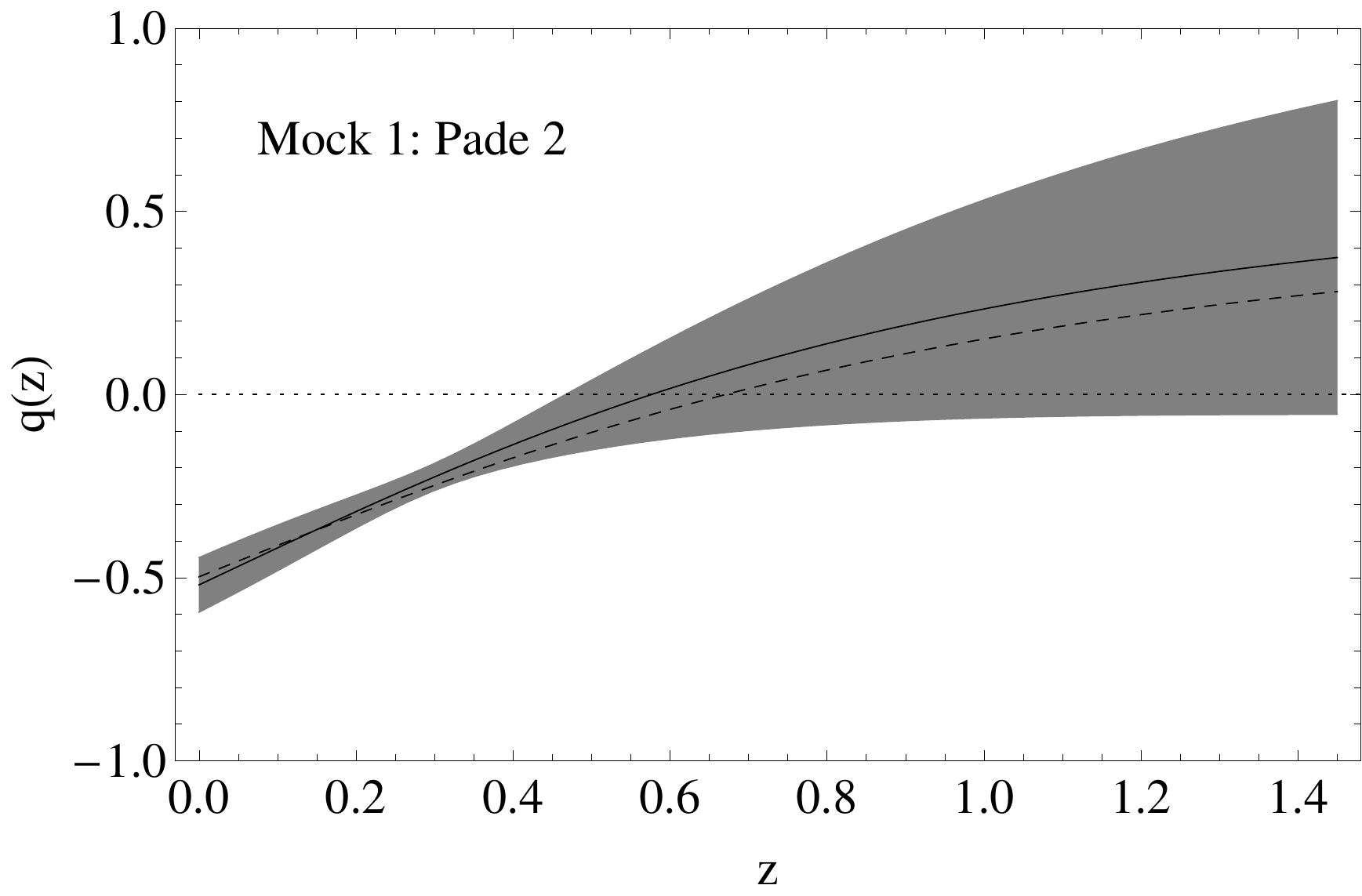}}}
\vspace{0cm}\rotatebox{0}{\vspace{0cm}\hspace{0cm}\resizebox{0.425\textwidth}{!}{\includegraphics{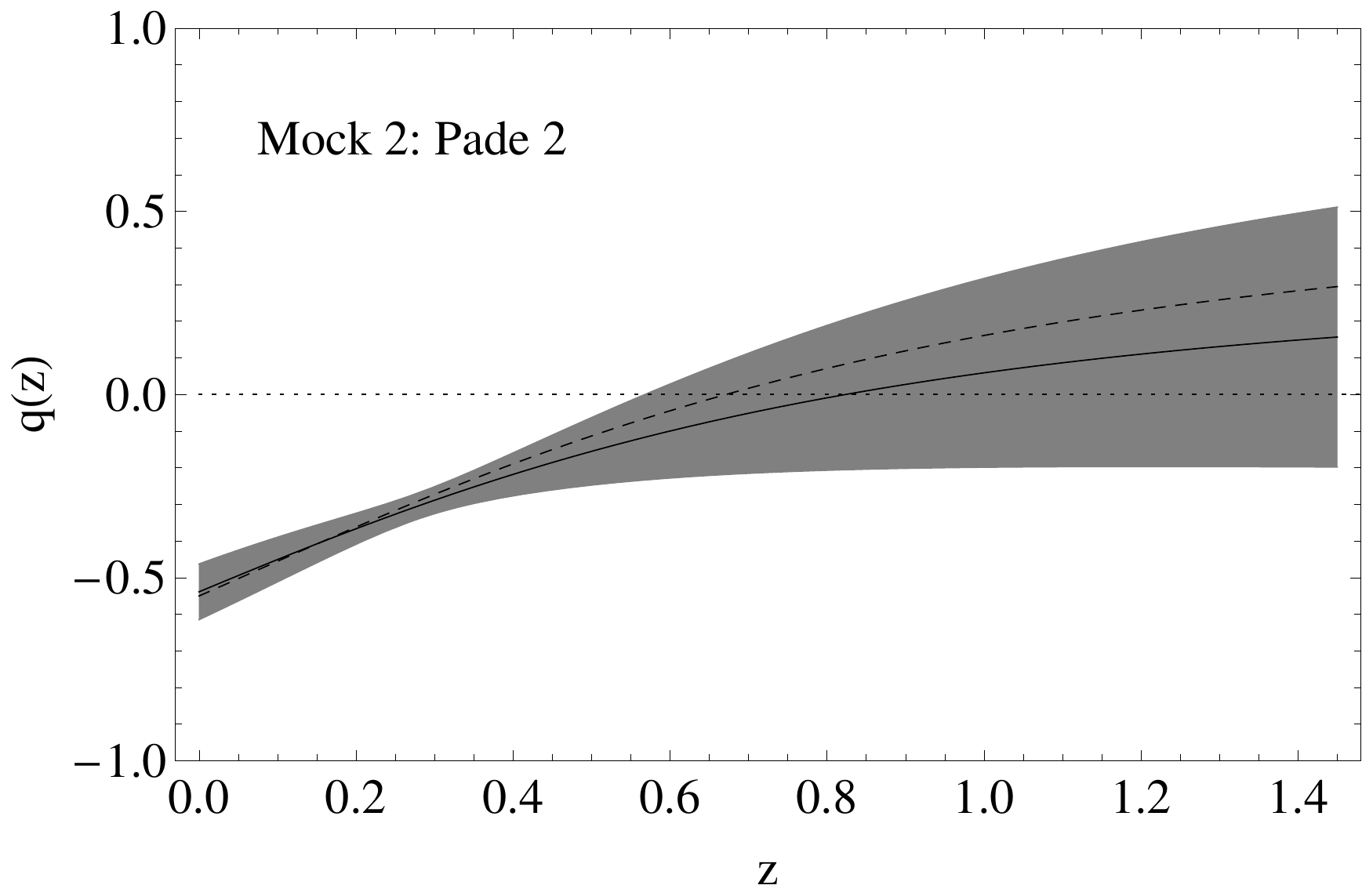}}}
\vspace{0cm}\rotatebox{0}{\vspace{0cm}\hspace{0cm}\resizebox{0.425\textwidth}{!}{\includegraphics{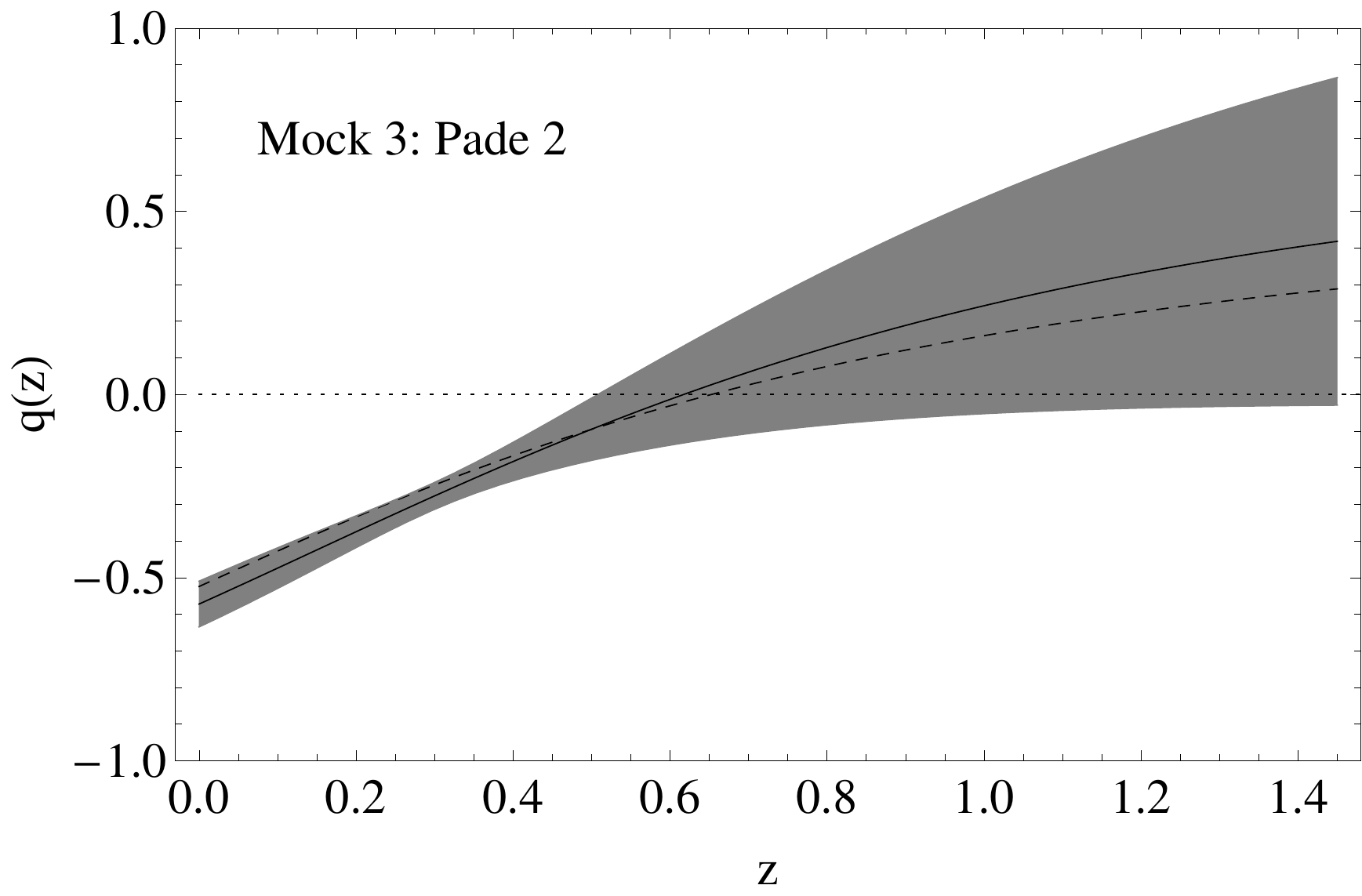}}}
\vspace{0cm}\rotatebox{0}{\vspace{0cm}\hspace{0cm}\resizebox{0.425\textwidth}{!}{\includegraphics{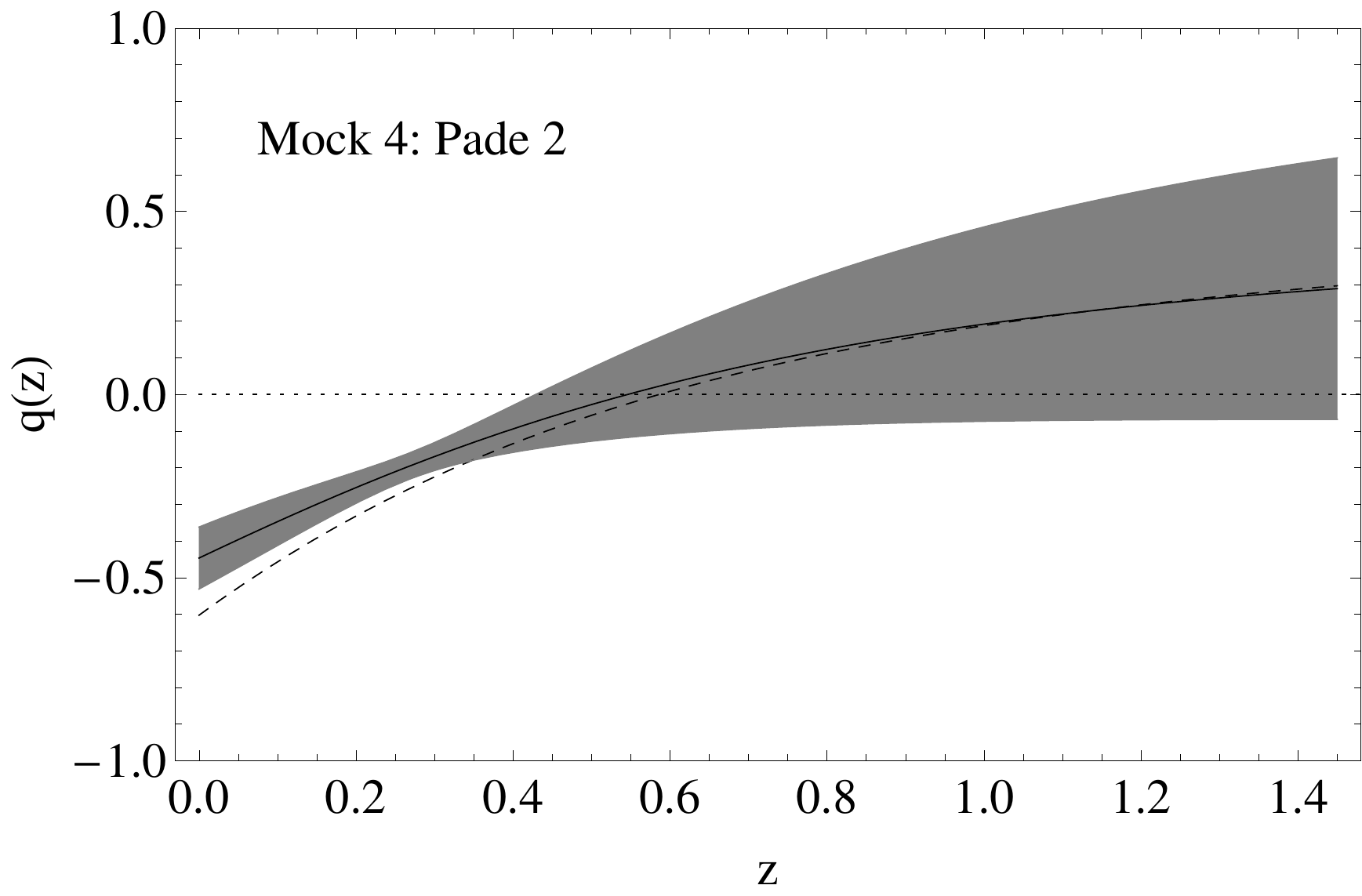}}}
\caption{The deceleration parameter $q(z)$ for all four mocks. The dashed line corresponds to the real model.\label{qzPade2}}
\end{figure*}
Even though the Pad\'e approximants mentioned previously work quite well, we found that a power-law approximant with fixed or variable exponents work even better. We will consider the former in this section and the latter in the next.

We can model the deceleration parameter with a two-parameter approximant with parameters $(a,b)$ as follows,
\be\label{qzp0}
q(z) = \frac{a(1+z)^3 - 1}{b(1+z)^3 +1}.
\ee
Then the Hubble parameter can be found to be
\be
H(z)=H_0 \left(\frac{1+b (z+1)^3}{1+b}\right)^{\frac{a+b}{3 b}},
\ee
from which we can find the luminosity distance as
\be
H_0 d_L(z)=(z+1) (1+b)^{\frac{a+b}{3 b}}\left((1+z) \, {}_2F_1\left[\frac{a+b}{3 b},\frac{1}{3};\frac{4}{3};-b (z+1)^3\right]-
{}_2F_1\left[\frac{a+b}{3 b},\frac{1}{3};\frac{4}{3};-b\right]\right)\,,
\ee
where ${}_2F_1[a,b,c;z]$ is the Gauss hypergeometric function. It should be noted that we can recover the \lcdm model in the limit $(a,b)=\left(\om/2\ol,\om/\ol\right)$. Also, it is easy to see from Eq.~(\ref{qzp0}) that the parameters $(a,b)$ can also be written in terms of the physically meaningful parameters $q_0\equiv q(z=0)=\frac{a-1}{b+1}$ and $q_\infty \equiv q(z\rightarrow\infty)=\frac{a}{b}$, as
\be
a = \frac{q_\infty(1+q_0)}{q_\infty-q_0} \,, \hspace{1.5cm}
b = \frac{1+q_0}{q_\infty -q_0}\,,
\ee
which gives a simple expression for
\be
q(z) = \frac{q_\infty(1+q_0)(1+z)^3+q_0-q_\infty}{(1+q_0)(1+z)^3+q_\infty-q_0}\,,
\ee
from which we recover the \lcdm result with $q_0=\frac{3}{2}\om - 1$ and $q_\infty = \frac{1}{2}$.

In Figs. \ref{dmuPade2} and \ref{qzPade2} we show the residues $\mu_{Pade}(z)-\mu_{real}(z)$ for all four mocks and the deceleration parameter $q(z)$ for all four mocks respectively. The dashed line corresponds to the real models and we have labeled this method as Pad\'e 2 in order to discriminate it from the simple linear Pad\'e mentioned earlier and the version with the variable exponent  variant we will mention later.

\subsubsection{Power law with variable exponent}
\begin{figure}[t!]
\centering
\vspace{0cm}\rotatebox{0}{\vspace{0cm}\hspace{0cm}\resizebox{0.45\textwidth}{!}{\includegraphics{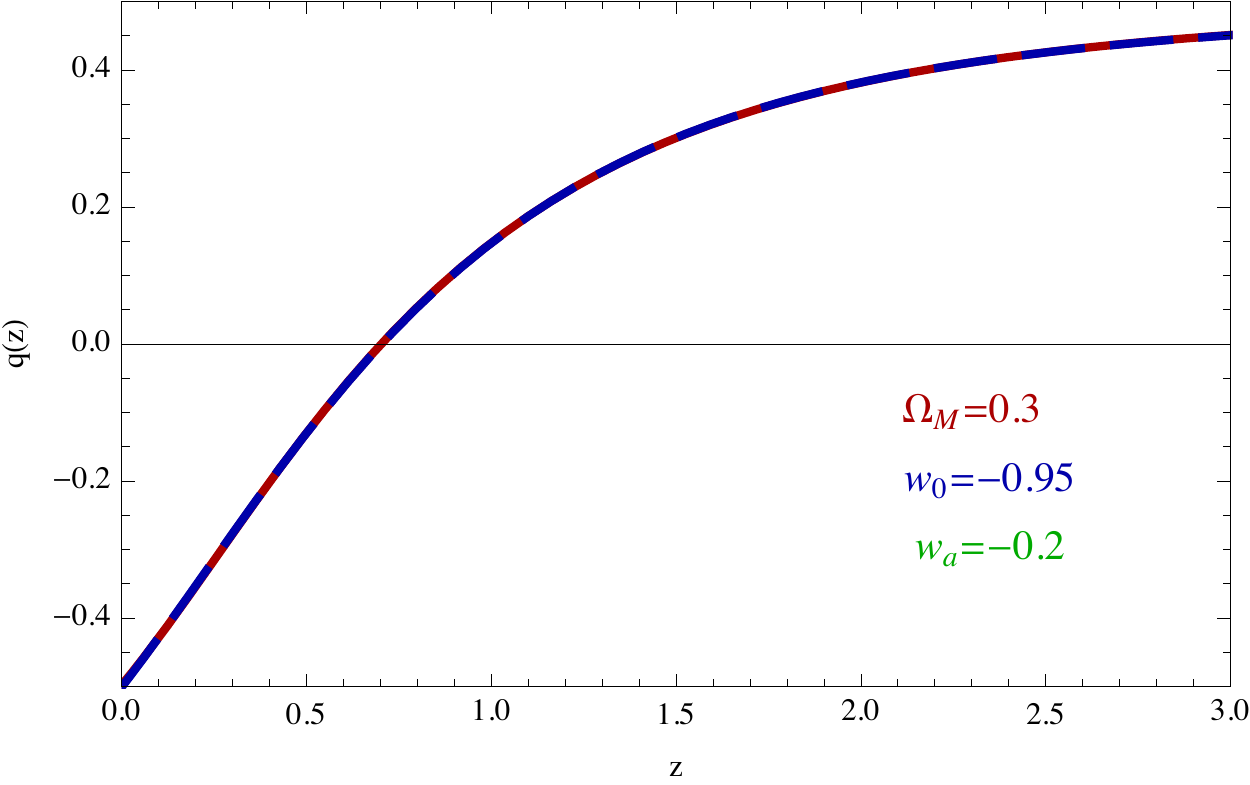}}}\
\vspace{0cm}\rotatebox{0}{\vspace{0cm}\hspace{0cm}\resizebox{0.45\textwidth}{!}{\includegraphics{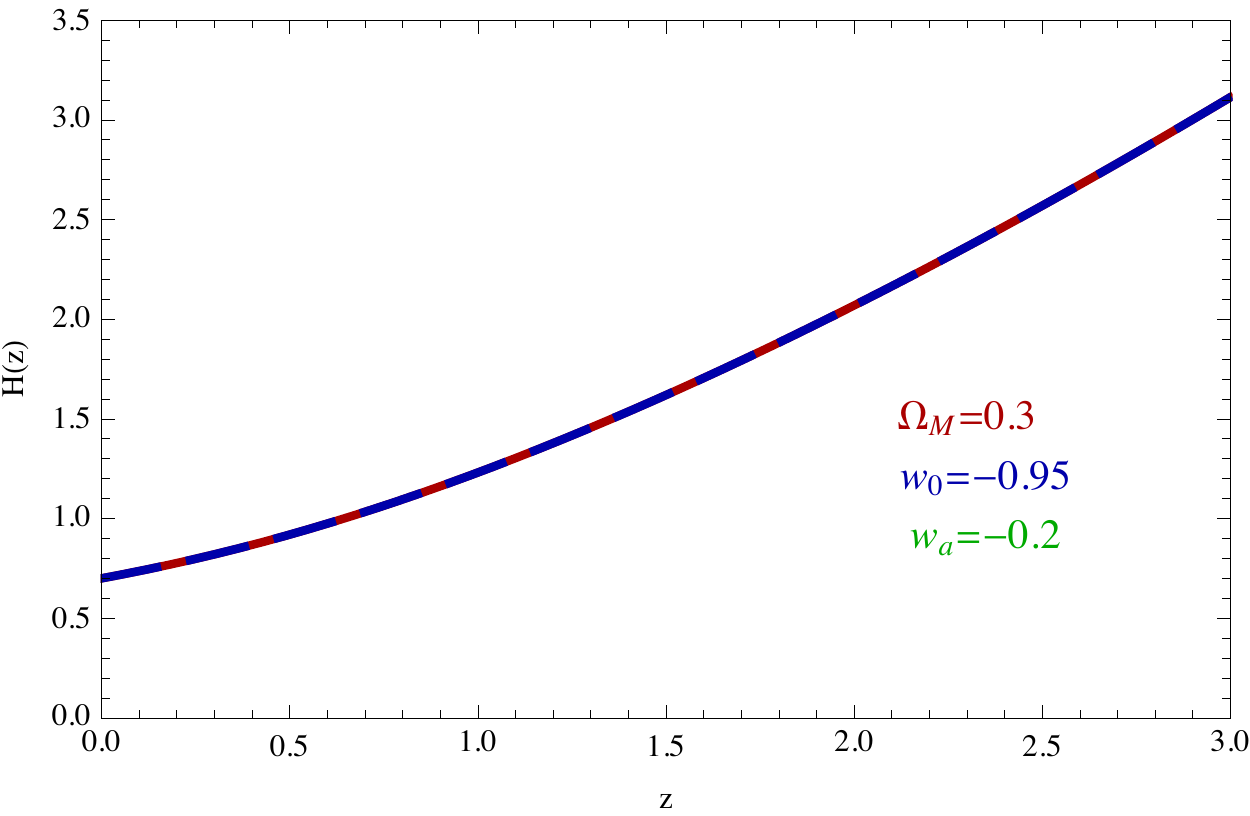}}}\\
\vspace{0cm}\rotatebox{0}{\vspace{0cm}\hspace{0cm}\resizebox{0.45\textwidth}{!}{\includegraphics{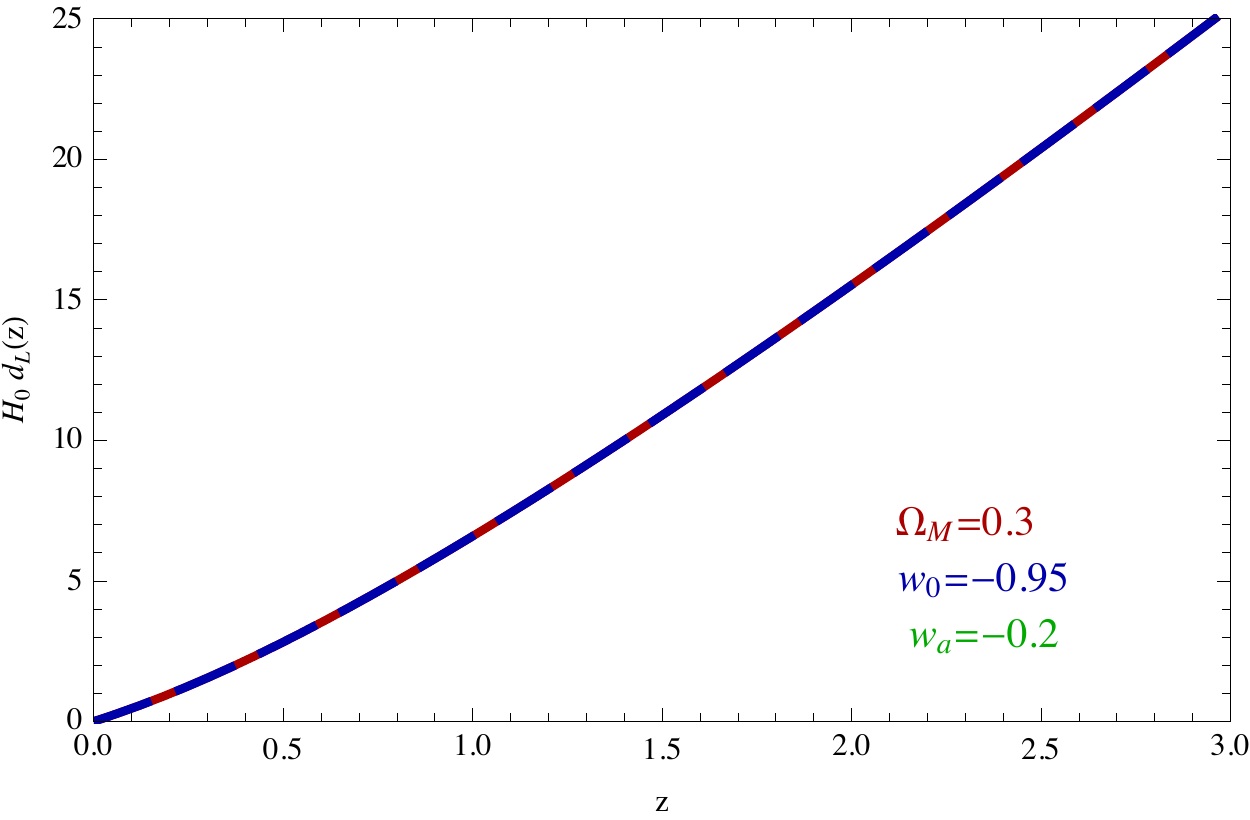}}}\
\vspace{0cm}\rotatebox{0}{\vspace{0cm}\hspace{0cm}\resizebox{0.45\textwidth}{!}{\includegraphics{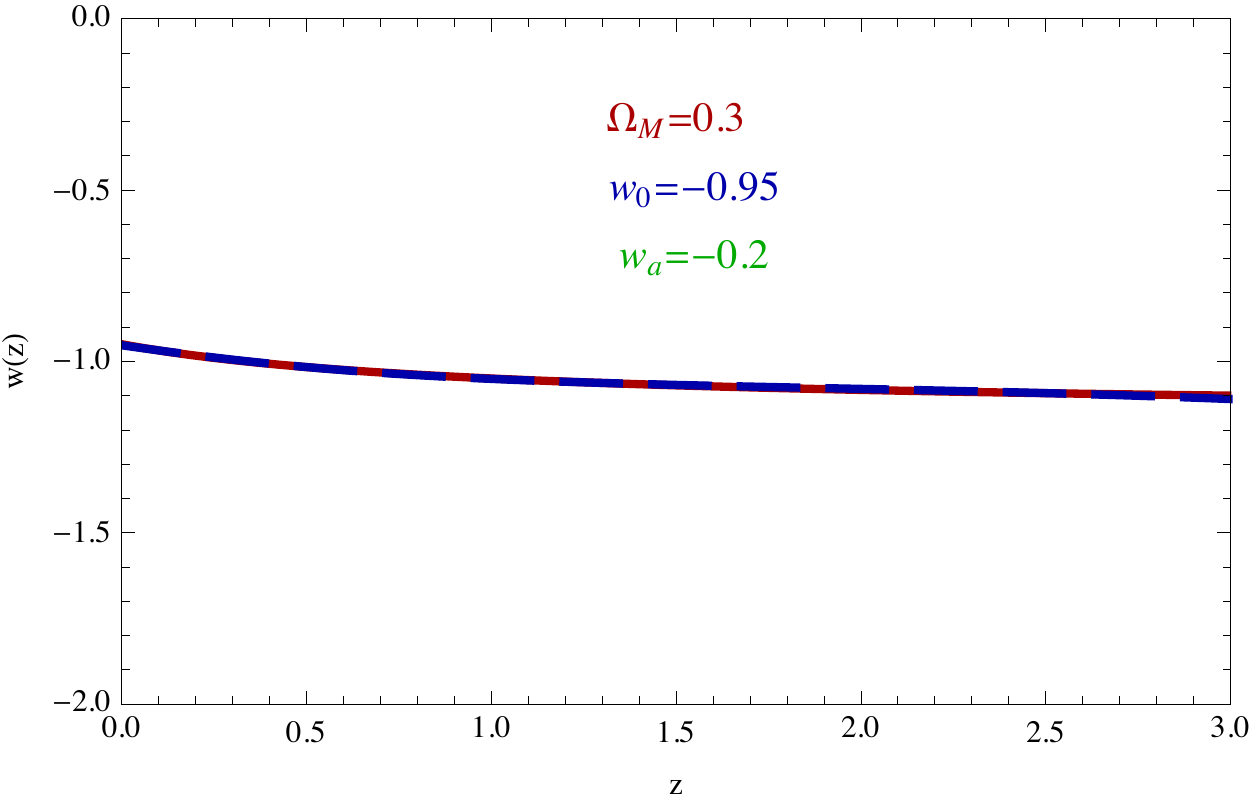}}}
\caption{The best-fit $q(z)$ (dashed line) compared to a $w_{0}w_{a}$CDM (continuous line). \label{goodfit}}
\end{figure}

\begin{figure*}[t!]
\centering
\vspace{0cm}\rotatebox{0}{\vspace{0cm}\hspace{0cm}\resizebox{0.43\textwidth}{!}{\includegraphics{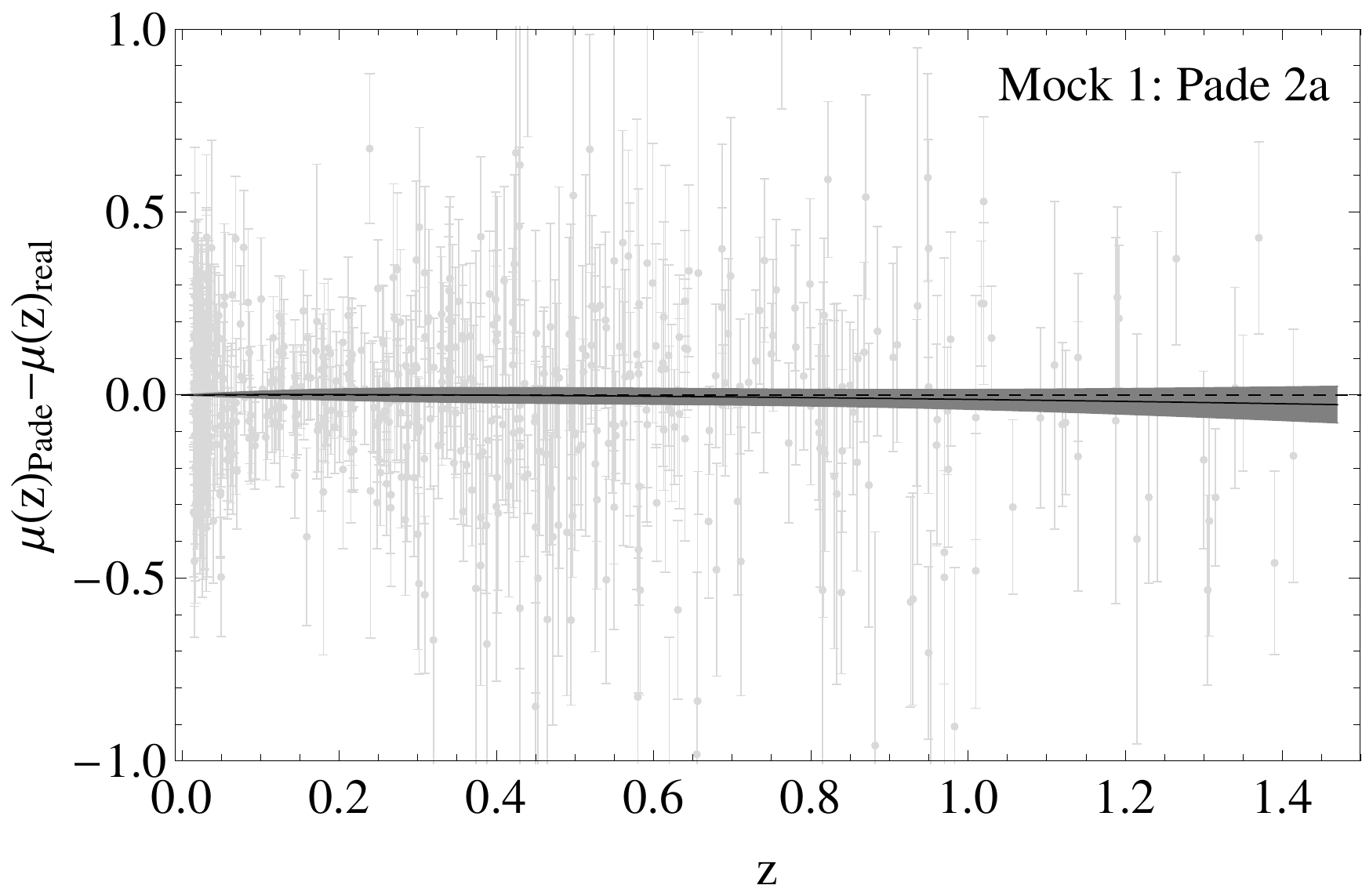}}}
\vspace{0cm}\rotatebox{0}{\vspace{0cm}\hspace{0cm}\resizebox{0.43\textwidth}{!}{\includegraphics{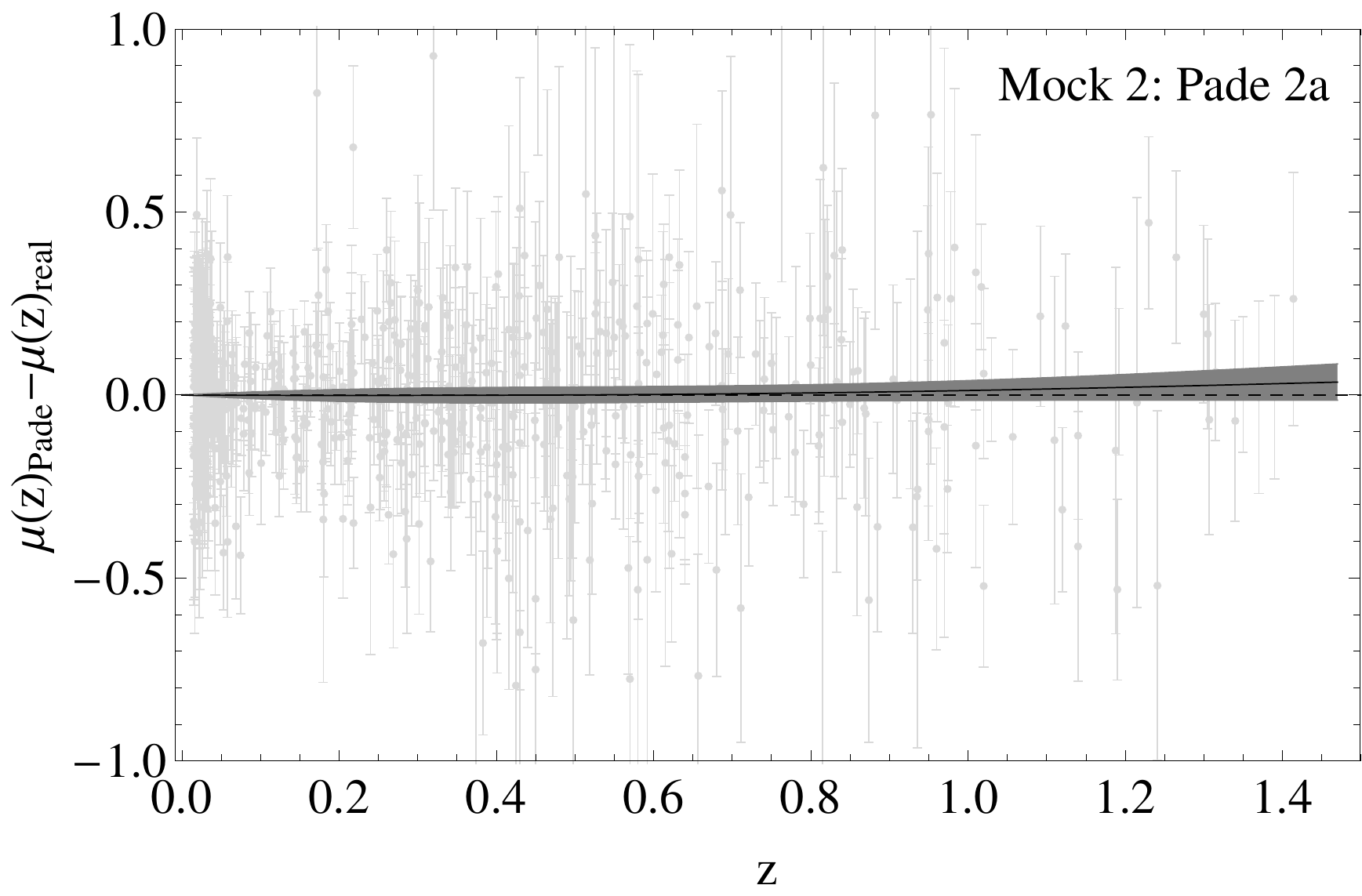}}}
\vspace{0cm}\rotatebox{0}{\vspace{0cm}\hspace{0cm}\resizebox{0.43\textwidth}{!}{\includegraphics{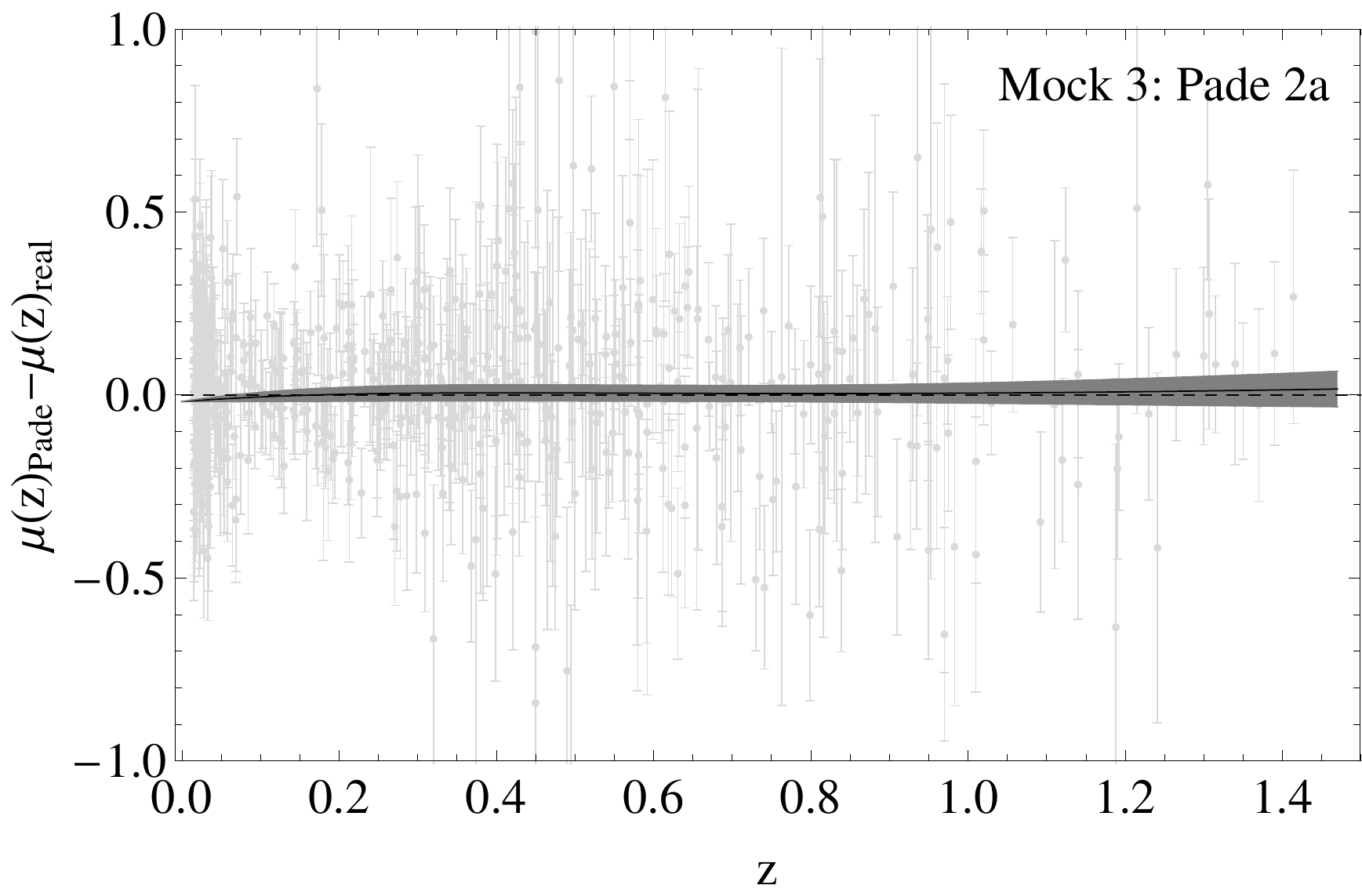}}}
\vspace{0cm}\rotatebox{0}{\vspace{0cm}\hspace{0cm}\resizebox{0.43\textwidth}{!}{\includegraphics{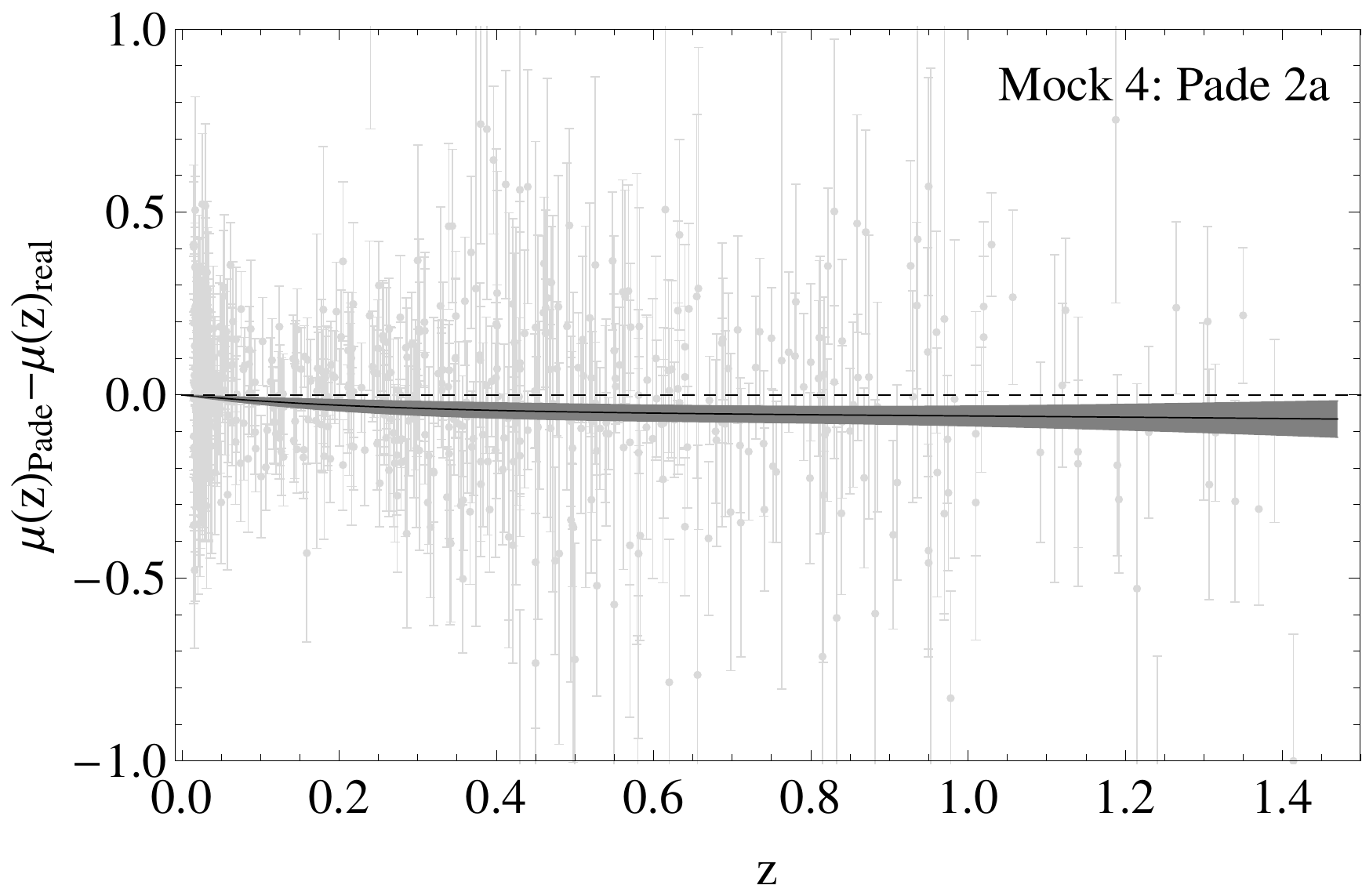}}}
\caption{The residues $\mu_{Pade}(z)-\mu_{real}(z)$ for all four mocks.\label{dmuPade2a}}
\vspace{1.0cm}
\centering
\vspace{0cm}\rotatebox{0}{\vspace{0cm}\hspace{0cm}\resizebox{0.43\textwidth}{!}{\includegraphics{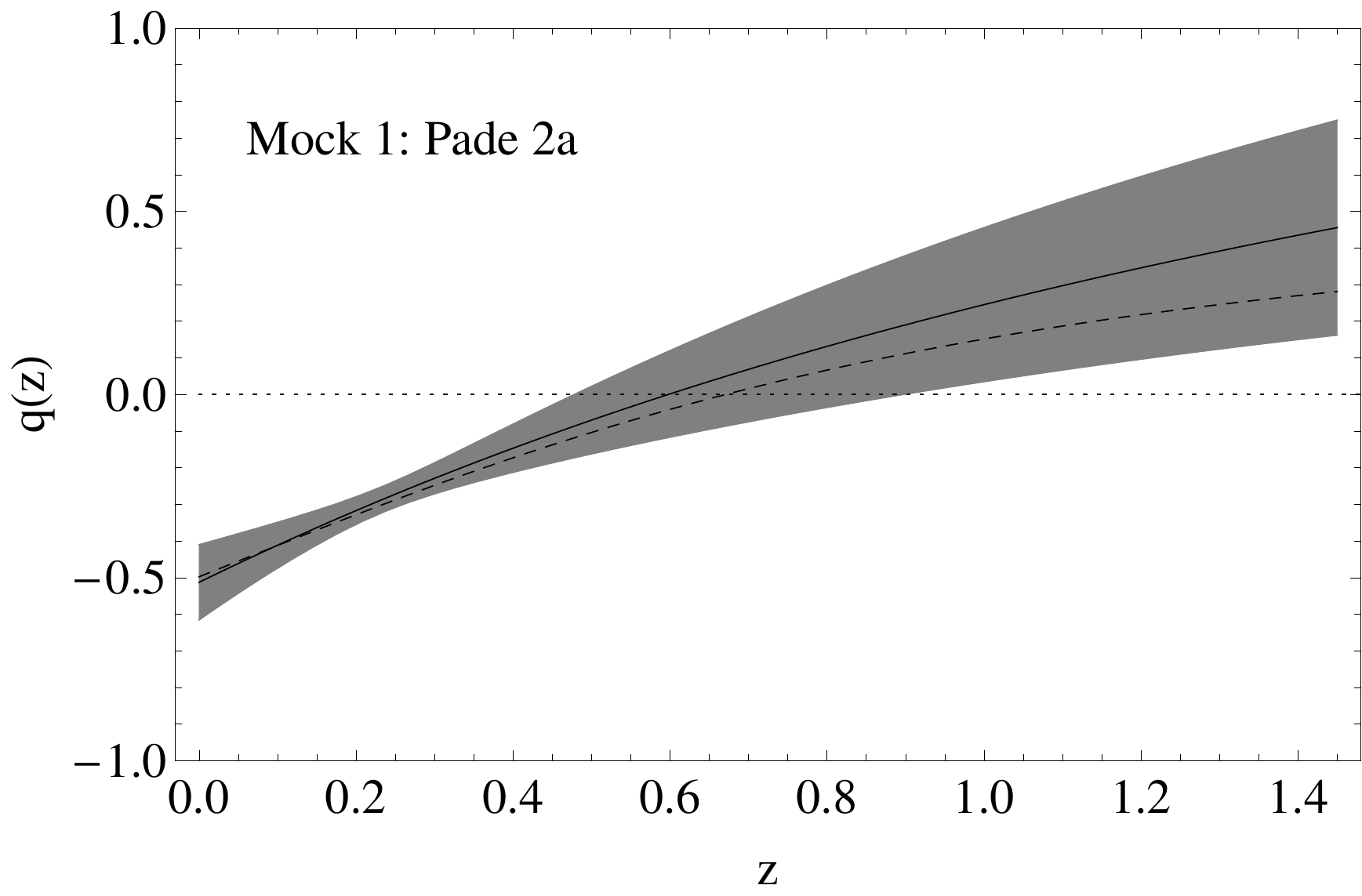}}}
\vspace{0cm}\rotatebox{0}{\vspace{0cm}\hspace{0cm}\resizebox{0.43\textwidth}{!}{\includegraphics{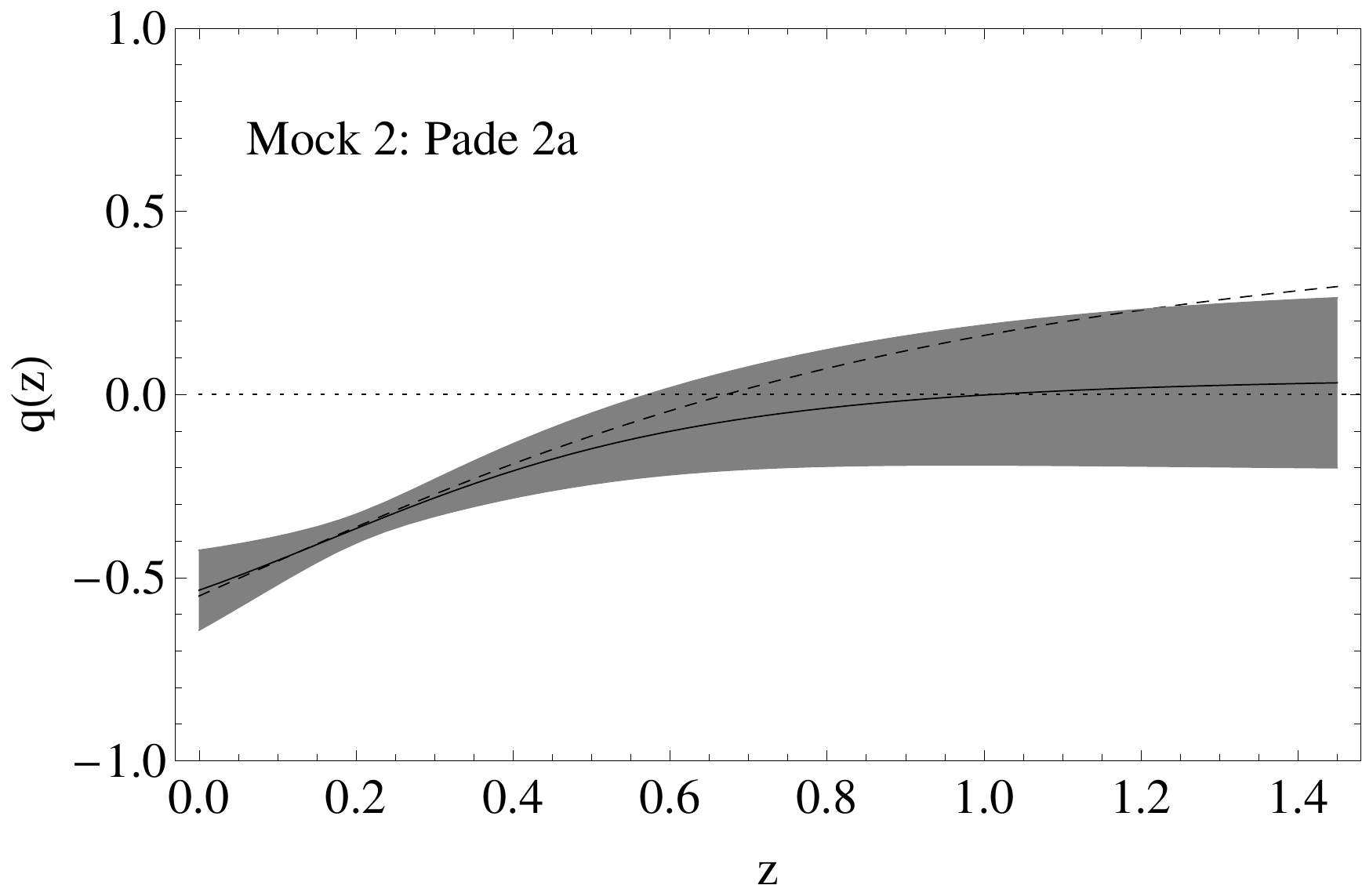}}}
\vspace{0cm}\rotatebox{0}{\vspace{0cm}\hspace{0cm}\resizebox{0.43\textwidth}{!}{\includegraphics{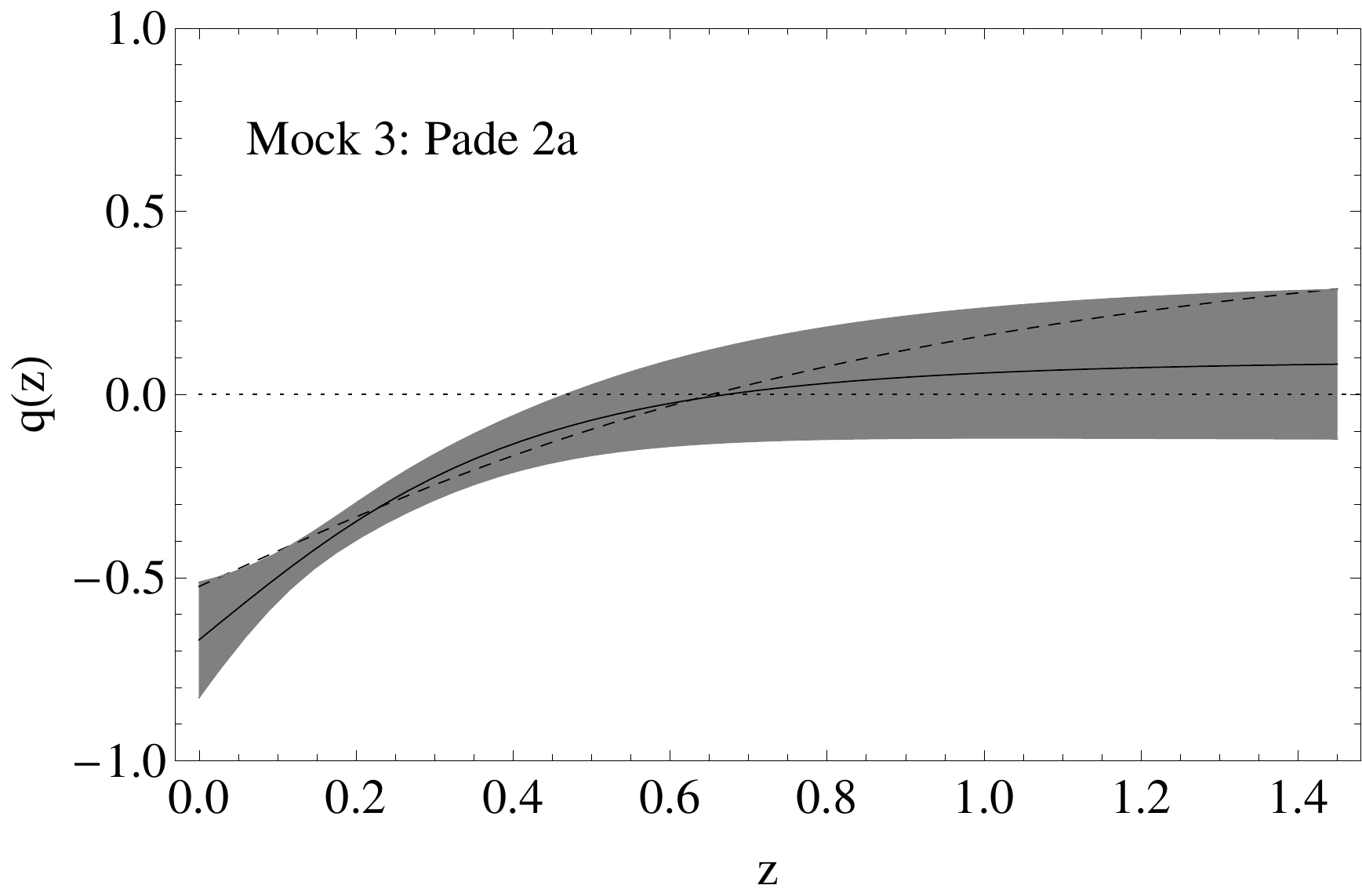}}}
\vspace{0cm}\rotatebox{0}{\vspace{0cm}\hspace{0cm}\resizebox{0.43\textwidth}{!}{\includegraphics{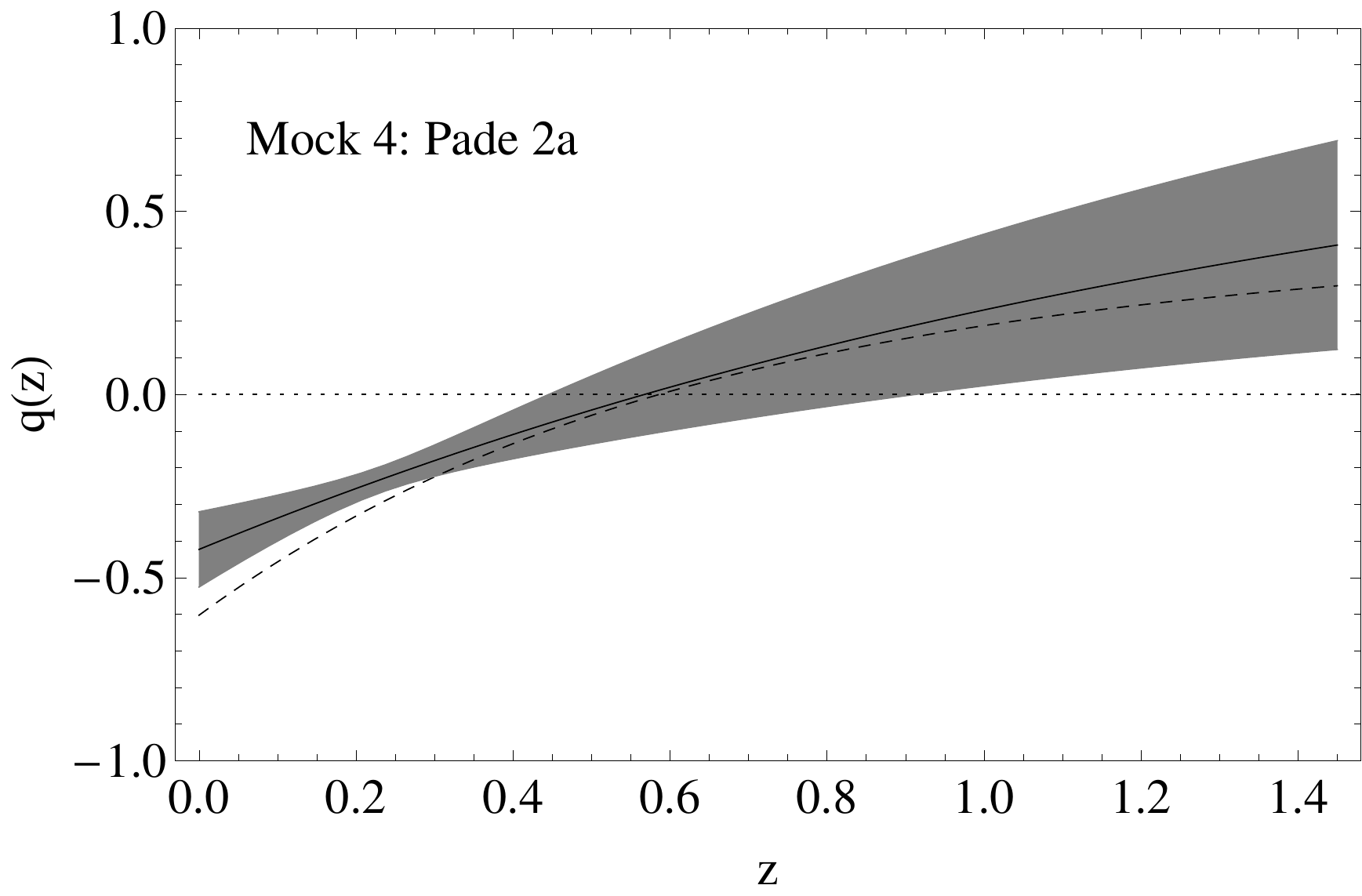}}}
\caption{The deceleration parameter $q(z)$ for all four mocks. The dashed line corresponds to the real model.\label{qzPade2a}}
\end{figure*}

Suppose we have a deceleration parameter written as a four-parameter fit,
\be\label{qzp}
q(z) = \frac{a(1+z)^d - c}{b(1+z)^d + 1}\,.
\ee
Then, the properly normalized rate of expansion can be obtained exactly,
\be\label{Hzp}
H(z) = H_0 (1+z)^{1-c}\left(\frac{1 + b(1+z)^d}{1 + b}\right)^\gamma\,,
\ee
where $\gamma = (a + b\,c)/(d\,b)$, from which the luminosity distance is obtained
\be\label{dLzp}
H_0\,d_L(z) = \frac{(1+b)^\gamma}{c}\,(1+z)\left((1+z)^c\,{}_2F_1\Big[\gamma,\frac{c}{d},1+\frac{c}{d},-b(1+z)^d\Big]-
{}_2F_1\Big[\gamma,\frac{c}{d},1+\frac{c}{d},-b\Big]\right)\,,
\ee
where ${}_2F_1[a,b,c;z]$ is the Gauss hypergeometric function.
And the equation of state parameter can be expressed as
\be\label{wzp}
3w(z) = \frac{2q(z) - 1}{1 - \om(1+z)^{1+2c}\left(\frac{1 + b(1+z)^d}{1 + b}\right)^{-2\gamma}}\,.
\ee
Note that with this parametrization, we recover the exact solution of $w$\lcdm with
\be\label{param}
b = 2a = \frac{\om}{1-\om}\,, \hspace{1cm}
c = - \frac{1+3w}{2}\,, \hspace{1cm} d = -3w\,, \hspace{1cm} \gamma = \frac{1}{2} \,.
\ee
However, we have tested that it works remarkably well for very distinct cosmologies. For example, with a
($w_0,\, w_a)$ \lcdm cosmology with parameters $h_0 = 0.7,\, \om = 0.3,\, w_0 = -0.95, w_a = -0.2$, we
find the corresponding best fit ($a=0.135,\, b = 0.271,\, c = 0.771,\, d = 3.271$), which gives surprisingly good results (see Fig.~\ref{goodfit}).

Also, it is easy to see from Eq.~(\ref{qzp}), that the parameters $(a,b,c)$ can also be written in terms of the physically meaningful parameters $q_0\equiv q(z=0)=\frac{a-c}{b+1}$, $q_1\equiv q'(z=0)=\frac{d (a+b c)}{(b+1)^2}$ and $q_\infty \equiv q(z\rightarrow\infty)=\frac{a}{b}$, as
\ba
a&=& \frac{q_\infty q_1}{d (q_\infty - q_0) - q_1}\,, \\
b&=& \frac{q_1}{d (q_\infty - q_0) - q_1}\,, \\
c&=& \frac{q_\infty (q_1 - d\,q_0)  +d\,q_0^2}{d (q_\infty - q_0) - q_1}\,.
\ea

In Figs. \ref{dmuPade2a} and \ref{qzPade2a} we show the residues $\mu_{Pade}(z)-\mu_{real}(z)$ for all four mocks and the deceleration parameter $q(z)$ for all four mocks, respectively. The dashed line corresponds to the real models and we have labeled this method as  Pade 2a in order to discriminate it from the simple linear Pad\'e and the version with the constant exponent variant we mentioned earlier.

\subsection{Pad\'e approximants for $d_L(z)$}
Another option is to use a Pad\'e approximant for the luminosity distance $d_L(z)$ or, equivalently, the comoving distance $r(z)$,
\be
r(z)=z\frac{1+a z}{1+ b z+c z^2}\,.
\ee
Such parametrizations were proposed in Refs. \cite{Huterer:2000mj} and \cite{Gerke:2002sx}, where it was found that they can range between challenging and even quite inaccurate. In what follows we will consider a Pad\'e approximant of higher order,

\begin{figure*}[t!]
\centering
\vspace{0cm}\rotatebox{0}{\vspace{0cm}\hspace{0cm}\resizebox{0.43\textwidth}{!}{\includegraphics{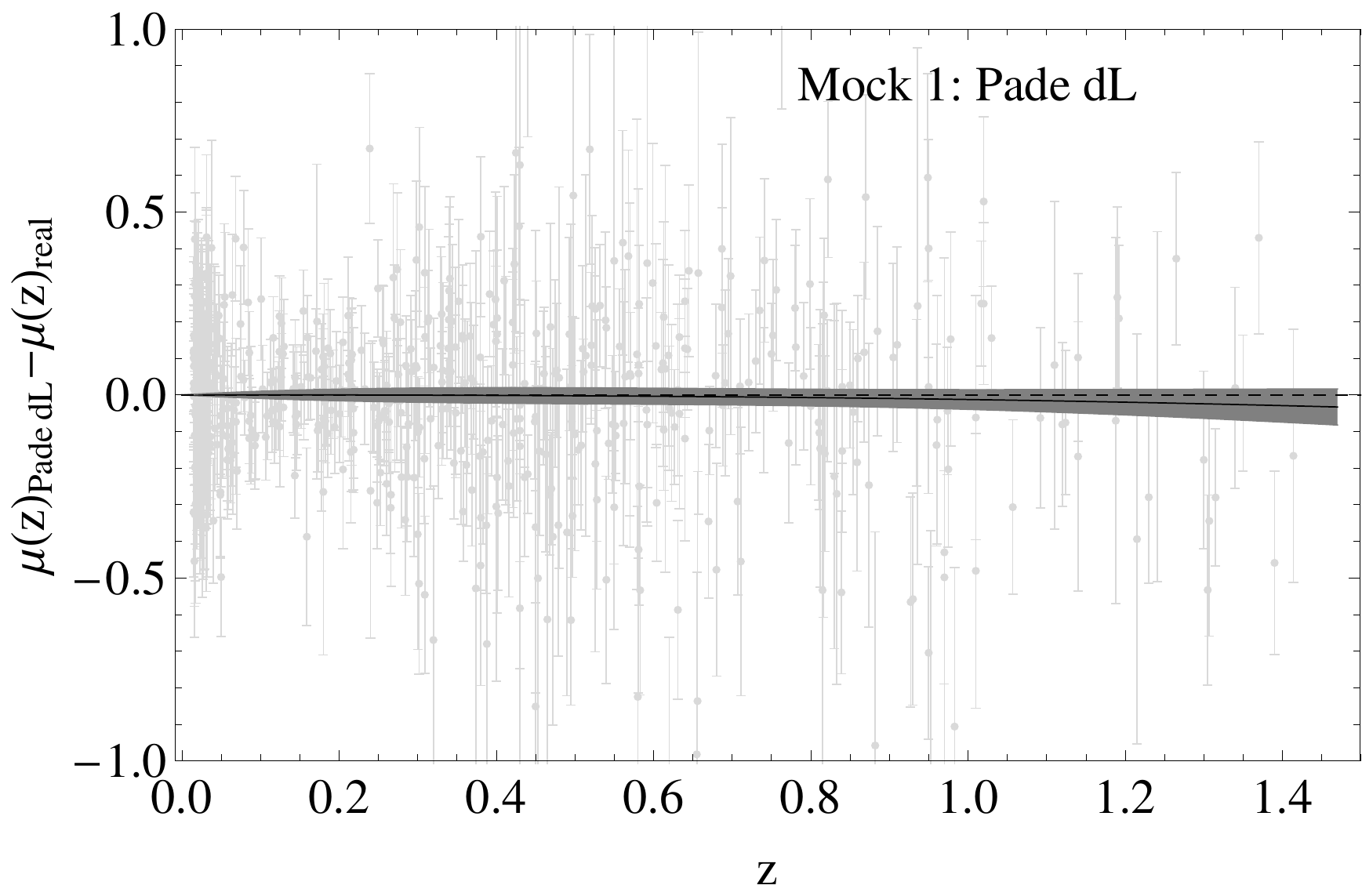}}}
\vspace{0cm}\rotatebox{0}{\vspace{0cm}\hspace{0cm}\resizebox{0.43\textwidth}{!}{\includegraphics{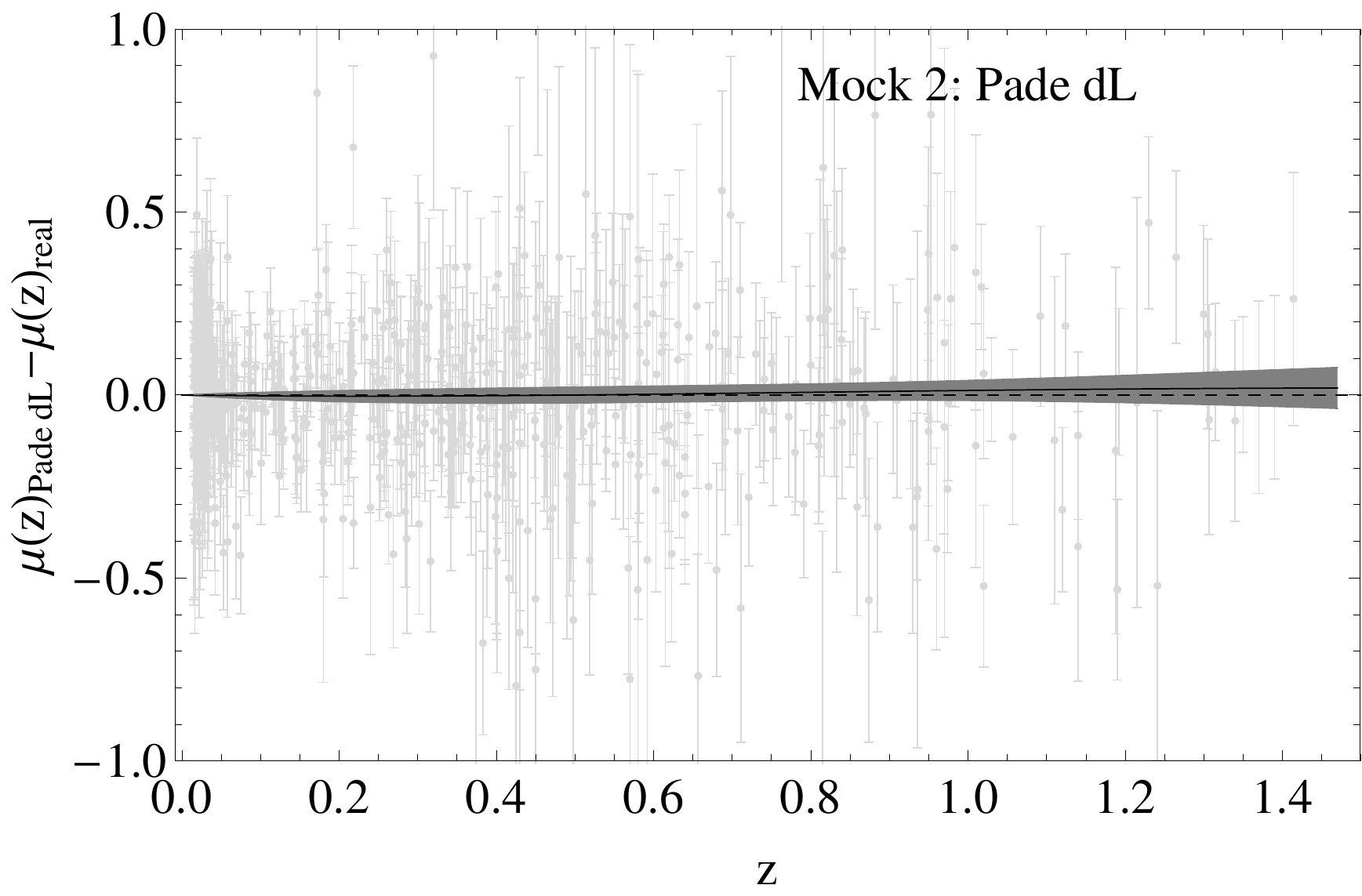}}}
\vspace{0cm}\rotatebox{0}{\vspace{0cm}\hspace{0cm}\resizebox{0.43\textwidth}{!}{\includegraphics{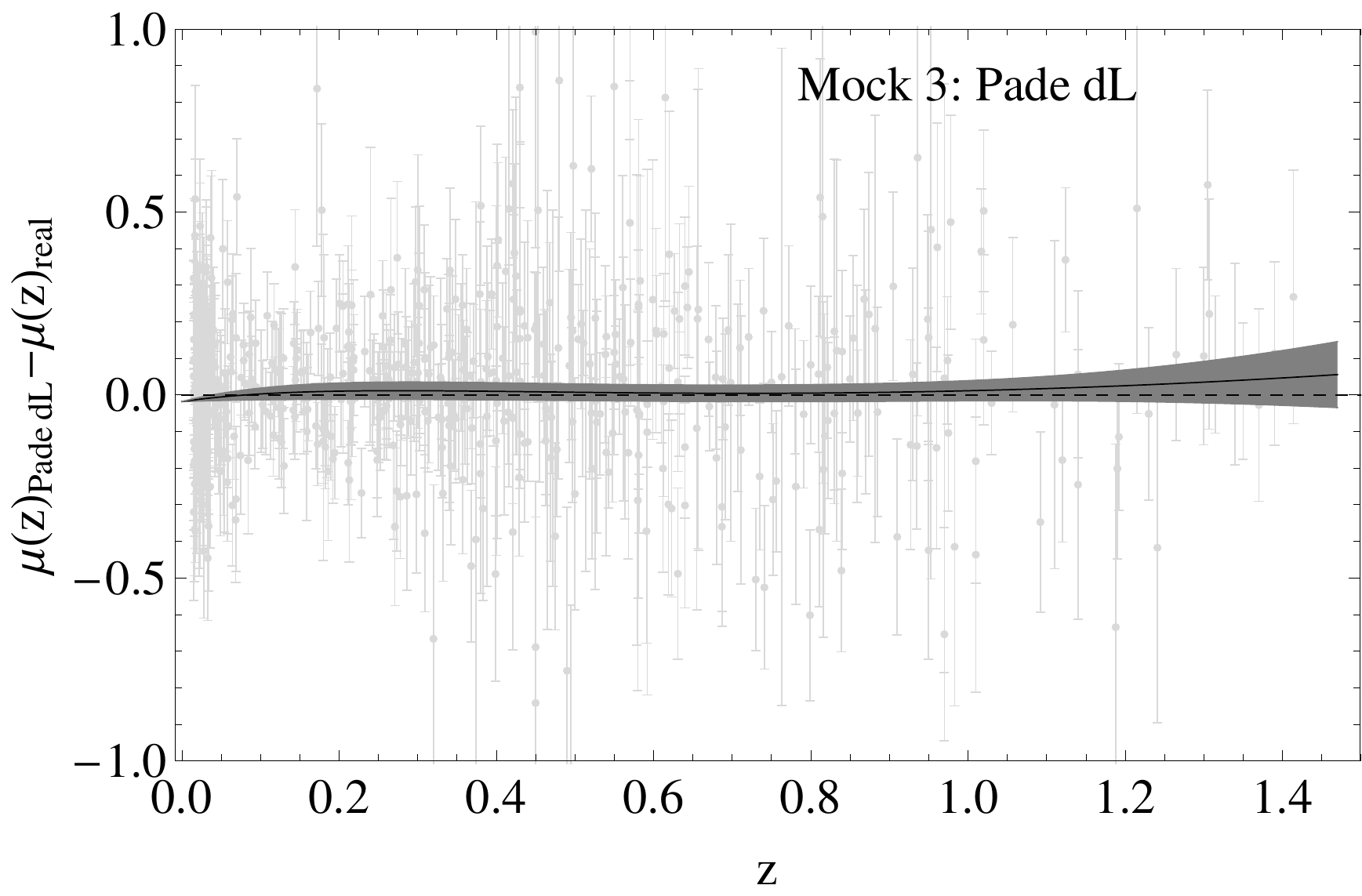}}}
\vspace{0cm}\rotatebox{0}{\vspace{0cm}\hspace{0cm}\resizebox{0.43\textwidth}{!}{\includegraphics{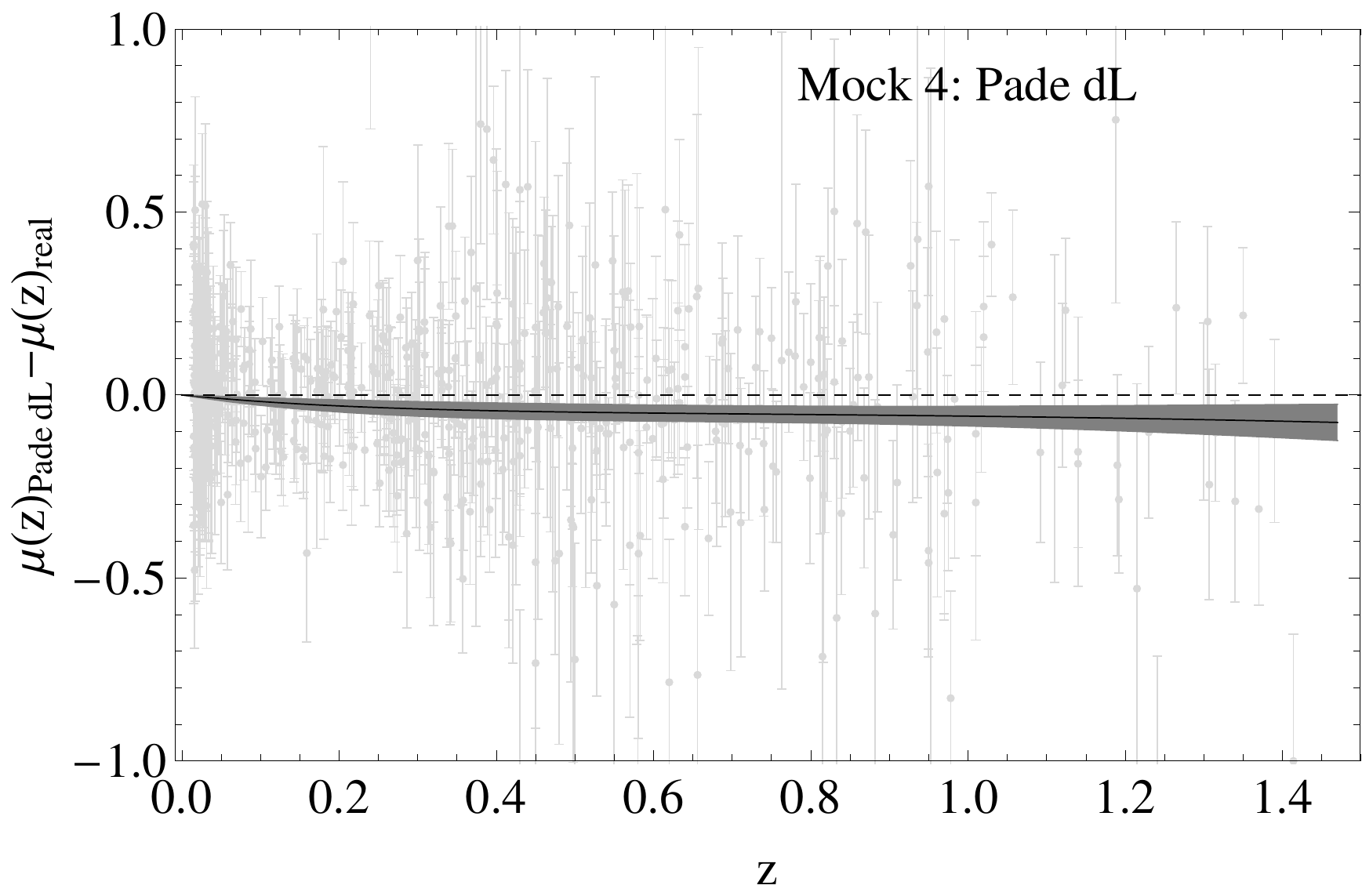}}}
\caption{The residues $\mu_{Pade}(z)-\mu_{real}(z)$ for all four mocks.\label{dmuPadedL}}
\vspace{1cm}
\centering
\vspace{0cm}\rotatebox{0}{\vspace{0cm}\hspace{0cm}\resizebox{0.43\textwidth}{!}{\includegraphics{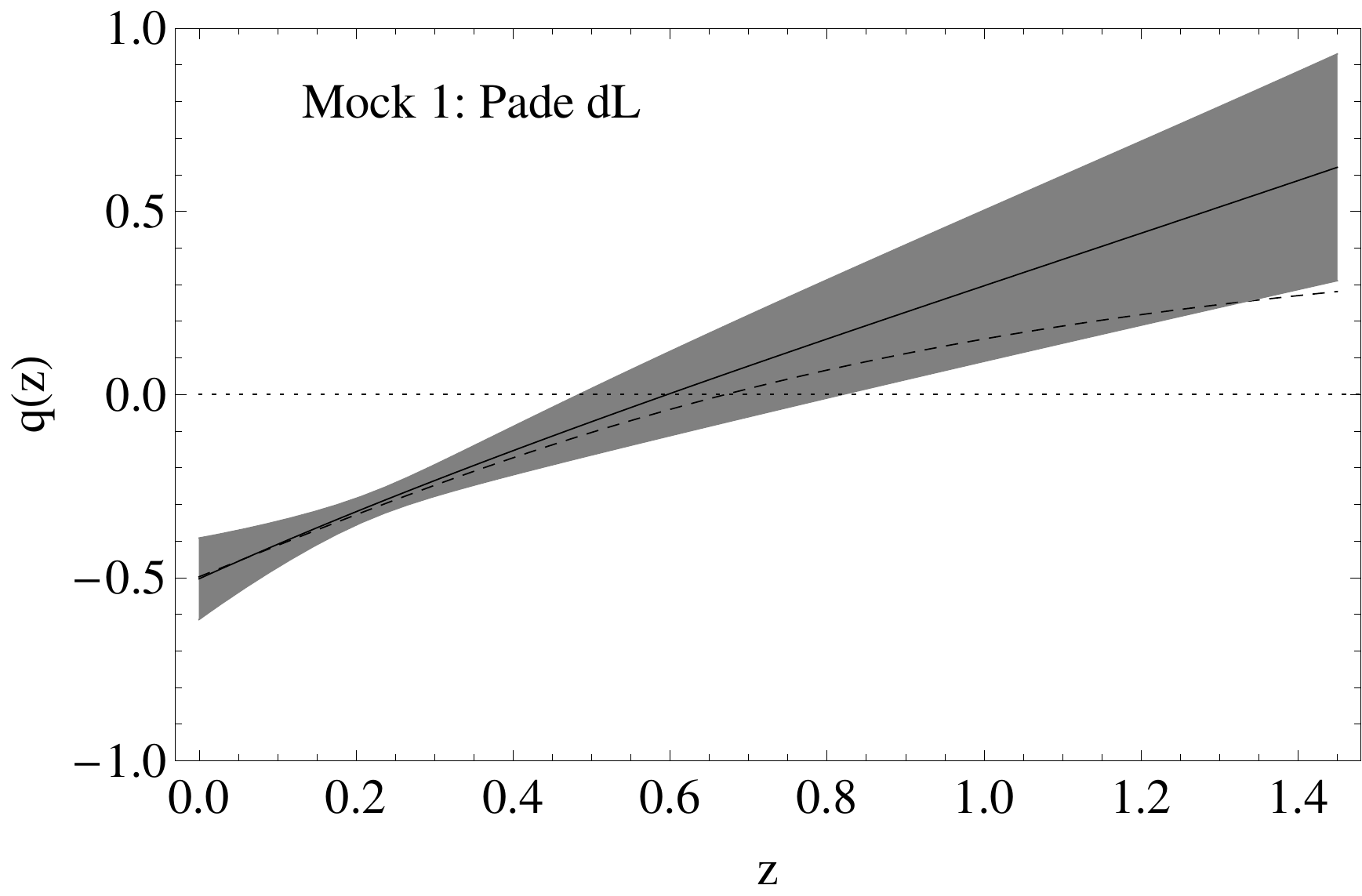}}}
\vspace{0cm}\rotatebox{0}{\vspace{0cm}\hspace{0cm}\resizebox{0.43\textwidth}{!}{\includegraphics{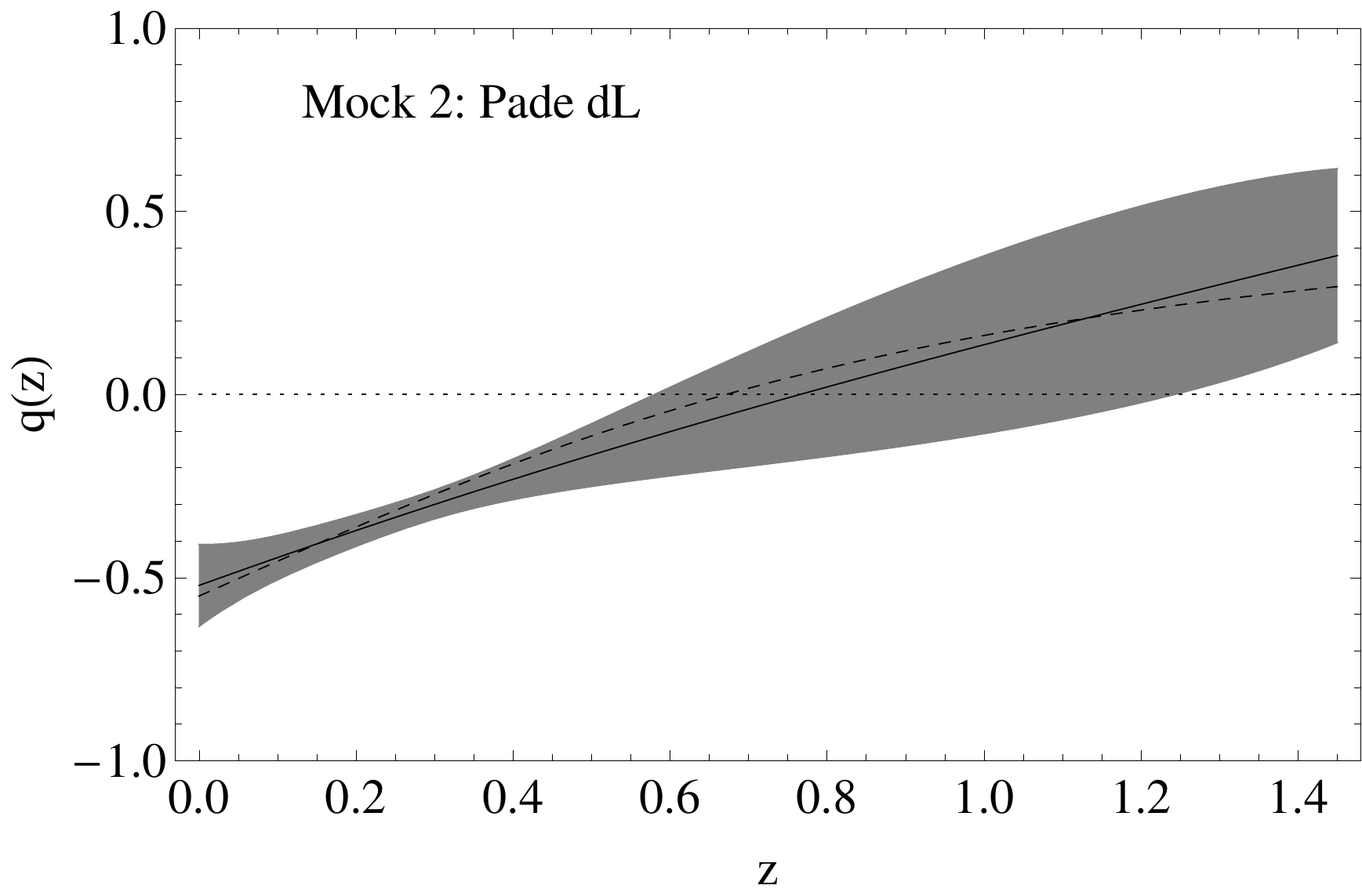}}}
\vspace{0cm}\rotatebox{0}{\vspace{0cm}\hspace{0cm}\resizebox{0.43\textwidth}{!}{\includegraphics{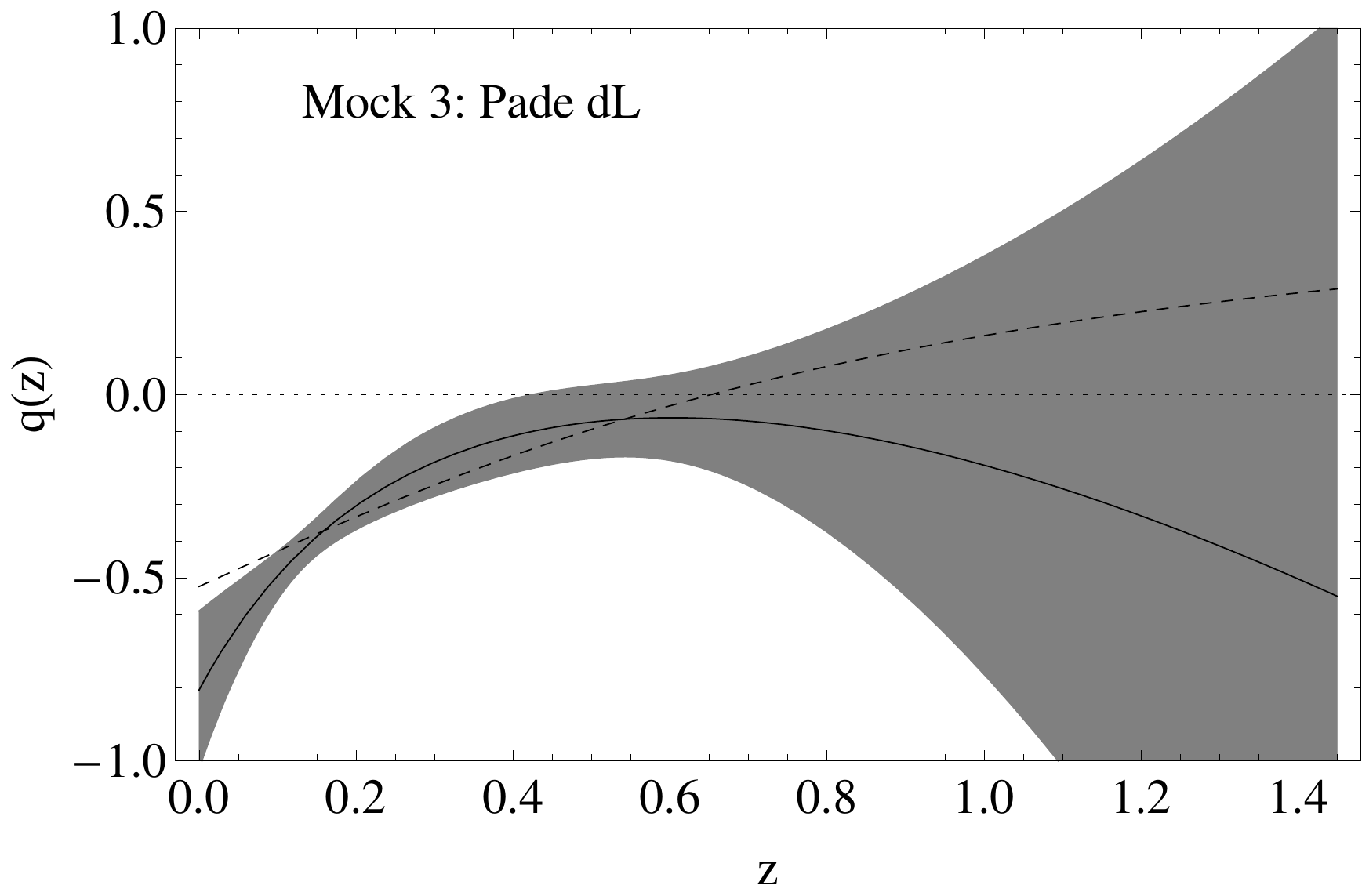}}}
\vspace{0cm}\rotatebox{0}{\vspace{0cm}\hspace{0cm}\resizebox{0.43\textwidth}{!}{\includegraphics{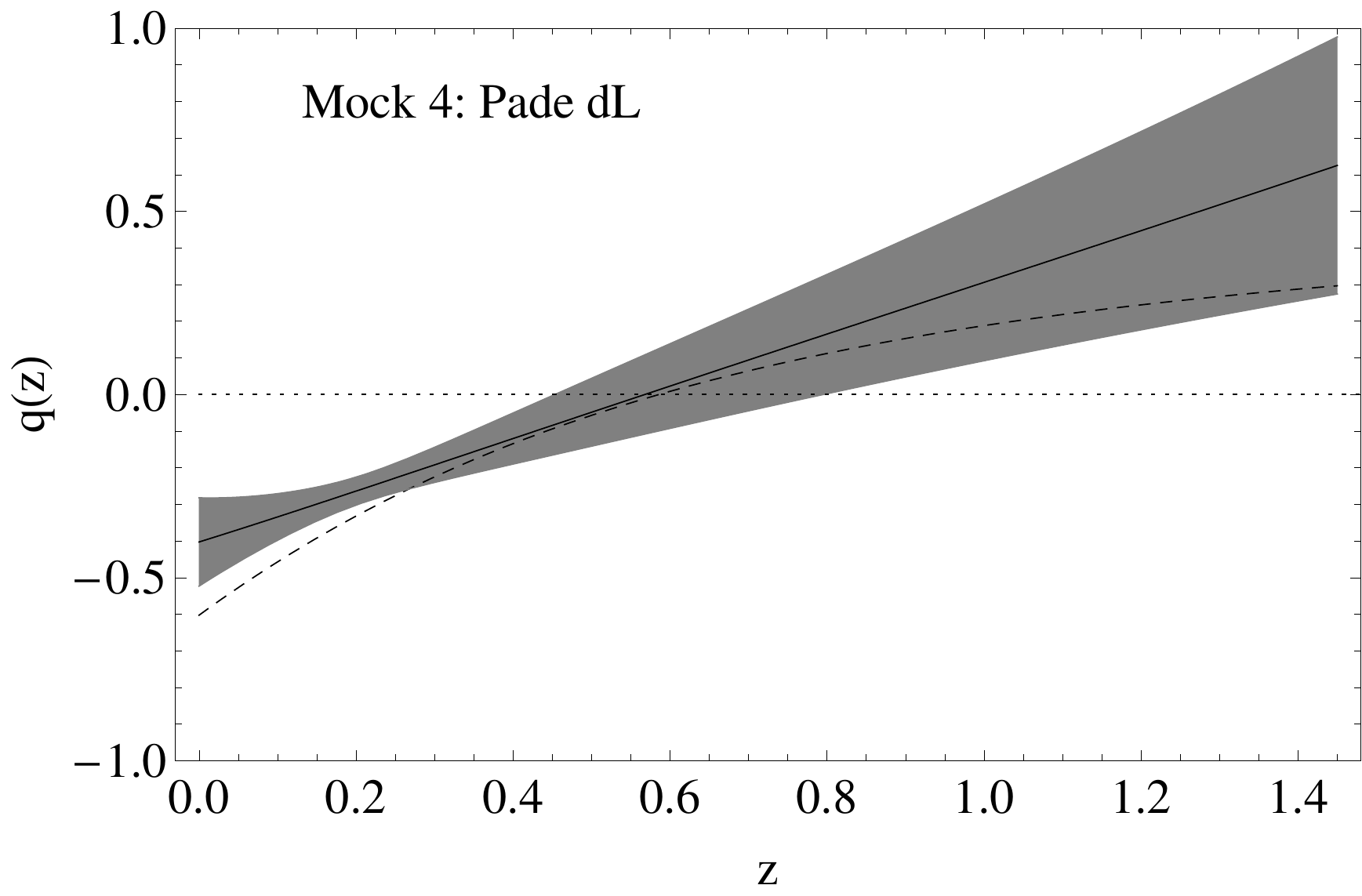}}}
\caption{The deceleration parameter $q(z)$ for all four mocks. The dashed line corresponds to the real model.\label{qzPadedL}}
\end{figure*}

\be
d_L(z)=z\frac{\prod_{i=1}^{i_{\rm max}}(1+a_i z)}{\prod_{j=1}^{j_{\rm max}}(1+b_j z)}
\ee
where we have chosen $i_{\rm max}=2$, $j_{\rm max}=3$ and $a_i,b_j$ are constants. By doing a Taylor expansion around $z=0$, it is easy to see that
\be
d_L(z)=z+(a_1+a_2-b_1-b_2-b_3)z^2+O(z^2)
\ee

In Figs. \ref{dmuPadedL} and \ref{qzPadedL} we show the residues $\mu_{Pade}(z)-\mu_{real}(z)$ for all four mocks and the deceleration parameter $q(z)$ for all four mocks respectively. The dashed line corresponds to the real models and we have labeled this method as Pade dL in order to discriminate it from the simple linear Pad\'e mentioned earlier and the version with the variable exponent  variant we will mention later.

\subsection{Taylor expansions for $\rho_{DE}(z)$}
Another commonly used method is to expand the dark energy density in Taylor series, usually around its value today \cite{Alam:2003fg}
\be
\Omega_{DE}(z)=A_0+A_1 (1+z)+A_2(1+z)^2+A_4(1+z)^4+A_5(1+z)^5+\cdots,
\ee
where $(A_0,A_1,A_2,A_3,A_5)$ are constants and $A_0$ can be fixed by using $H(z=0)=H_0$. We didn't include a term like $A_3 (1+z)^3$ as it would be degenerate with the matter density $\om (1+z)^3$.

However, we found that a fit to the four mock data gave completely unphysical results, with $\om$ being negative or much bigger than 1 in all of the cases and even in the relatively simple case where only $(\om,A_1,A_2)$ are free to vary. Thus, we will no longer discuss this case.

\subsection{Taylor expansions for $d_L(z)$}

\begin{figure*}[t!]
\centering
\vspace{0cm}\rotatebox{0}{\vspace{0cm}\hspace{0cm}\resizebox{0.43\textwidth}{!}{\includegraphics{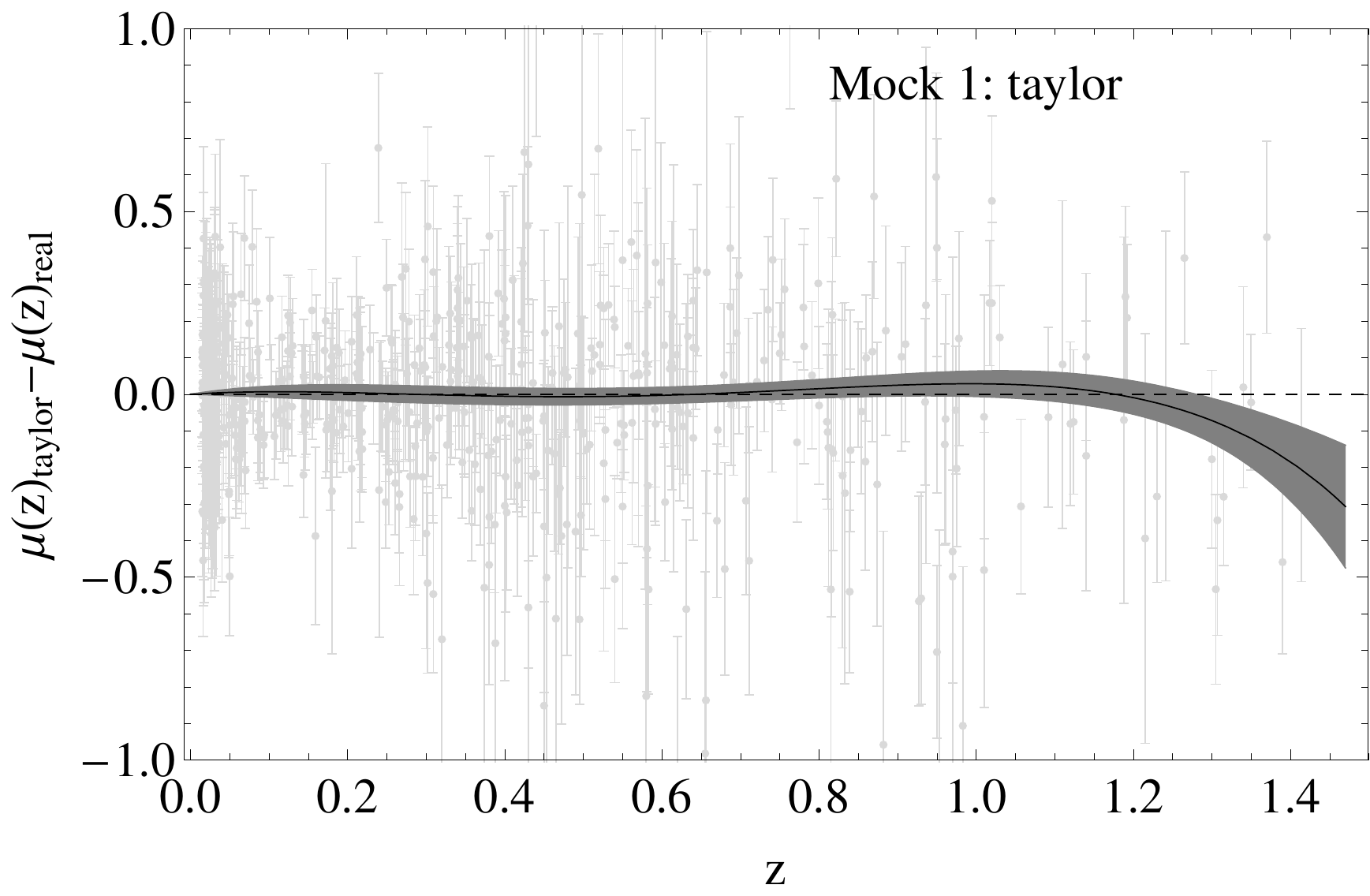}}}
\vspace{0cm}\rotatebox{0}{\vspace{0cm}\hspace{0cm}\resizebox{0.43\textwidth}{!}{\includegraphics{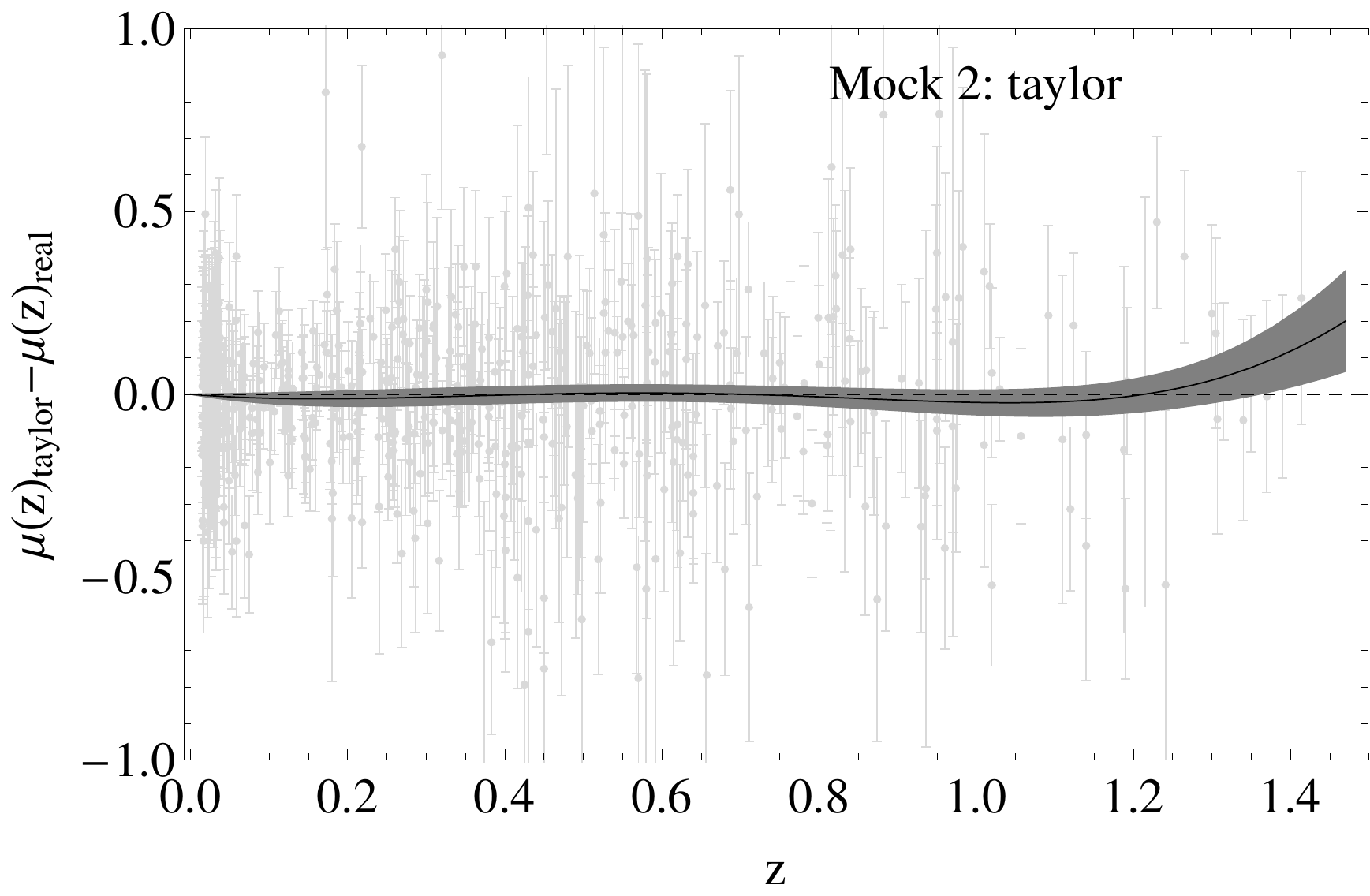}}}
\vspace{0cm}\rotatebox{0}{\vspace{0cm}\hspace{0cm}\resizebox{0.43\textwidth}{!}{\includegraphics{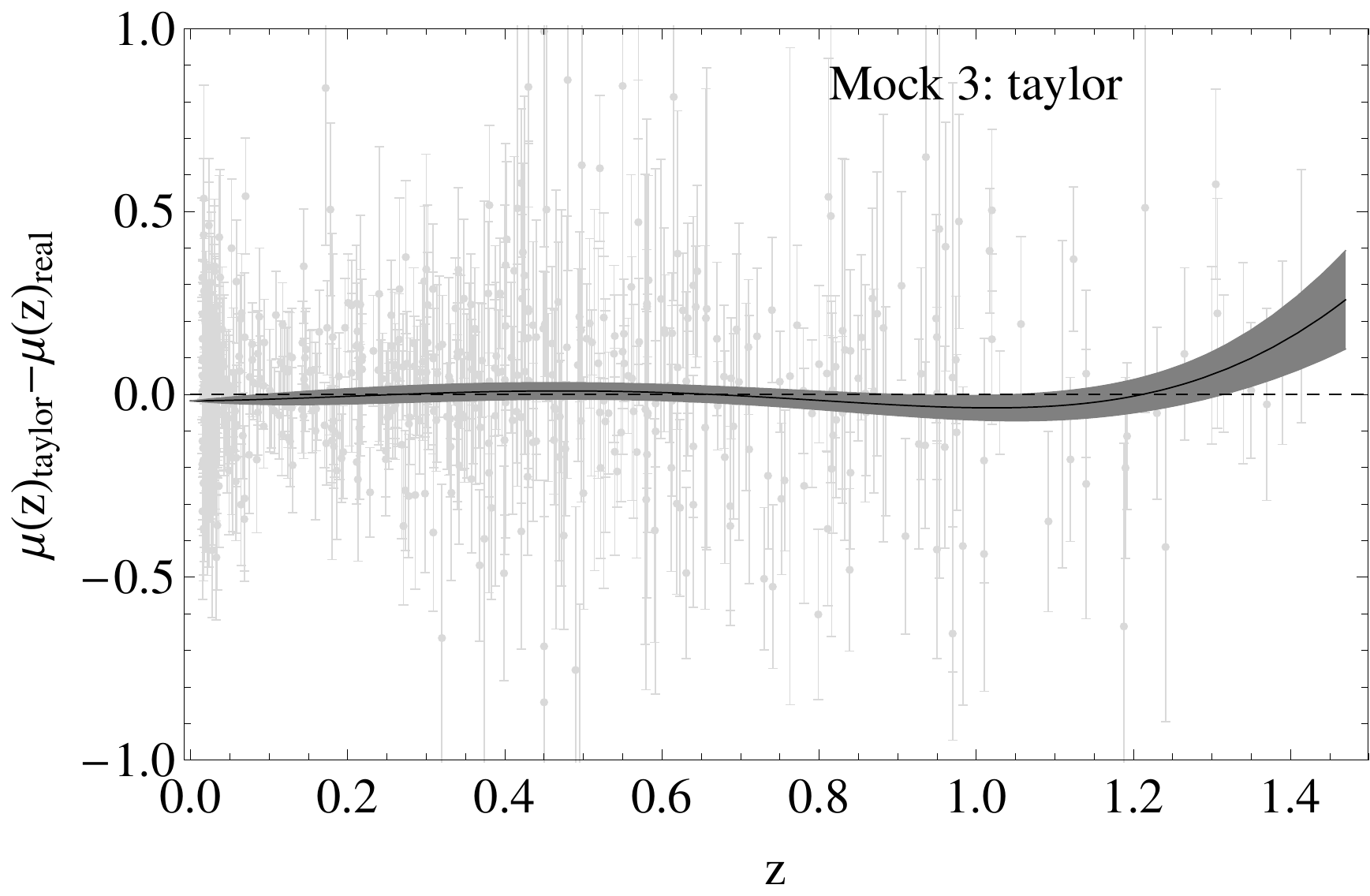}}}
\vspace{0cm}\rotatebox{0}{\vspace{0cm}\hspace{0cm}\resizebox{0.43\textwidth}{!}{\includegraphics{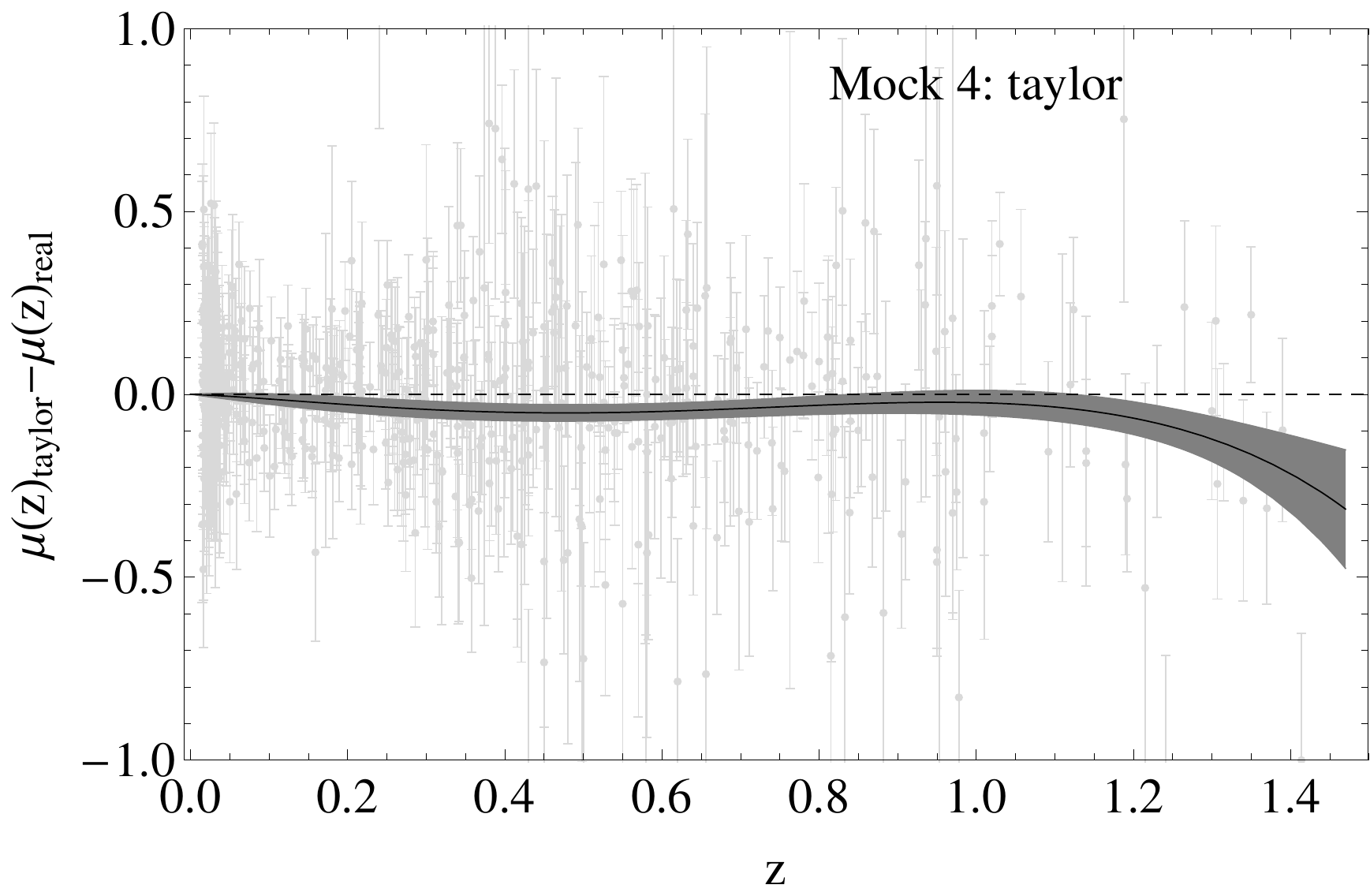}}}
\caption{The residues $\mu_{taylor}(z)-\mu_{real}(z)$ for all four mocks.\label{dmutaylor}}
\vspace{1.0cm}
\centering
\vspace{0cm}\rotatebox{0}{\vspace{0cm}\hspace{0cm}\resizebox{0.43\textwidth}{!}{\includegraphics{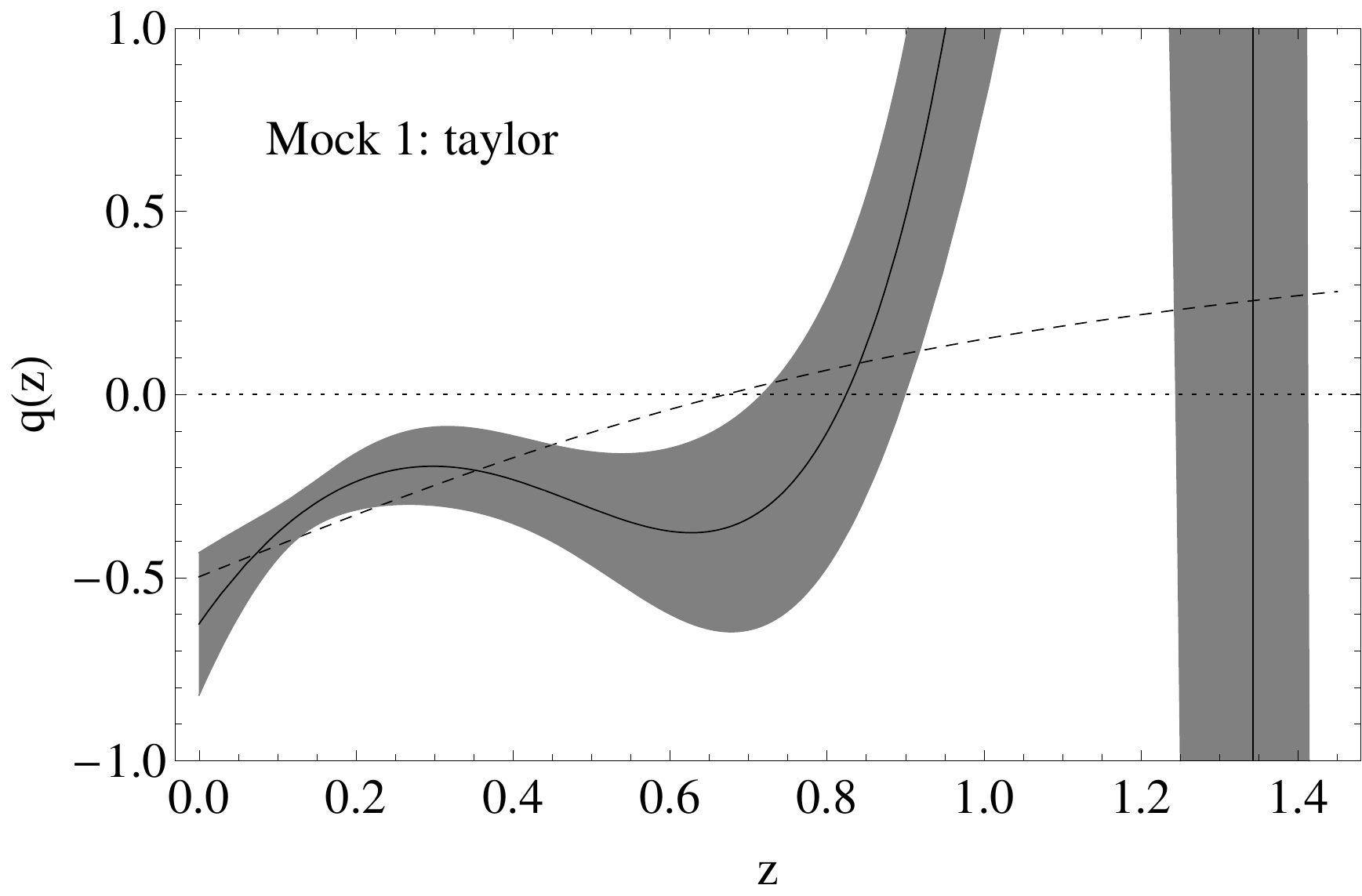}}}
\vspace{0cm}\rotatebox{0}{\vspace{0cm}\hspace{0cm}\resizebox{0.43\textwidth}{!}{\includegraphics{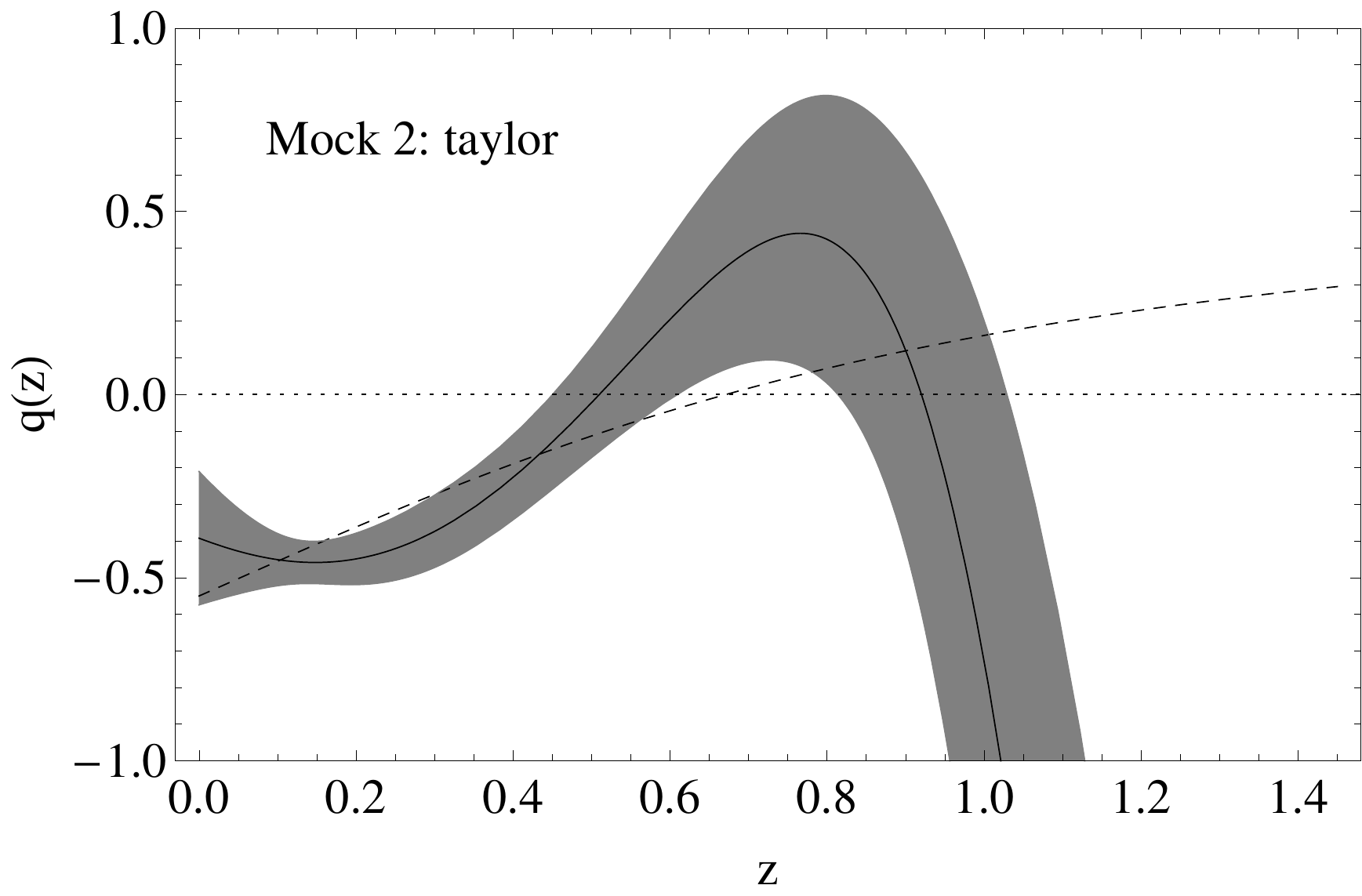}}}
\vspace{0cm}\rotatebox{0}{\vspace{0cm}\hspace{0cm}\resizebox{0.43\textwidth}{!}{\includegraphics{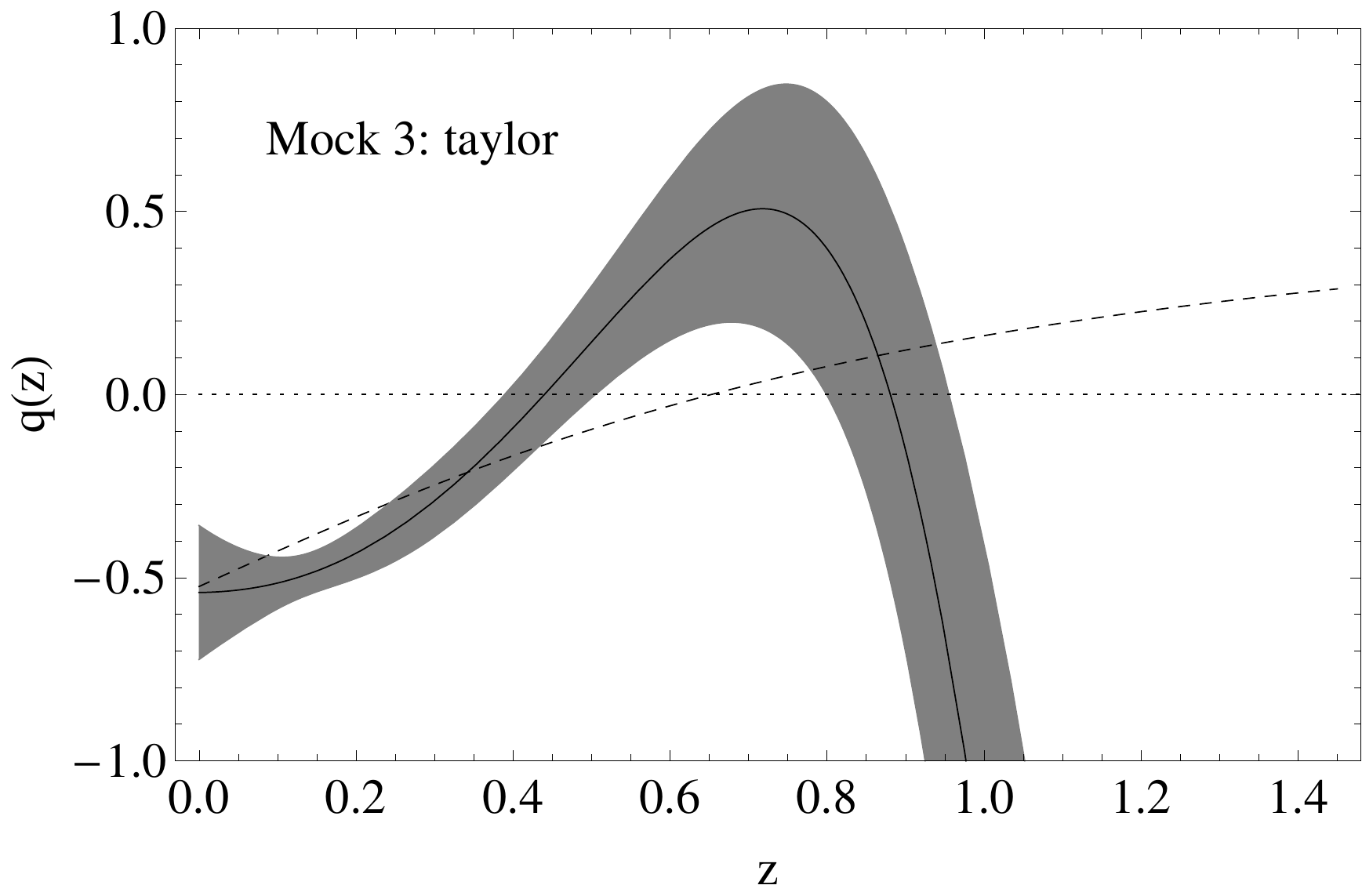}}}
\vspace{0cm}\rotatebox{0}{\vspace{0cm}\hspace{0cm}\resizebox{0.43\textwidth}{!}{\includegraphics{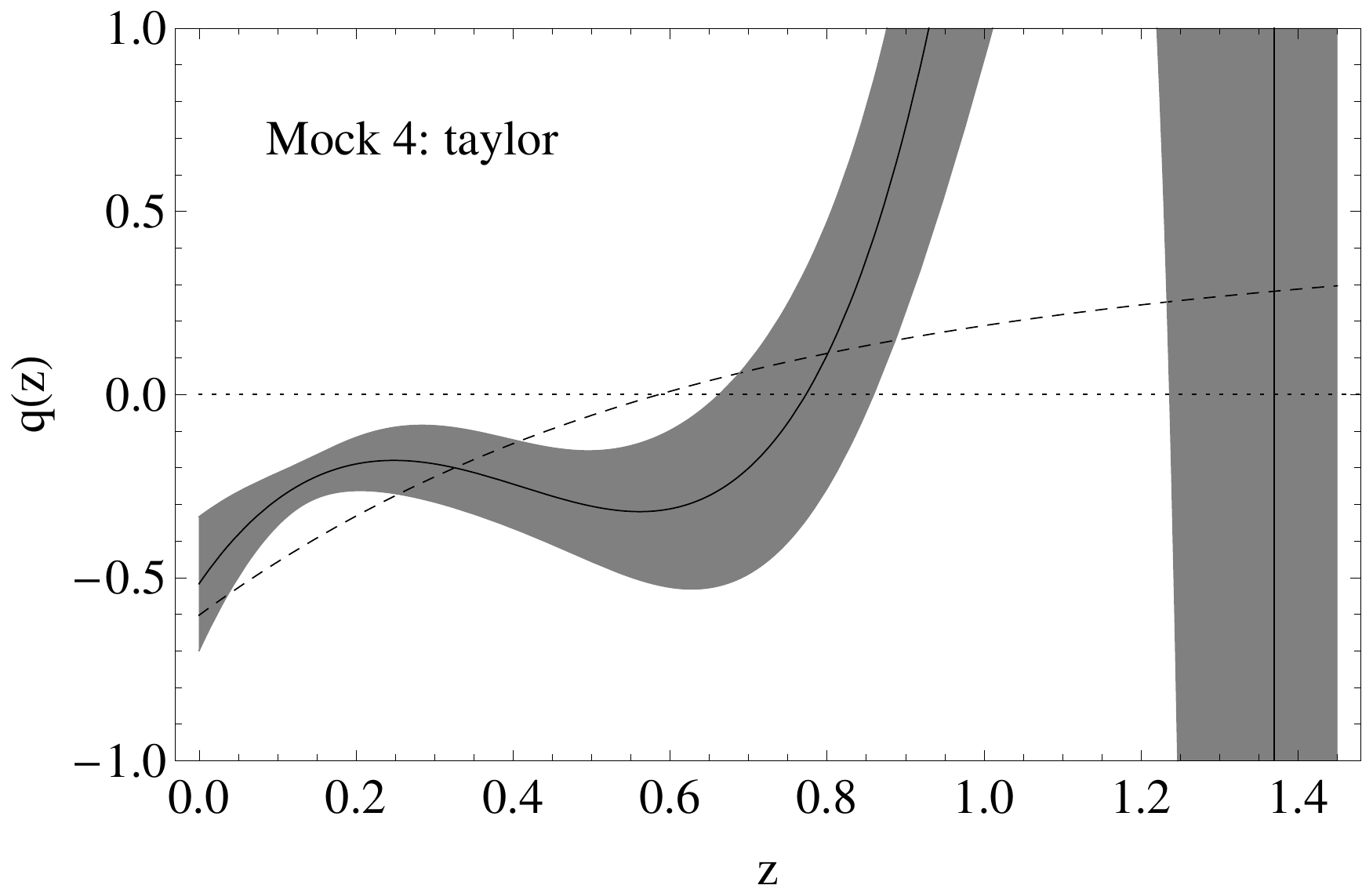}}}
\caption{The deceleration parameter $q(z)$ for all four mocks. The dashed line corresponds to the real model.\label{qztaylor}}
\end{figure*}

Instead of Taylor expanding $\Omega_{DE}(z)$, one could Taylor expand the luminosity distance instead,
\be
d_L(z)=z+A_2 z^2+A_3 z^3+A_4 z^4+A_5 z^5+A_6 z^6+\cdots.
\ee
In this case we expect the series expansion to fail at high $z$, but it should work reasonably well for small redshifts, especially since we have many more data in that range.

In Figs. \ref{dmutaylor} and \ref{qztaylor} we show the residues $\mu_{taylor}(z)-\mu_{real}(z)$ for all four mocks and the deceleration parameter $q(z)$ for all four mocks, respectively. The dashed line corresponds to the real models and we have labeled this method ``taylor".

As can be seen in Fig. \ref{qztaylor}, there is a big discrepancy between the Taylor expansion and the real models at high redshift, just as we expected, but also there seem to be singularities in the deceleration parameter that make these models unphysical.

\subsection{Chebyshev polynomials for $q(z)$}
An interesting alternative is to expand the deceleration parameter $q(z)$ in terms of Chebychev polynomials $\{T_i(z)\}_{i=0}^{M-1}$ of order $M$. The latter are a set of orthogonal polynomials that can act as a base of functions with the property that when $z\in[-1,1]$ they have the smallest maximum deviation from the true function at any given order $M$. The first few Chebyshev polynomials are $T_0(z)=1,~T_1(z)=z,~T_2(z)=-1+2 z^2,~T_3(z)=-3 z+4 z^3$. When $z\in[-1,1]$, the variable $z$ can be written as $z=\cos(\theta)$, and the polynomials can also be expressed as $T_n(\cos(\theta))=\cos(n\theta)=\cos(n\arccos(z))$, which implies that $|T_n(z)|\leq1$. Since in general our data are not in the range $[-1,1]$, we can normalize $z$ by using $\tilde{z}=\frac{2z}{z_{\rm max}}-1$ and using instead the basis $T_n\left(\tilde{z}\right)\equiv T_n(\frac{2z}{z_{\rm max}}-1)$, where $z_{\rm max}$ is the maximum value of the $N$ data $z_i$. Finally, we will mostly follow the notation of Ref. \cite{press92}.

\begin{figure*}[t!]
\centering
\vspace{0cm}\rotatebox{0}{\vspace{0cm}\hspace{0cm}\resizebox{0.43\textwidth}{!}{\includegraphics{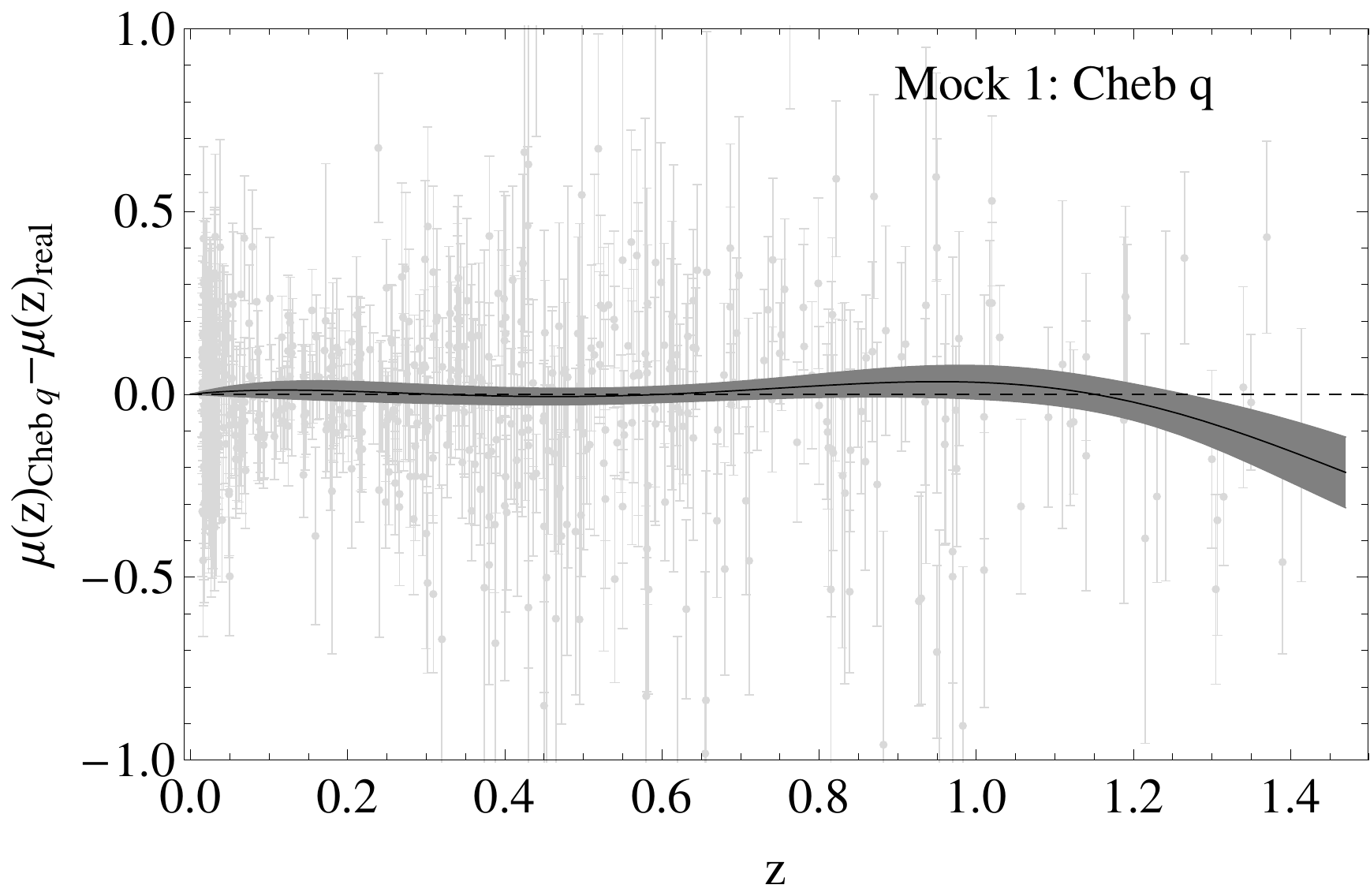}}}
\vspace{0cm}\rotatebox{0}{\vspace{0cm}\hspace{0cm}\resizebox{0.43\textwidth}{!}{\includegraphics{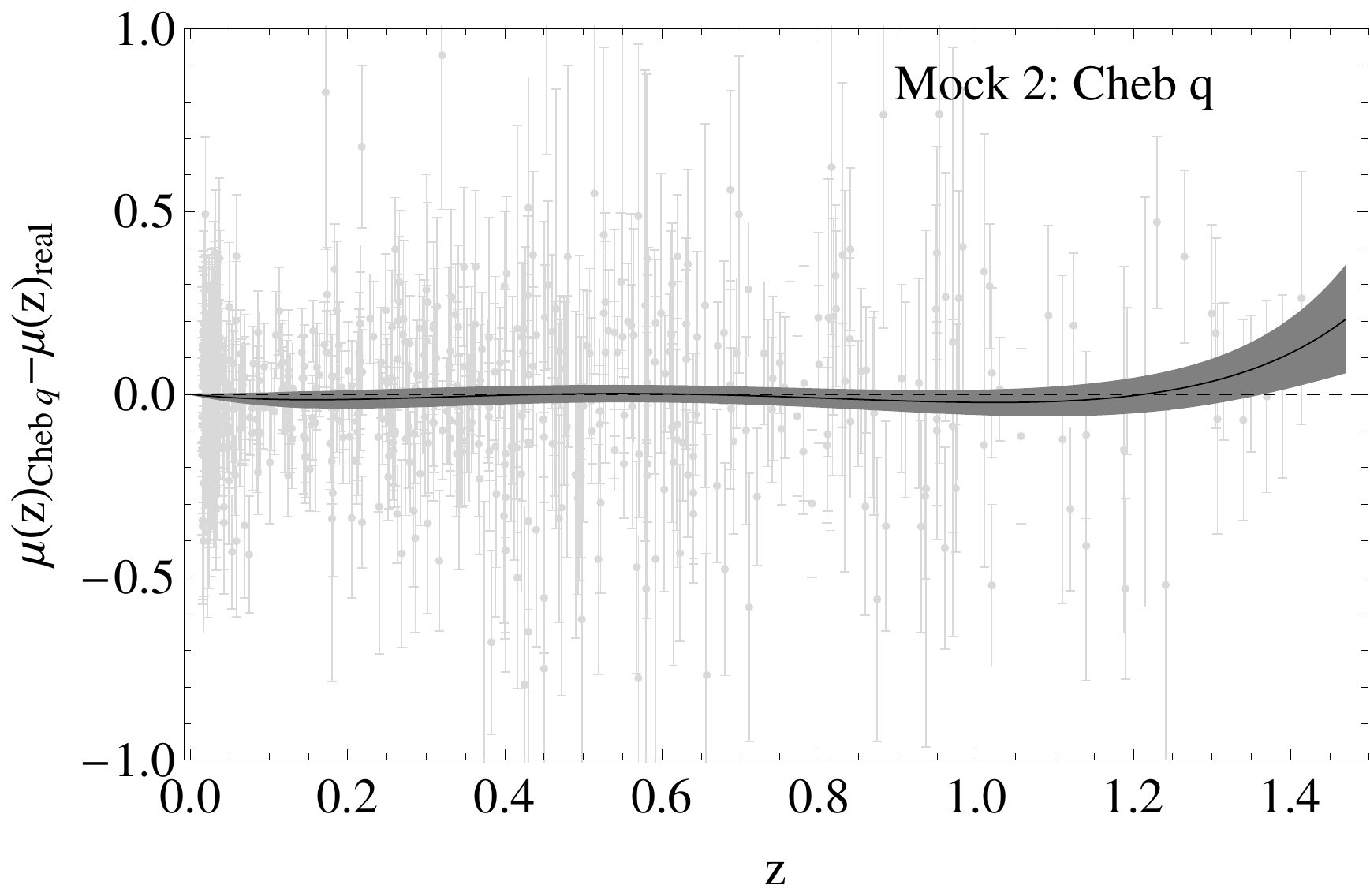}}}
\vspace{0cm}\rotatebox{0}{\vspace{0cm}\hspace{0cm}\resizebox{0.43\textwidth}{!}{\includegraphics{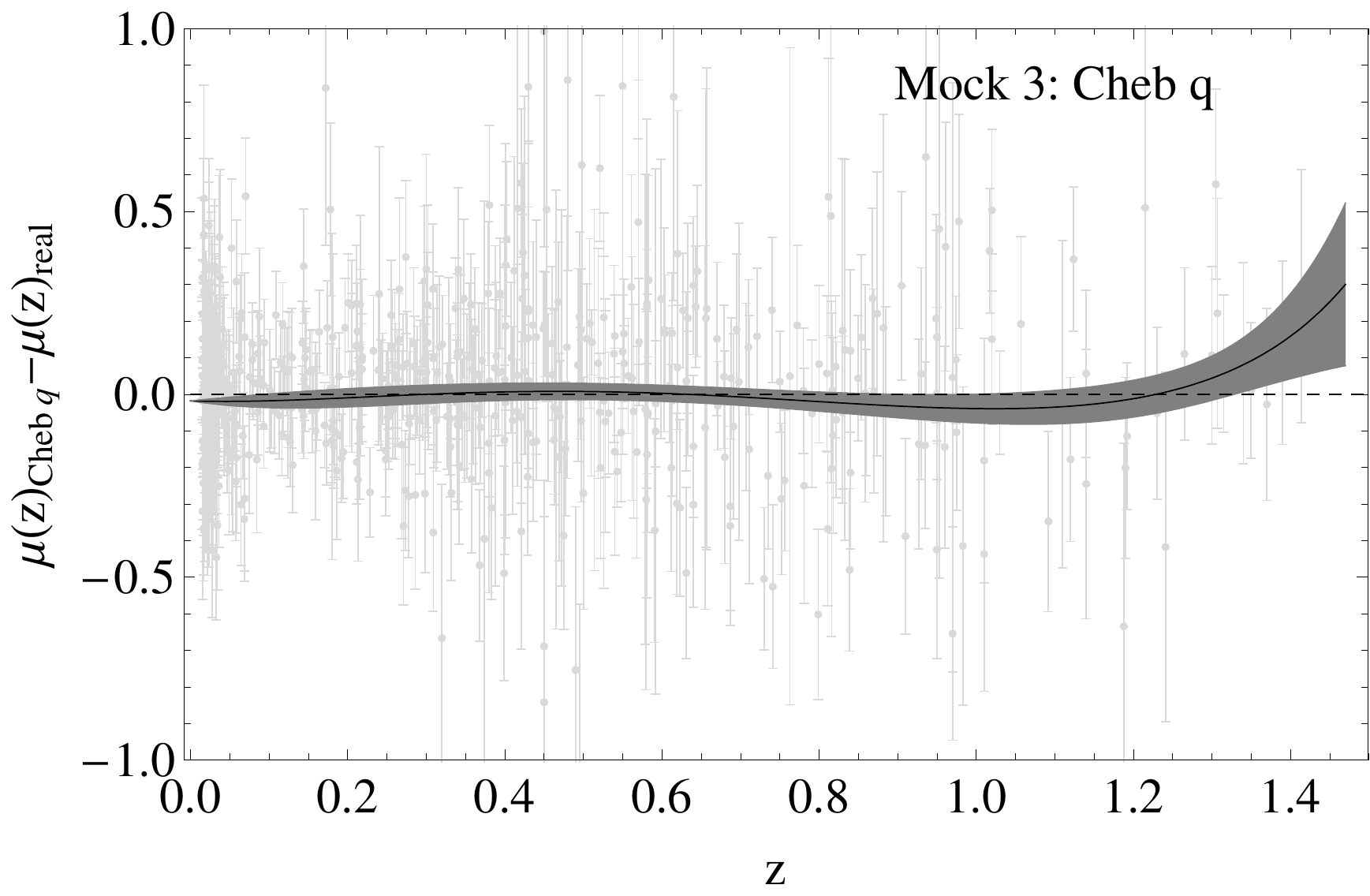}}}
\vspace{0cm}\rotatebox{0}{\vspace{0cm}\hspace{0cm}\resizebox{0.43\textwidth}{!}{\includegraphics{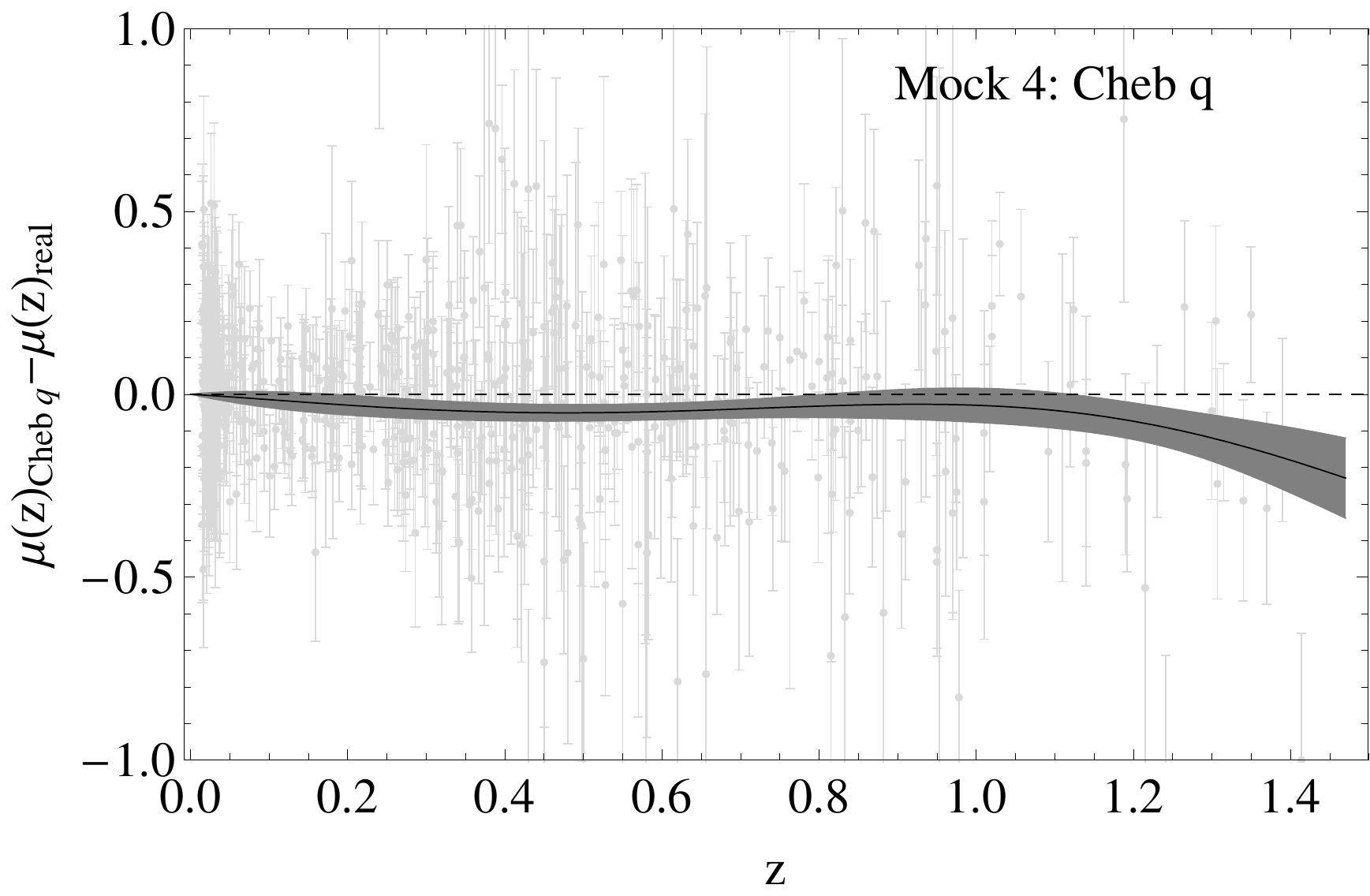}}}
\caption{The residues $\mu_{Cheb q}(z)-\mu_{real}(z)$ for all four mocks.\label{dmuChebq}}
\vspace{1cm}
\centering
\vspace{0cm}\rotatebox{0}{\vspace{0cm}\hspace{0cm}\resizebox{0.43\textwidth}{!}{\includegraphics{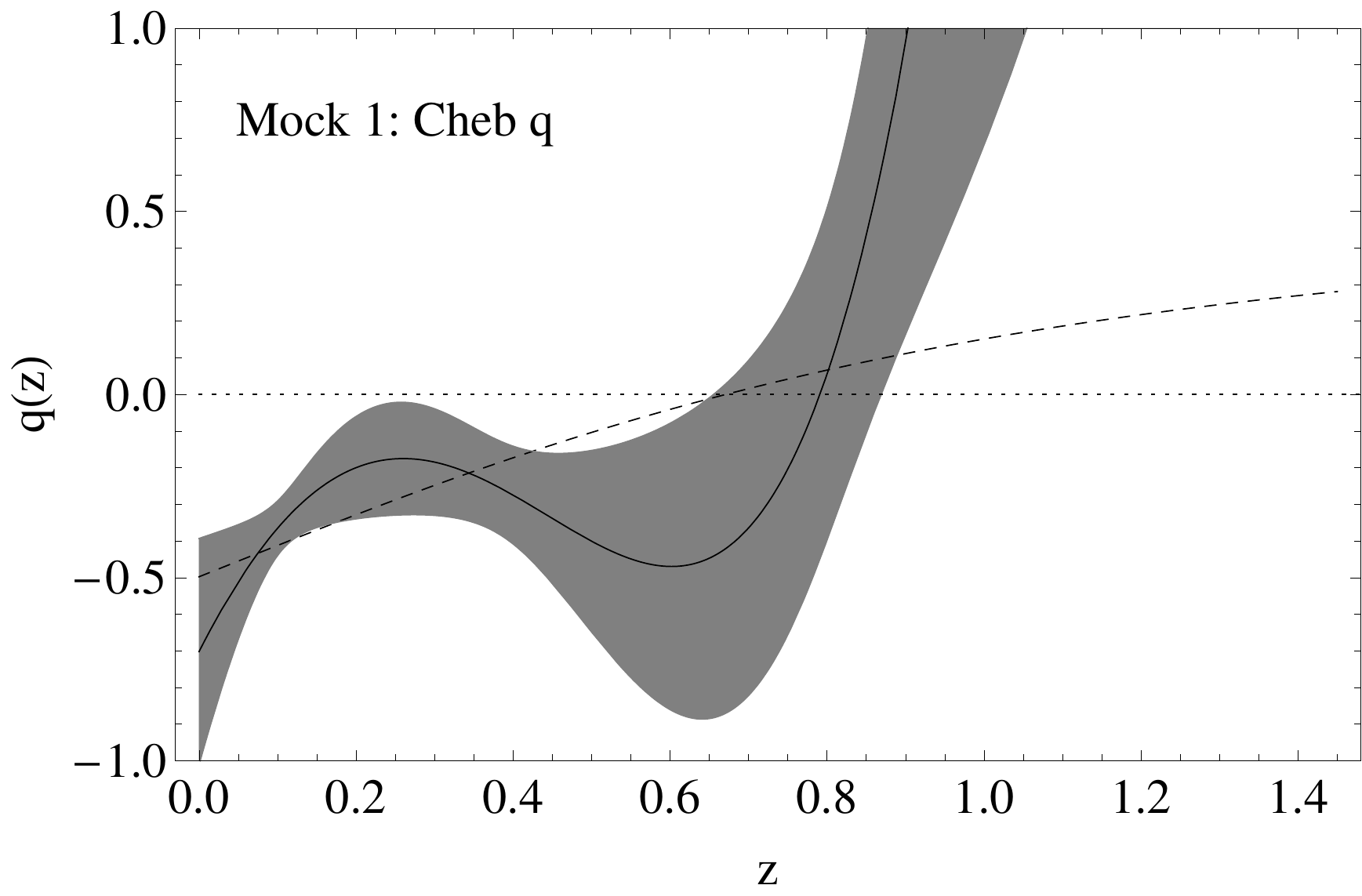}}}
\vspace{0cm}\rotatebox{0}{\vspace{0cm}\hspace{0cm}\resizebox{0.43\textwidth}{!}{\includegraphics{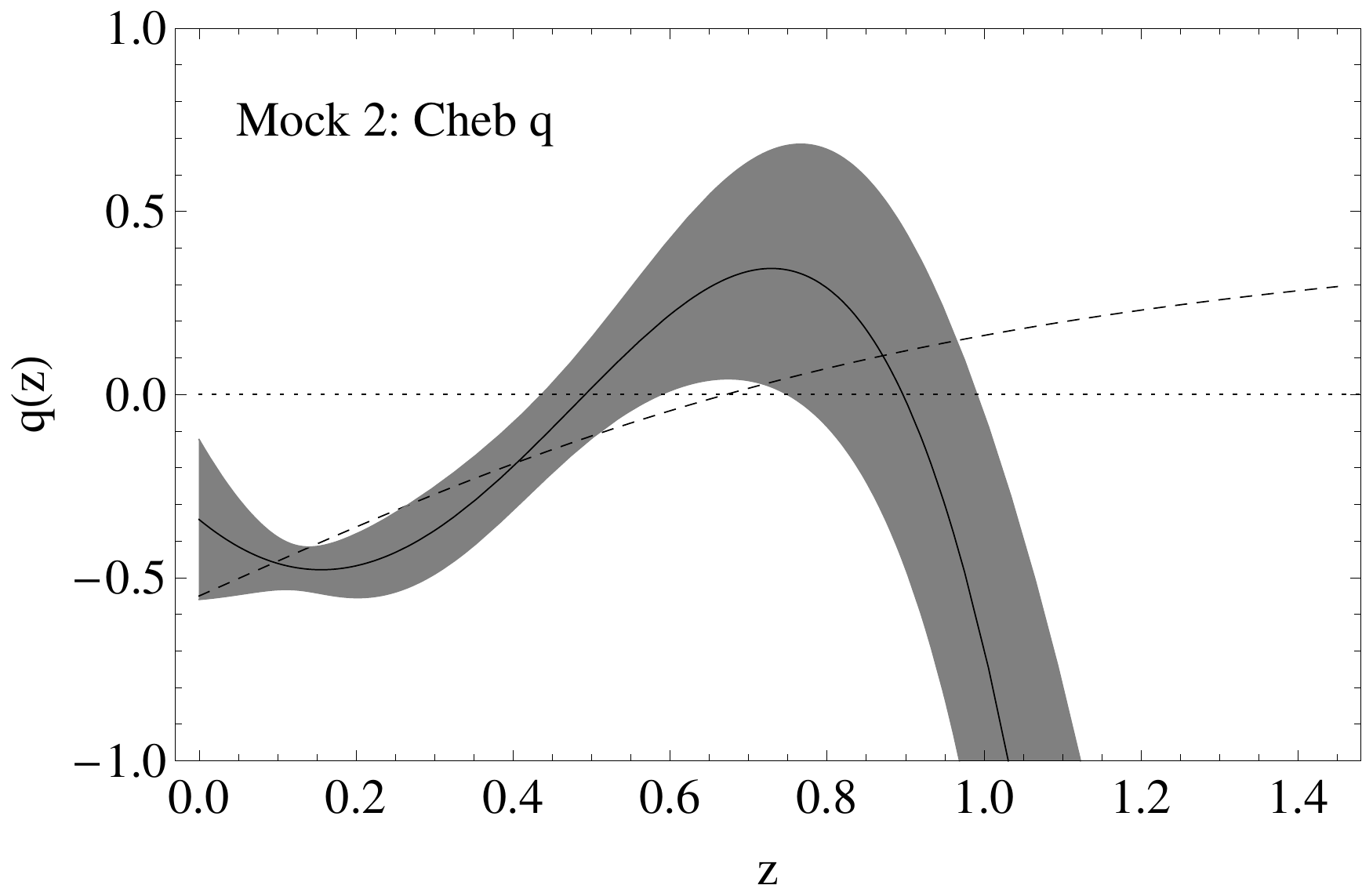}}}
\vspace{0cm}\rotatebox{0}{\vspace{0cm}\hspace{0cm}\resizebox{0.43\textwidth}{!}{\includegraphics{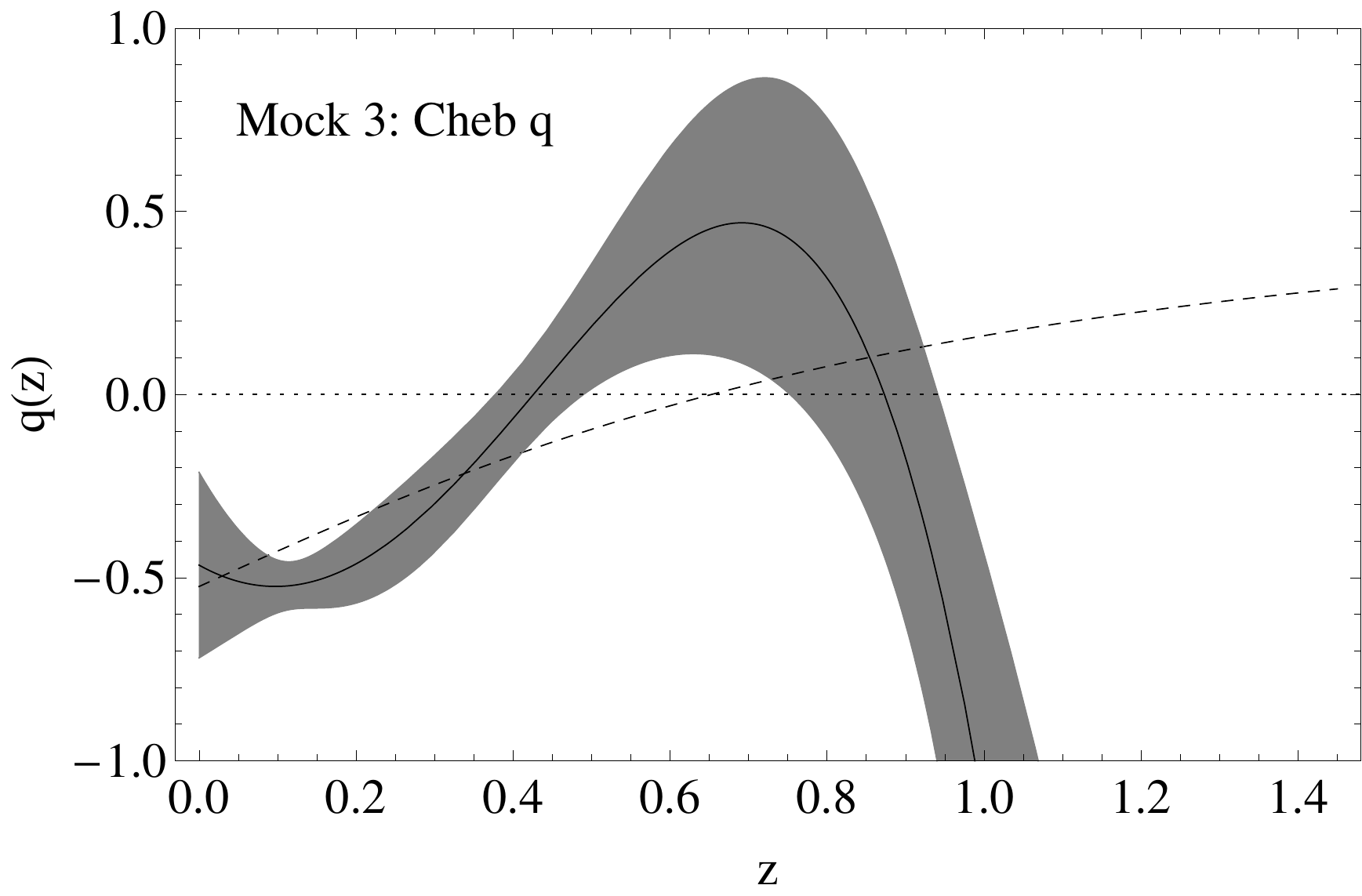}}}
\vspace{0cm}\rotatebox{0}{\vspace{0cm}\hspace{0cm}\resizebox{0.43\textwidth}{!}{\includegraphics{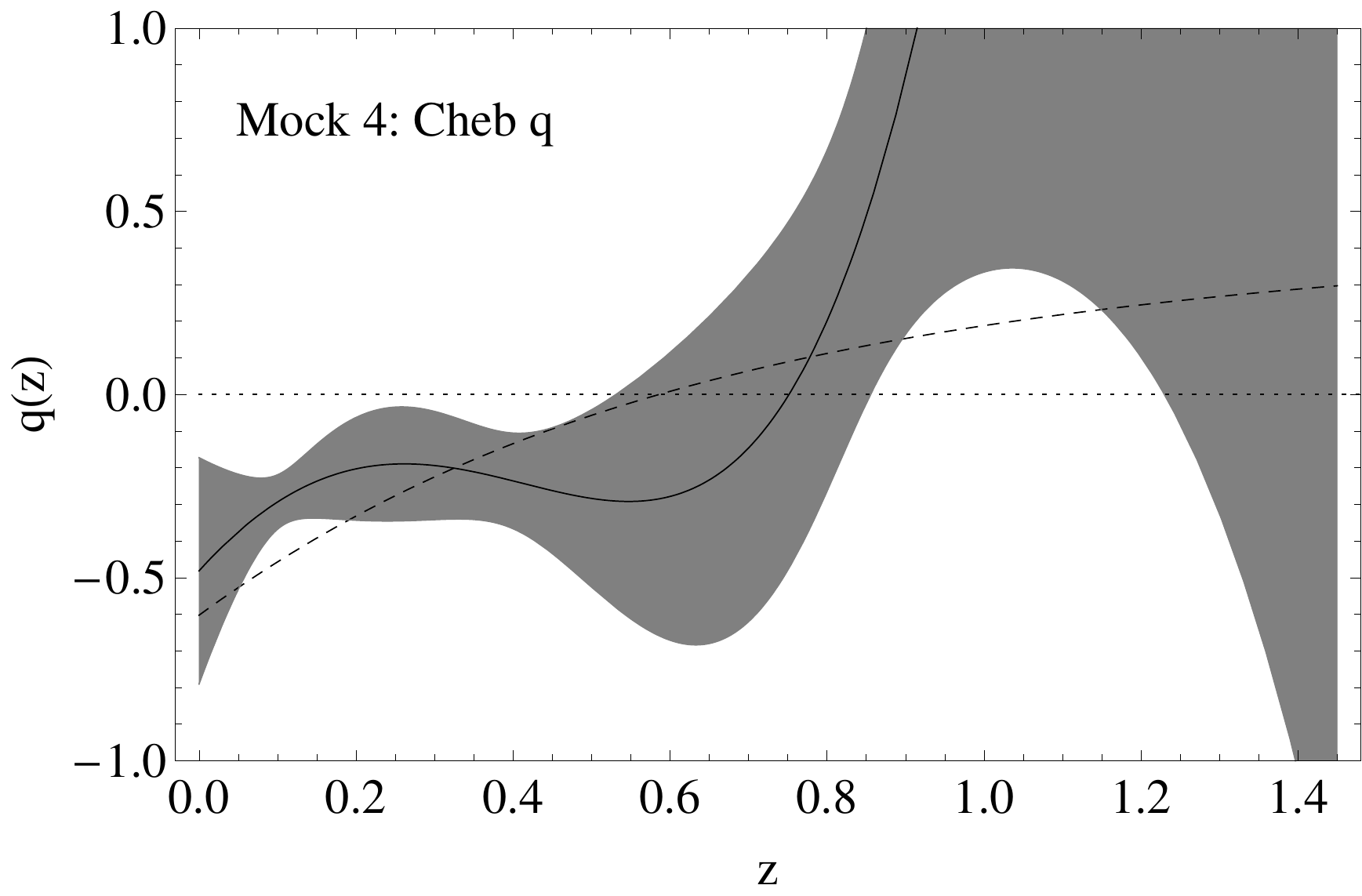}}}
\caption{The deceleration parameter $q(z)$ for all four mocks. The dashed line corresponds to the real model.\label{qzChebq}}
\end{figure*}

With these in mind, we can write the deceleration parameter as
\be
q(z)=-1+(1+z)\sum_{n=0}^M q_n T(n,\tilde{z}),
\ee
where the $q_n$ are constants. Then by keeping the first four terms we can find the Hubble parameter as
\ba
H(z)/H_0&=&e^{\int_0^z \frac{1+q(x)}{1+x} dx} \nn \\
&=&e^{\frac{8 q_3 z^4}{z_{\max }^3}+\frac{8 \left(q_2-6 q_3\right) z^3}{3 z_{\max}^2}+\frac{\left(q_1-4 q_2+9 q_3\right) z^2}{z_{\max}}+\left(q_0-q_1+q_2-q_3\right) z}\nn \\
&=& e^{A_1 z+A_2 z^2+A_3 z^3+A_4 z^4}, \label{Hcheb}
\ea
where
\ba
A_1&\equiv& q_0-q_1+q_2-q_3 \\
A_2&\equiv& \frac{q_1-4 q_2+9 q_3}{z_{\max }}\\
A_3&\equiv& \frac{8 \left(q_2-6 q_3\right)}{3 z_{\max }^2}\\
A_4&\equiv& \frac{8 q_3}{z_{\max }^3}
\ea
From Eq.(\ref{Hcheb}) it is easy to calculate the luminosity distance and fit the mock SnIa data. In Figs. \ref{dmuChebq} and \ref{qzChebq} we show the residues $\mu_{Cheb}(z)-\mu_{real}(z)$ for all four mocks and the deceleration parameter $q(z)$ for all four mocks, respectively. The dashed line corresponds to the real models and we have labeled this method as ``Cheb q".

Again, as can be seen in Fig. \ref{qzChebq}, there is a big discrepancy between the Chebyshev expansions and the real models at high redshift and again there seem to be singularities in the deceleration parameter at $z\sim1$ that make these models unphysical.

\subsection{Chebyshev polynomials for $d_L(z)$}
Similarly to the previous case, we can also expand the luminosity distance in terms of Chebyshev polynomials of up to sixth order
\be
d_L(z)=\sum_{n=0}^M A_n T(n,\tilde{z})\label{dLcheb}
\ee
where $M=6$ and $A_n$ are constants. By taking into account that
\ba
d_L(z=0)&=&0\\
d_L'(z=0)&=&1
\ea
\begin{figure*}[t!]
\centering
\vspace{0cm}\rotatebox{0}{\vspace{0cm}\hspace{0cm}\resizebox{0.43\textwidth}{!}{\includegraphics{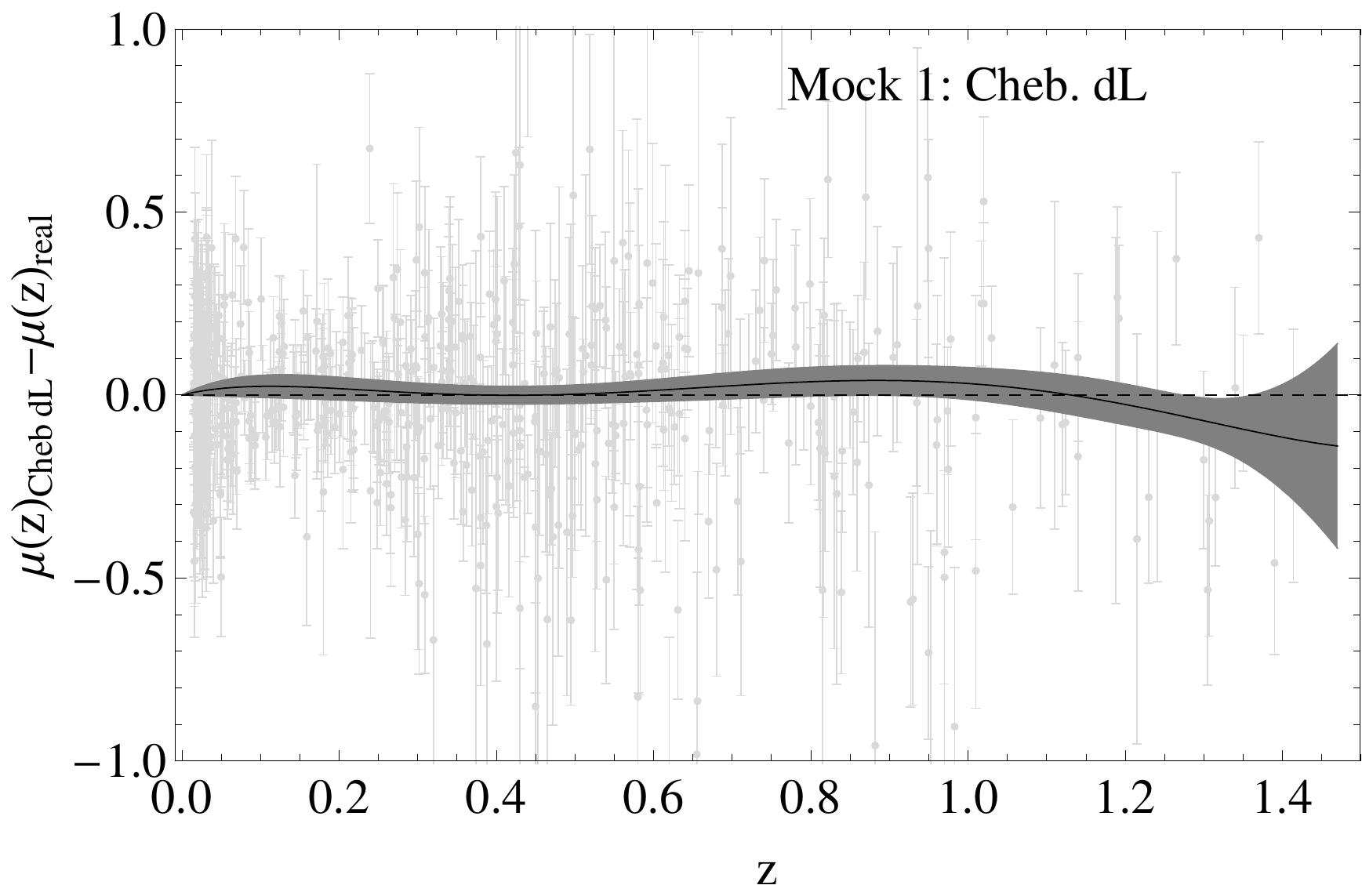}}}
\vspace{0cm}\rotatebox{0}{\vspace{0cm}\hspace{0cm}\resizebox{0.43\textwidth}{!}{\includegraphics{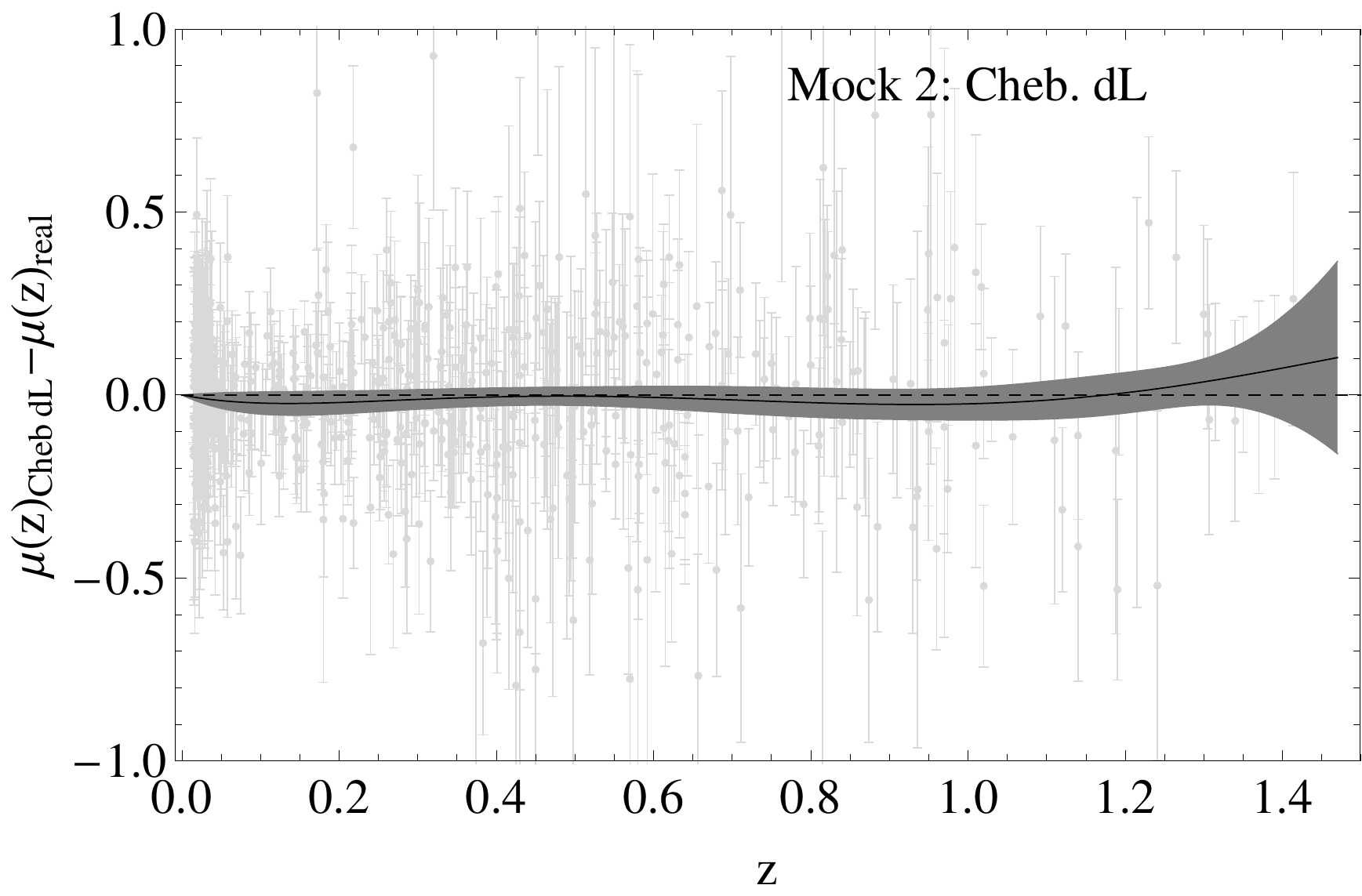}}}
\vspace{0cm}\rotatebox{0}{\vspace{0cm}\hspace{0cm}\resizebox{0.43\textwidth}{!}{\includegraphics{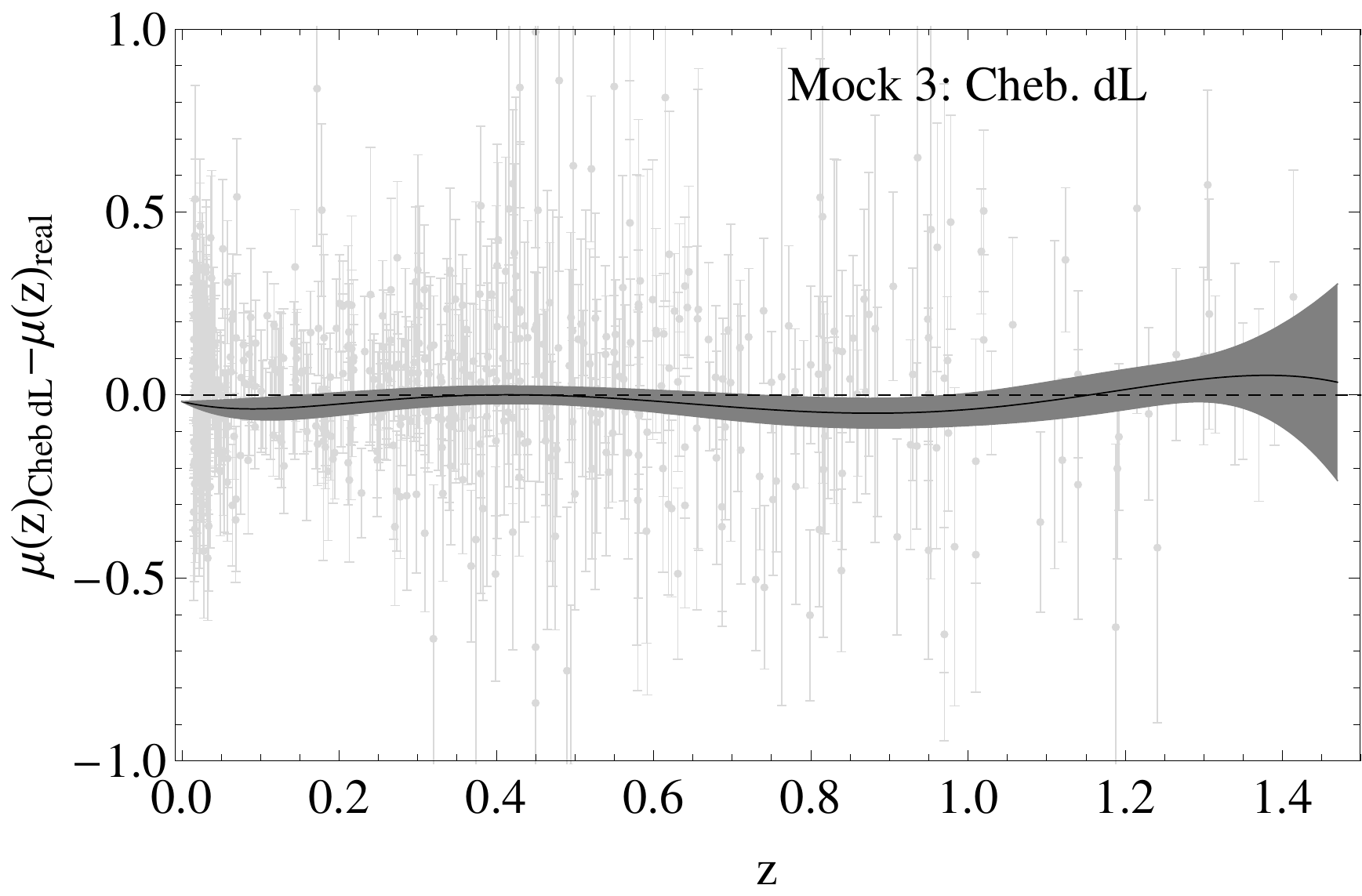}}}
\vspace{0cm}\rotatebox{0}{\vspace{0cm}\hspace{0cm}\resizebox{0.43\textwidth}{!}{\includegraphics{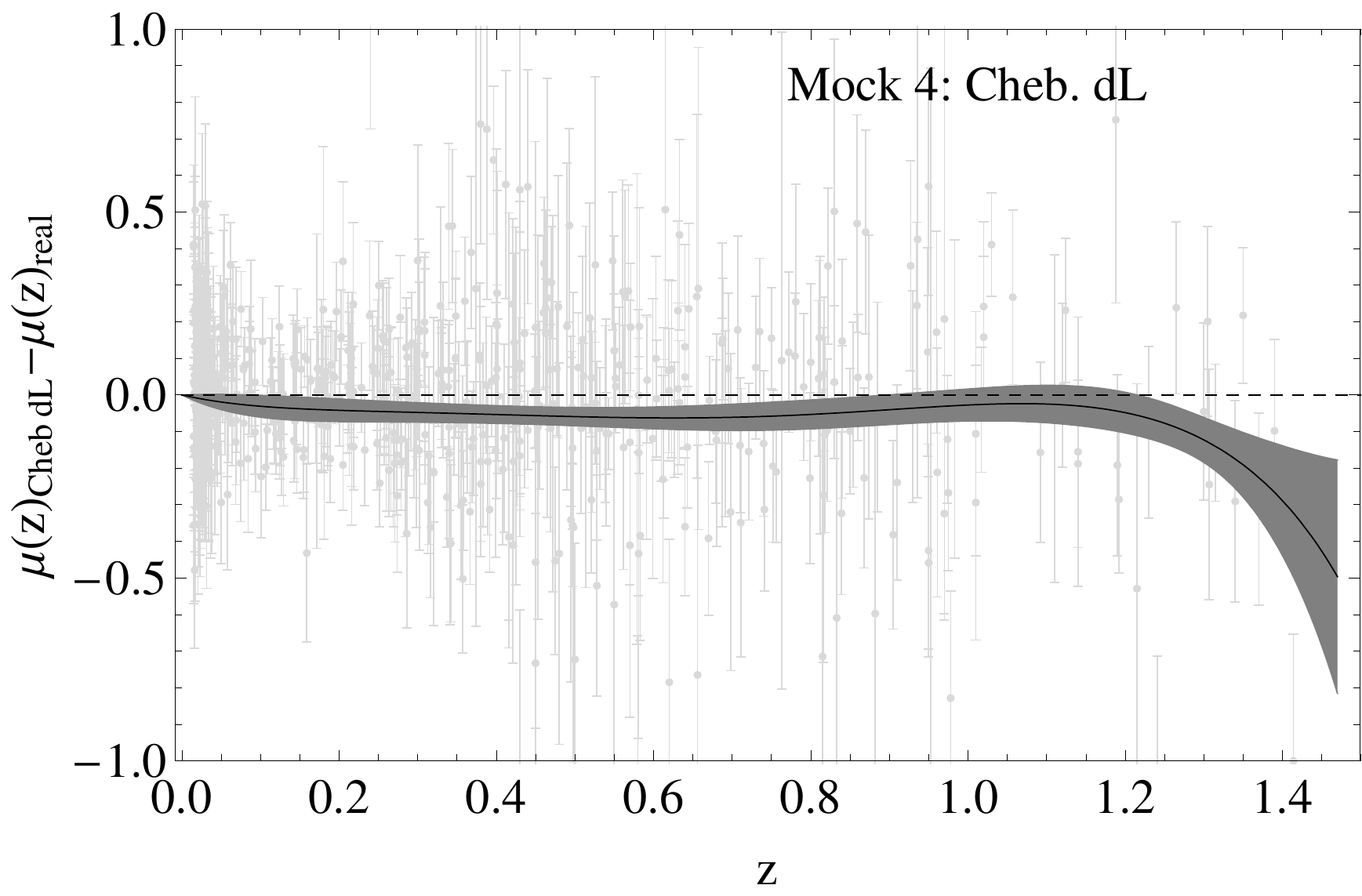}}}
\caption{The residues $\mu_{Cheb d_L}(z)-\mu_{real}(z)$ for all four mocks.\label{dmudLcheb}}
\vspace{1cm}
\centering
\vspace{0cm}\rotatebox{0}{\vspace{0cm}\hspace{0cm}\resizebox{0.43\textwidth}{!}{\includegraphics{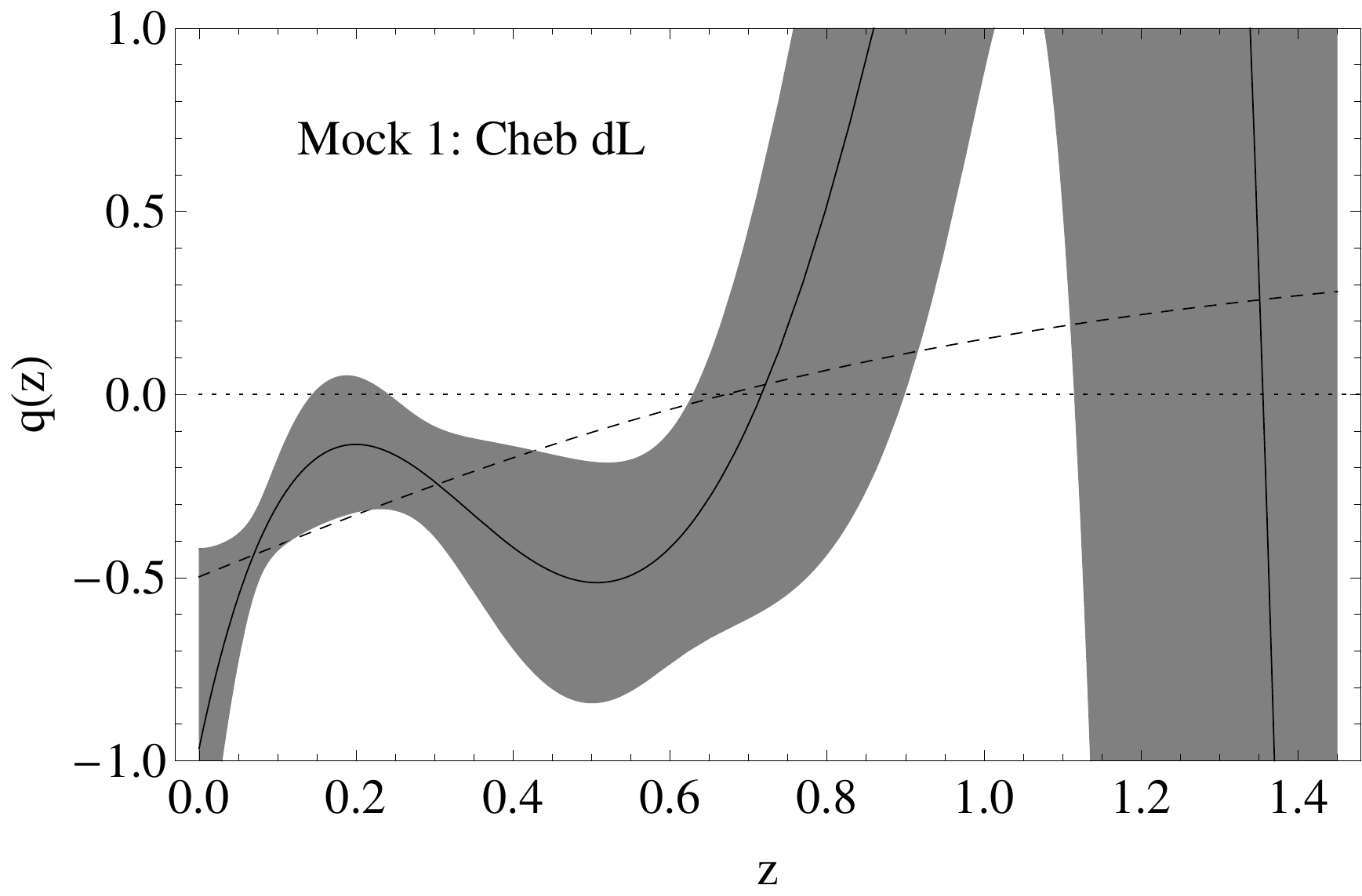}}}
\vspace{0cm}\rotatebox{0}{\vspace{0cm}\hspace{0cm}\resizebox{0.43\textwidth}{!}{\includegraphics{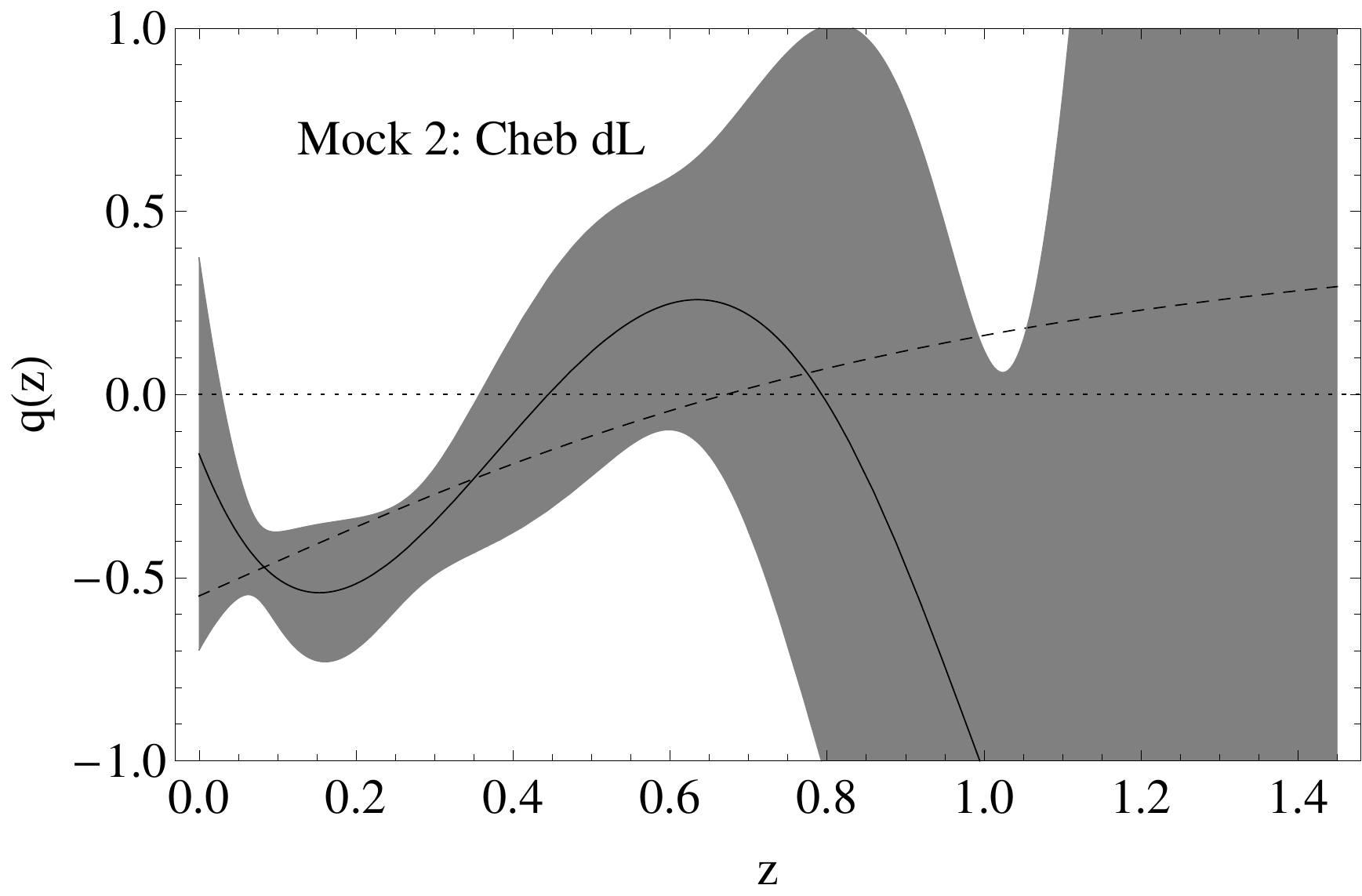}}}
\vspace{0cm}\rotatebox{0}{\vspace{0cm}\hspace{0cm}\resizebox{0.43\textwidth}{!}{\includegraphics{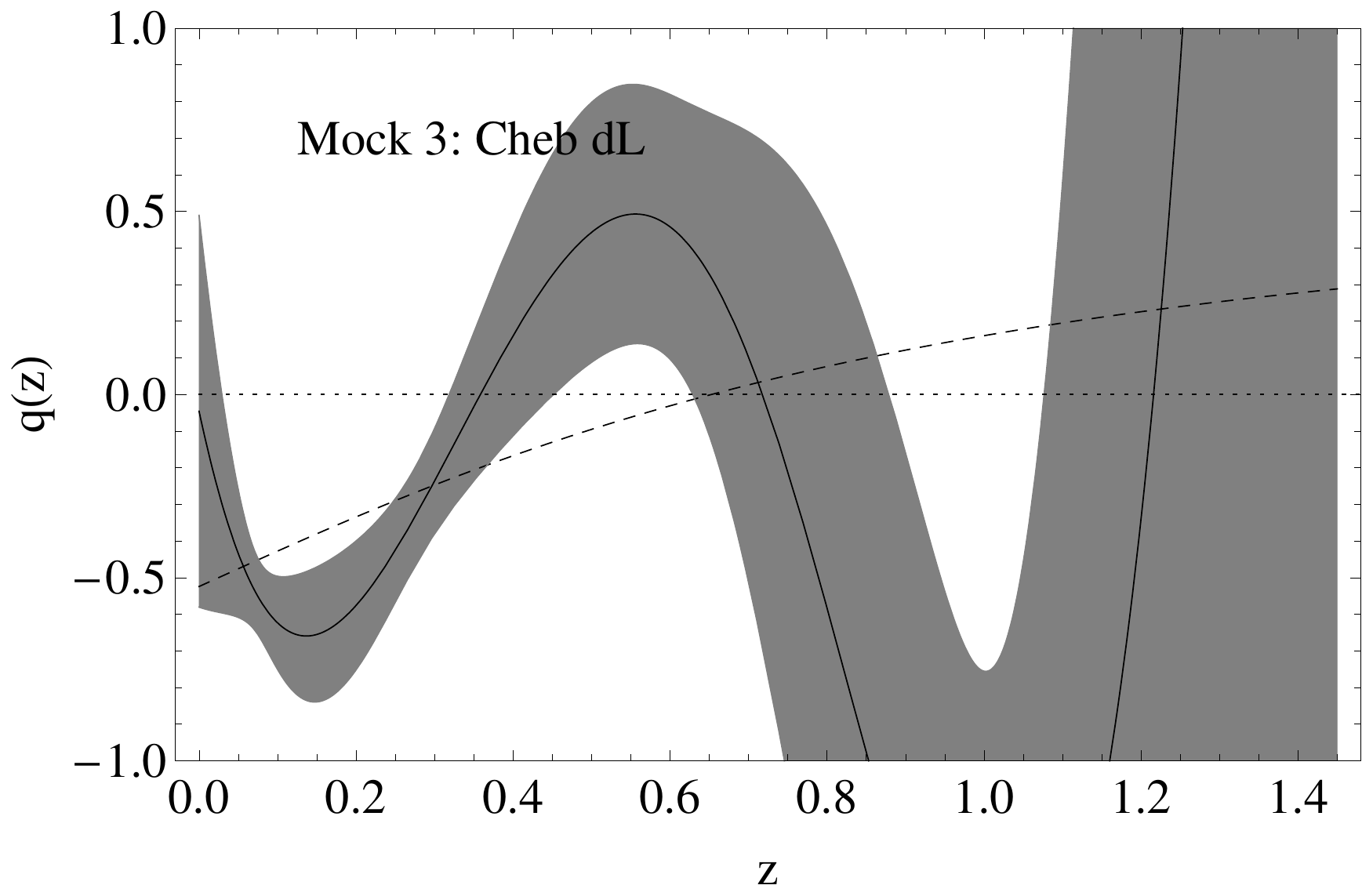}}}
\vspace{0cm}\rotatebox{0}{\vspace{0cm}\hspace{0cm}\resizebox{0.43\textwidth}{!}{\includegraphics{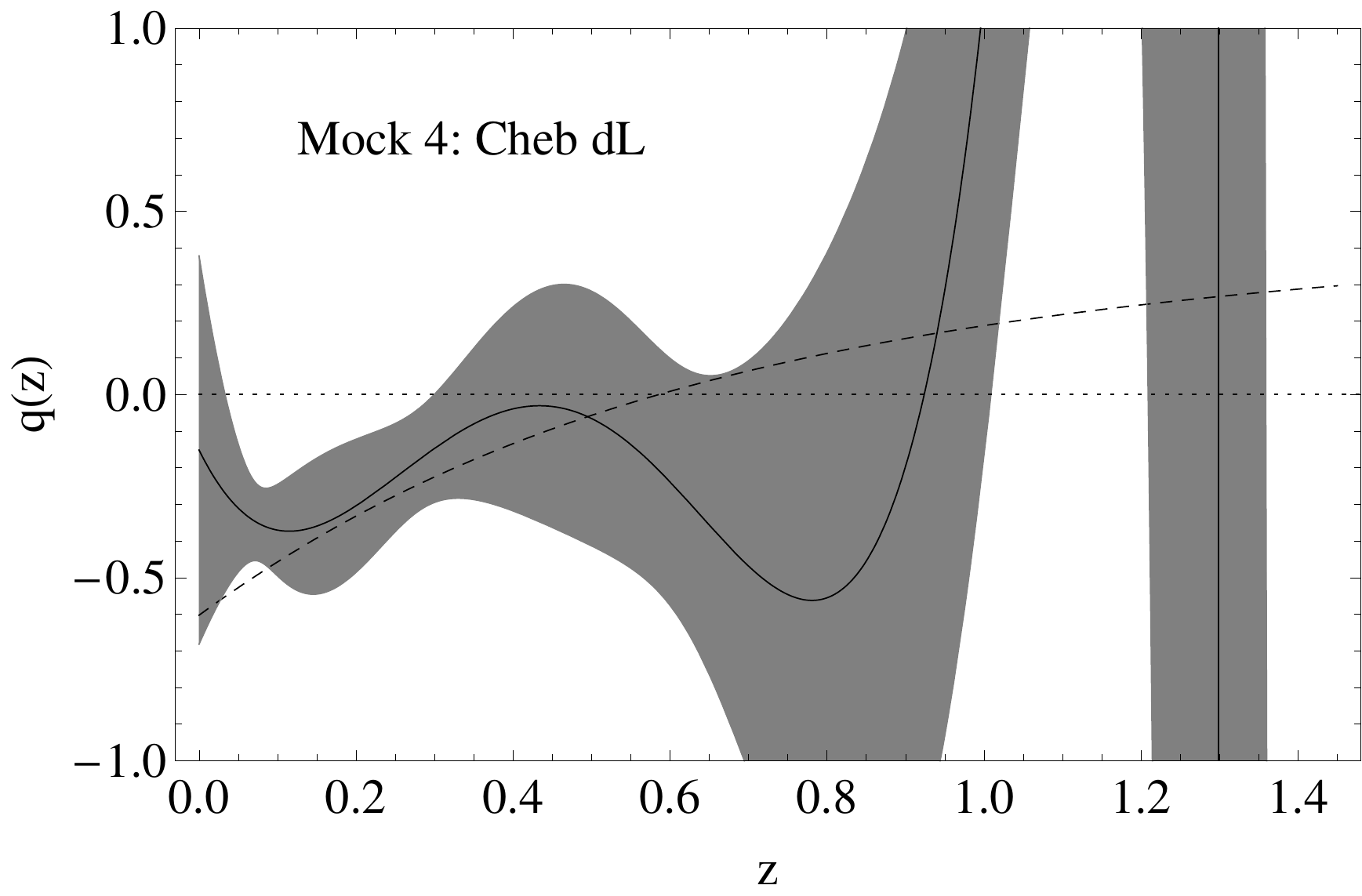}}}
\caption{The deceleration parameter $q(z)$ for all four mocks. The dashed line corresponds to the real model.\label{qzdLcheb}}
\end{figure*}
and the fact that $\tilde{z}=2\frac{z}{z_{\rm max}}-1$, Eq.~(\ref{dLcheb}) can be rewritten simply in terms of $z$ as
\ba
d_L(z)&=&z+\frac{8 \left(A_2-6 A_3+20 A_4-50 A_5+105 A_6\right) z^2}{z_{\max }^2}+\frac{32 \left(A_3-8 A_4+35 A_5-112 A_6\right) z^3}{z_{\max }^3}\nn \\&+&\frac{128 \left(A_4-10 A_5+54 A_6\right) z^4}{z_{\max }^4}+\frac{512 \left(A_5-12 A_6\right) z^5}{z_{\max }^5}+\frac{2048 A_6 z^6}{z_{\max }^6}\label{dLcheb1}
\ea
From Eq.(\ref{dLcheb1}) it is easy to calculate the luminosity distance and fit the mock SnIa data.

In Figs. \ref{dmudLcheb} and \ref{qzdLcheb} we show the residues $\mu_{chebdL}(z)-\mu_{real}(z)$ for all four mocks and the deceleration parameter $q(z)$ for all four mocks, respectively. The dashed line corresponds to the real models and we have labeled this method as ``Cheb dL".

Again, as can be seen in Fig. \ref{qzdLcheb}, there is a big discrepancy between the Chebyshev expansions and the real models at high redshift, and again there seem to be singularities in the deceleration parameter at $z\sim1$ that make these models unphysical.

\subsection{Cosmography}
One of the most commonly used approaches in the literature is to model the luminosity distance solely based on the kinematics of the expansion a method known as cosmography \cite{Visser:2004bf}. This is done by considering the higher derivatives of the scale factor up to sixth order as follows,

\ba
H(t)& \equiv &+\frac{1}{a}\frac{da}{dt}   \\
q(t)& \equiv &-\frac{1}{a H(t)^2}\frac{d^2a}{dt^2}   \\
j(t)& \equiv &+\frac{1}{a H(t)^3}\frac{d^3a}{dt^3}   \\
s(t)& \equiv &+\frac{1}{a H(t)^4}\frac{d^4a}{dt^4}   \\
l(t)& \equiv &+\frac{1}{a H(t)^5}\frac{d^5a}{dt^5}   \\
m(t)& \equiv &+\frac{1}{a H(t)^6}\frac{d^6a}{dt^6}.
\ea
Then it can be shown that the luminosity distance can be written as \cite{Aviles:2012ay}
\ba
H_0 d_L(z)&=&z+\frac{1}{2} \left(1-q_0\right) z^2+\frac{1}{6} z^3 \left(-j_0+3 q_0^2+q_0-1\right)\nn \\&+&\frac{1}{24} z^4 \left(5 j_0 \left(2 q_0+1\right)-q_0 \left(15 q_0 \left(q_0+1\right)+2\right)+s_0+2\right)\nn \\&+&\frac{1}{120} z^5 (-j_0 \left(5 q_0 \left(21 q_0+22\right)+27\right)+10 j_0^2-l_0+3 q_0 \left(q_0 \left(5 q_0 \left(7 q_0+11\right)+27\right)-5 s_0+2\right)\nn \\&-&11 s_0-6)+\frac{1}{720} z^6 (j_0 \left(5 q_0 \left(21 q_0 \left(12 q_0+19\right)+208\right)-35 s_0+168\right)-10 j_0^2 \left(28 q_0+19\right)\nn \\&-&3 q_0 \left(-7 l_0+q_0 \left(5 q_0 \left(7 q_0 \left(9 q_0+19\right)+104\right)-70 s_0+168\right)-95 s_0\right)\nn\\&+&19 l_0+m_0-24 q_0+104 s_0+24).
\ea

\begin{figure*}[t!]
\centering
\vspace{0cm}\rotatebox{0}{\vspace{0cm}\hspace{0cm}\resizebox{0.42\textwidth}{!}{\includegraphics{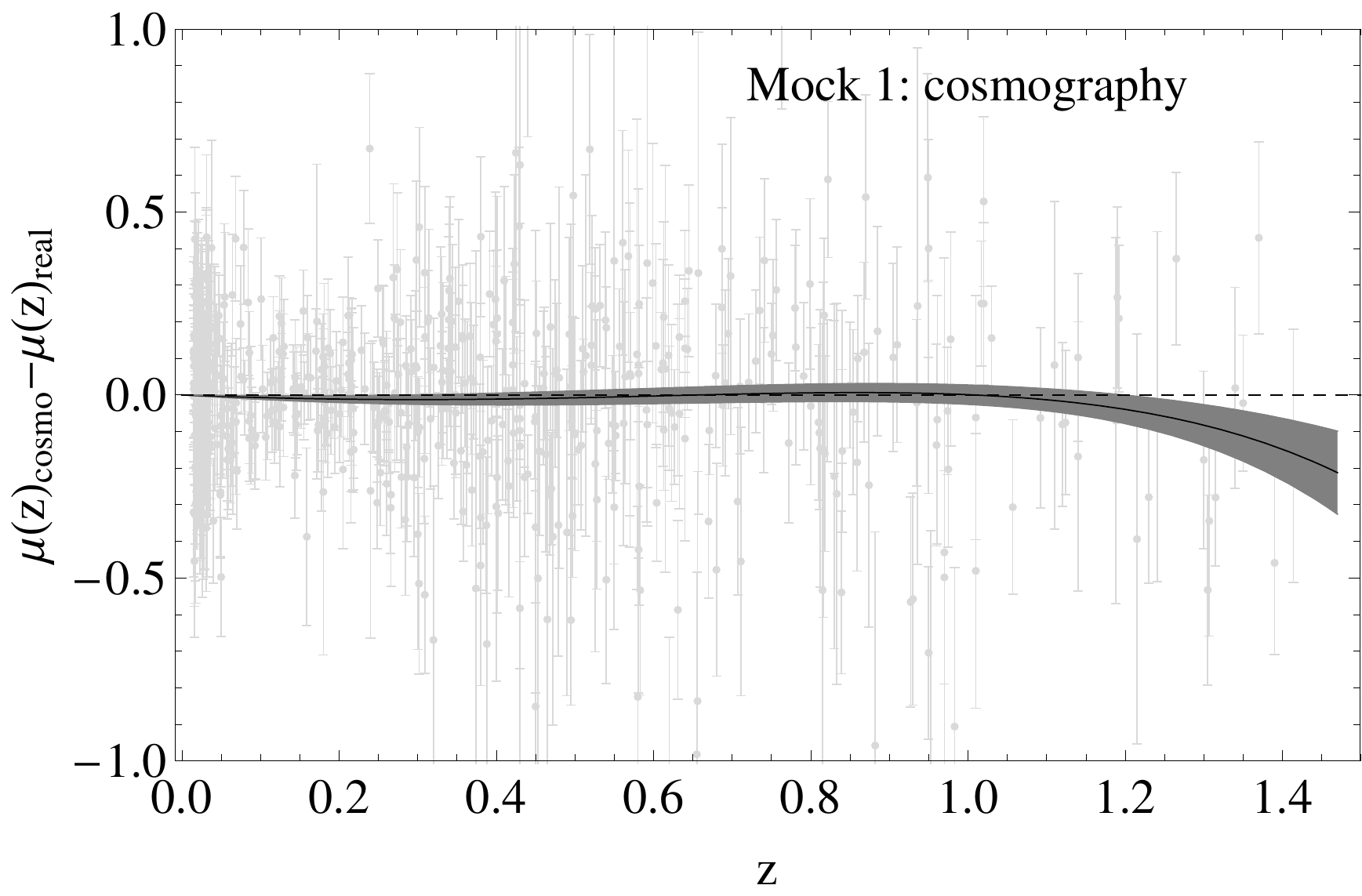}}}
\vspace{0cm}\rotatebox{0}{\vspace{0cm}\hspace{0cm}\resizebox{0.42\textwidth}{!}{\includegraphics{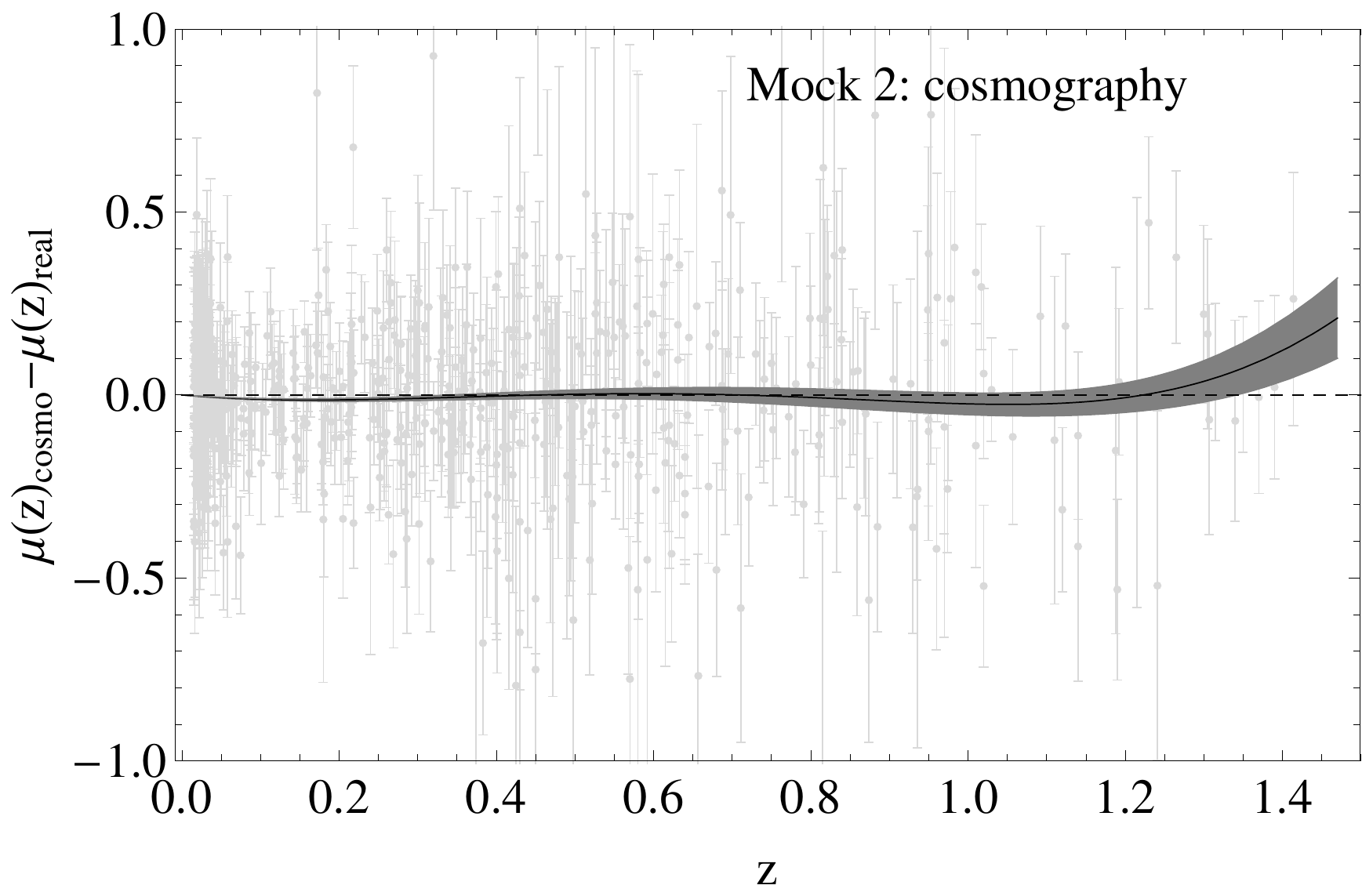}}}
\vspace{0cm}\rotatebox{0}{\vspace{0cm}\hspace{0cm}\resizebox{0.42\textwidth}{!}{\includegraphics{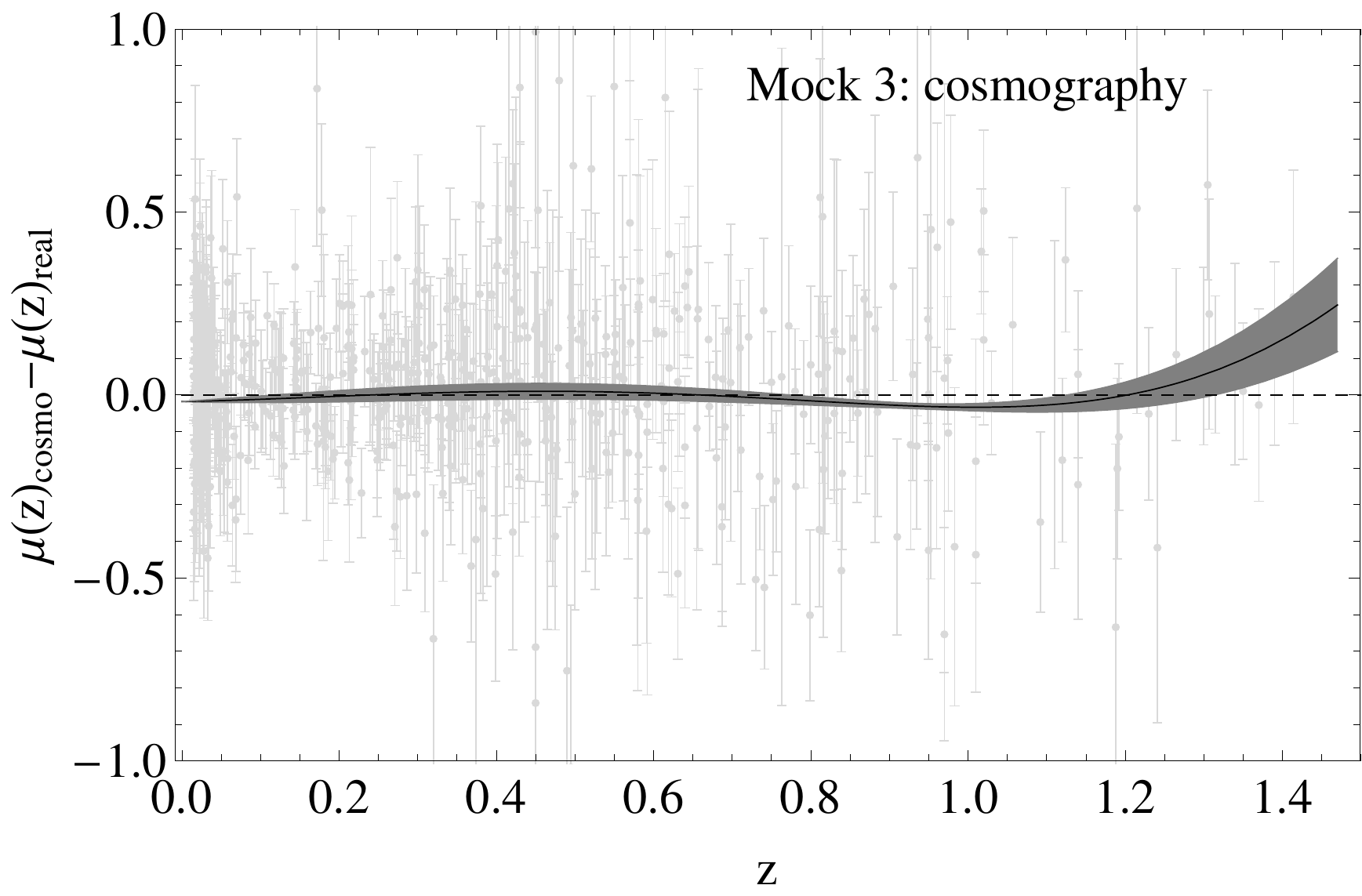}}}
\vspace{0cm}\rotatebox{0}{\vspace{0cm}\hspace{0cm}\resizebox{0.42\textwidth}{!}{\includegraphics{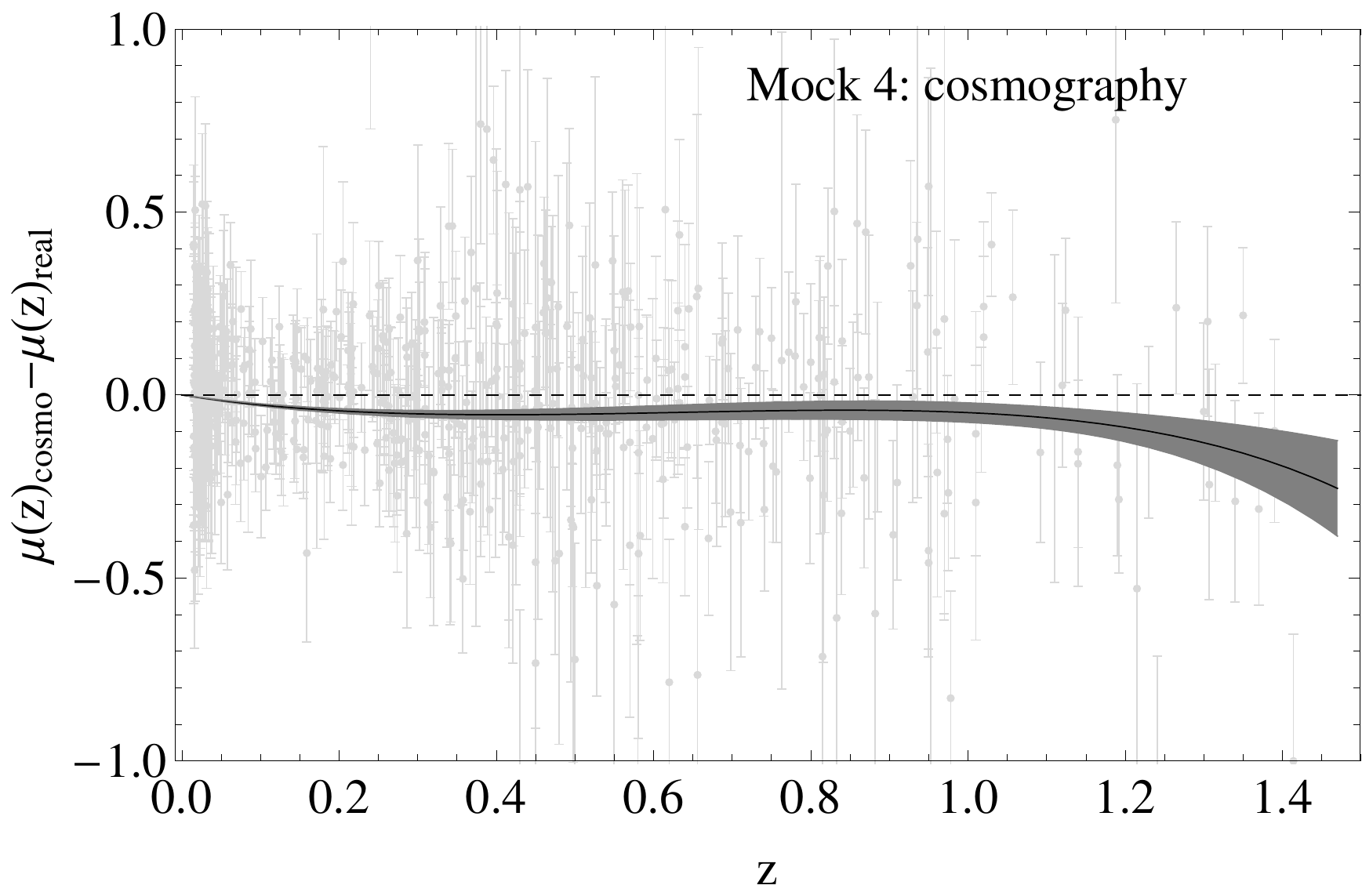}}}
\caption{The residues $\mu_{cosmo}(z)-\mu_{real}(z)$ for all four mocks.\label{dmucosmo}}
\vspace{1.5cm}
\centering
\vspace{0cm}\rotatebox{0}{\vspace{0cm}\hspace{0cm}\resizebox{0.42\textwidth}{!}{\includegraphics{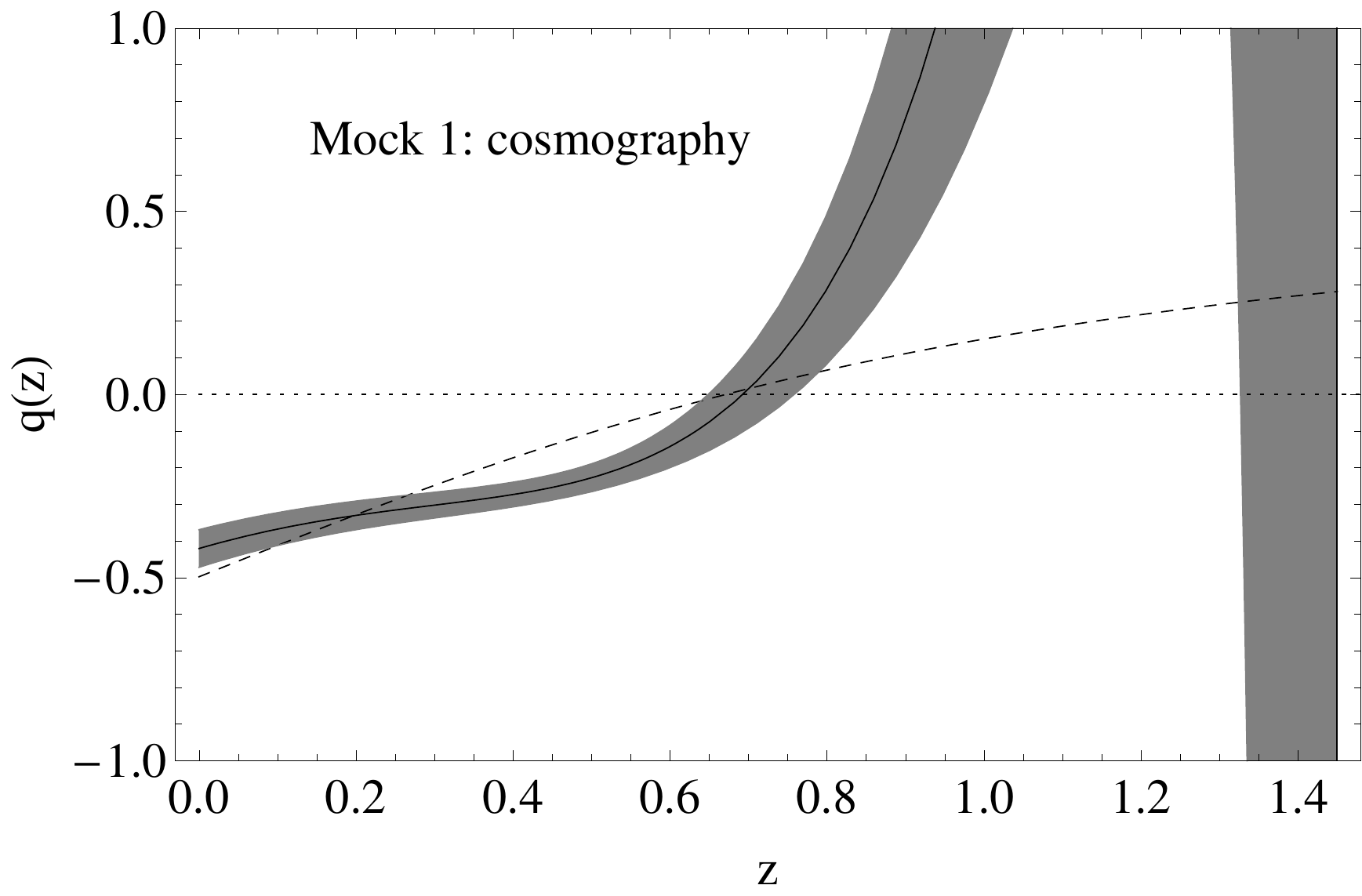}}}
\vspace{0cm}\rotatebox{0}{\vspace{0cm}\hspace{0cm}\resizebox{0.42\textwidth}{!}{\includegraphics{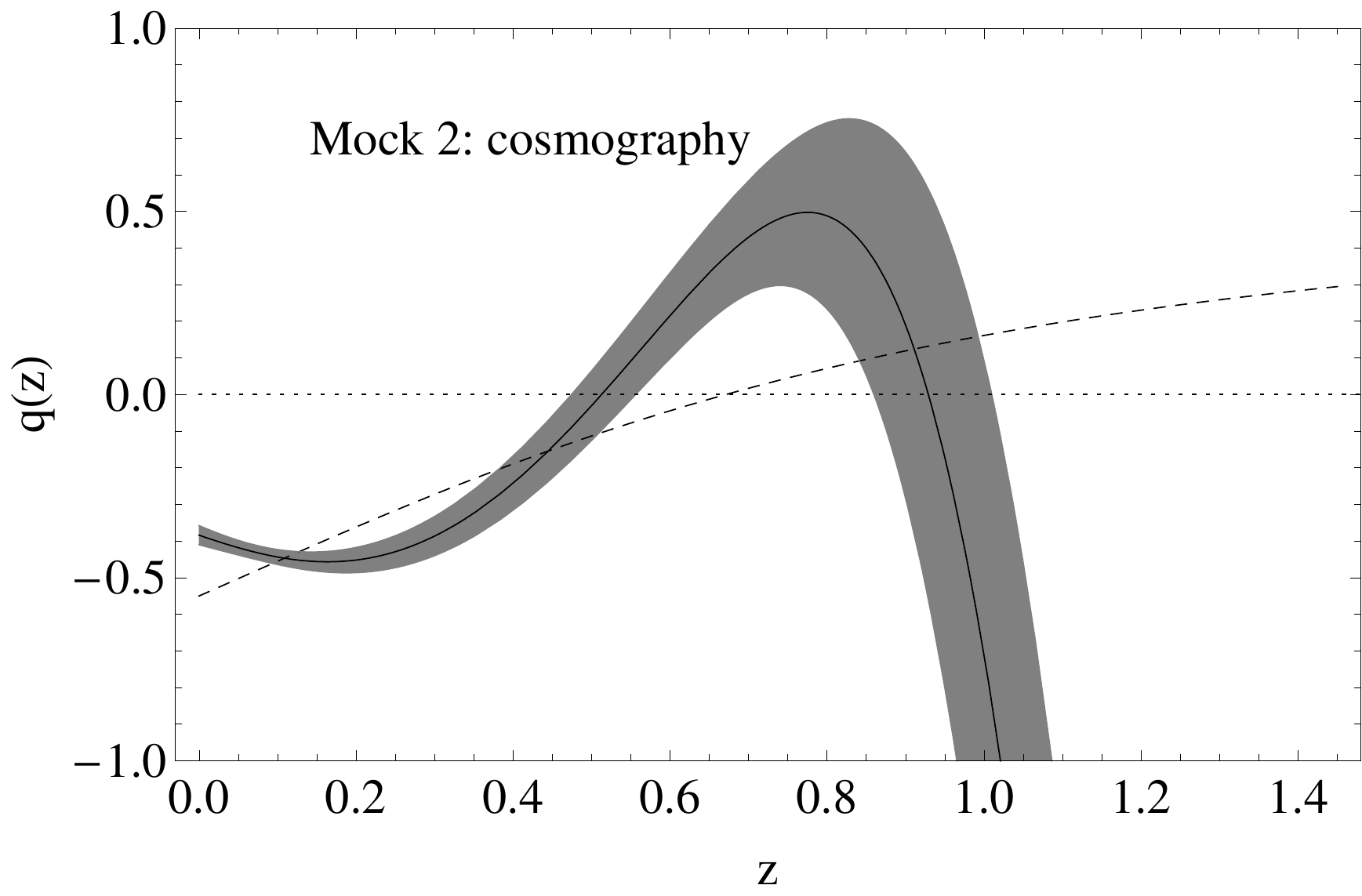}}}
\vspace{0cm}\rotatebox{0}{\vspace{0cm}\hspace{0cm}\resizebox{0.42\textwidth}{!}{\includegraphics{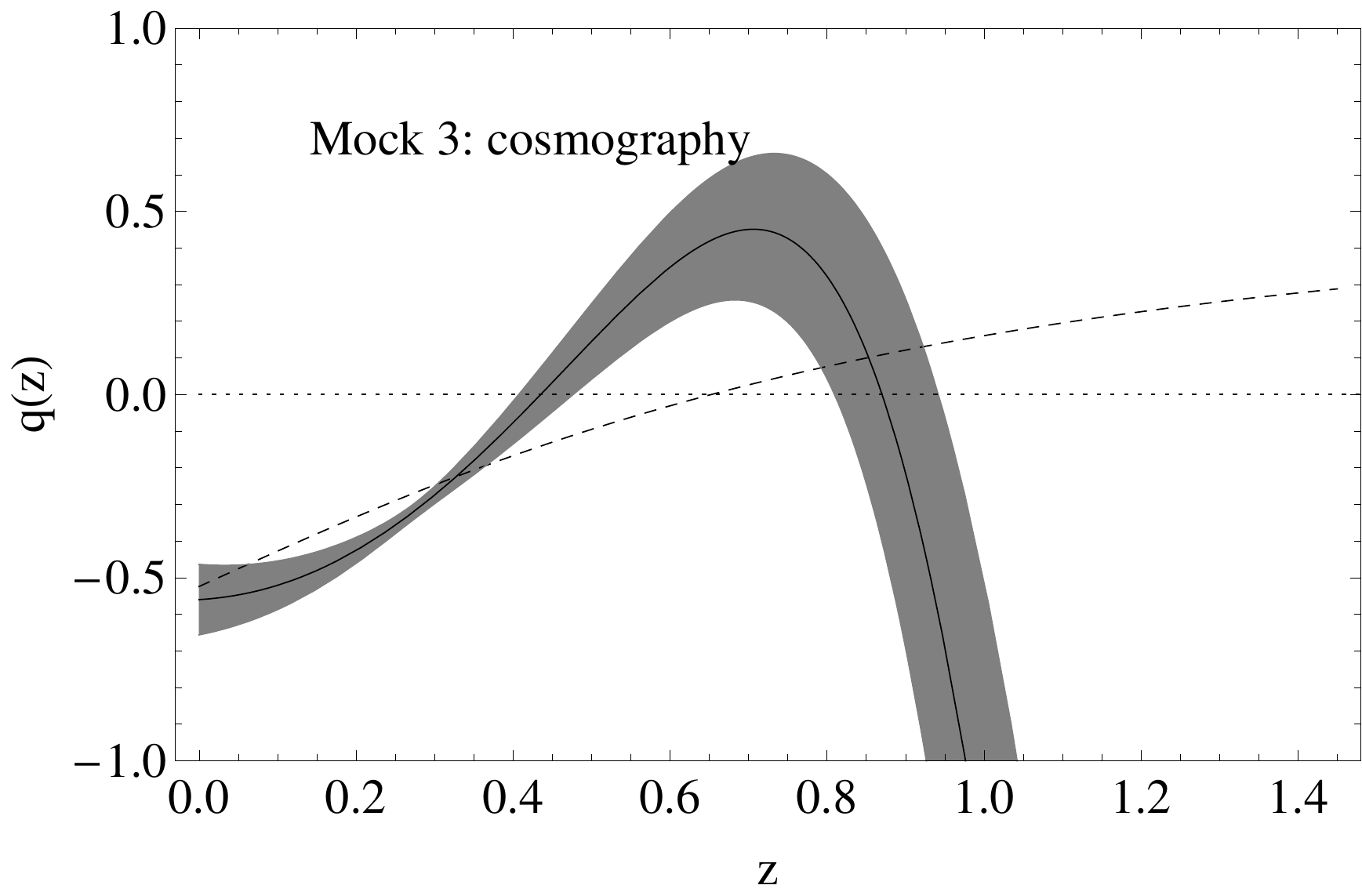}}}
\vspace{0cm}\rotatebox{0}{\vspace{0cm}\hspace{0cm}\resizebox{0.42\textwidth}{!}{\includegraphics{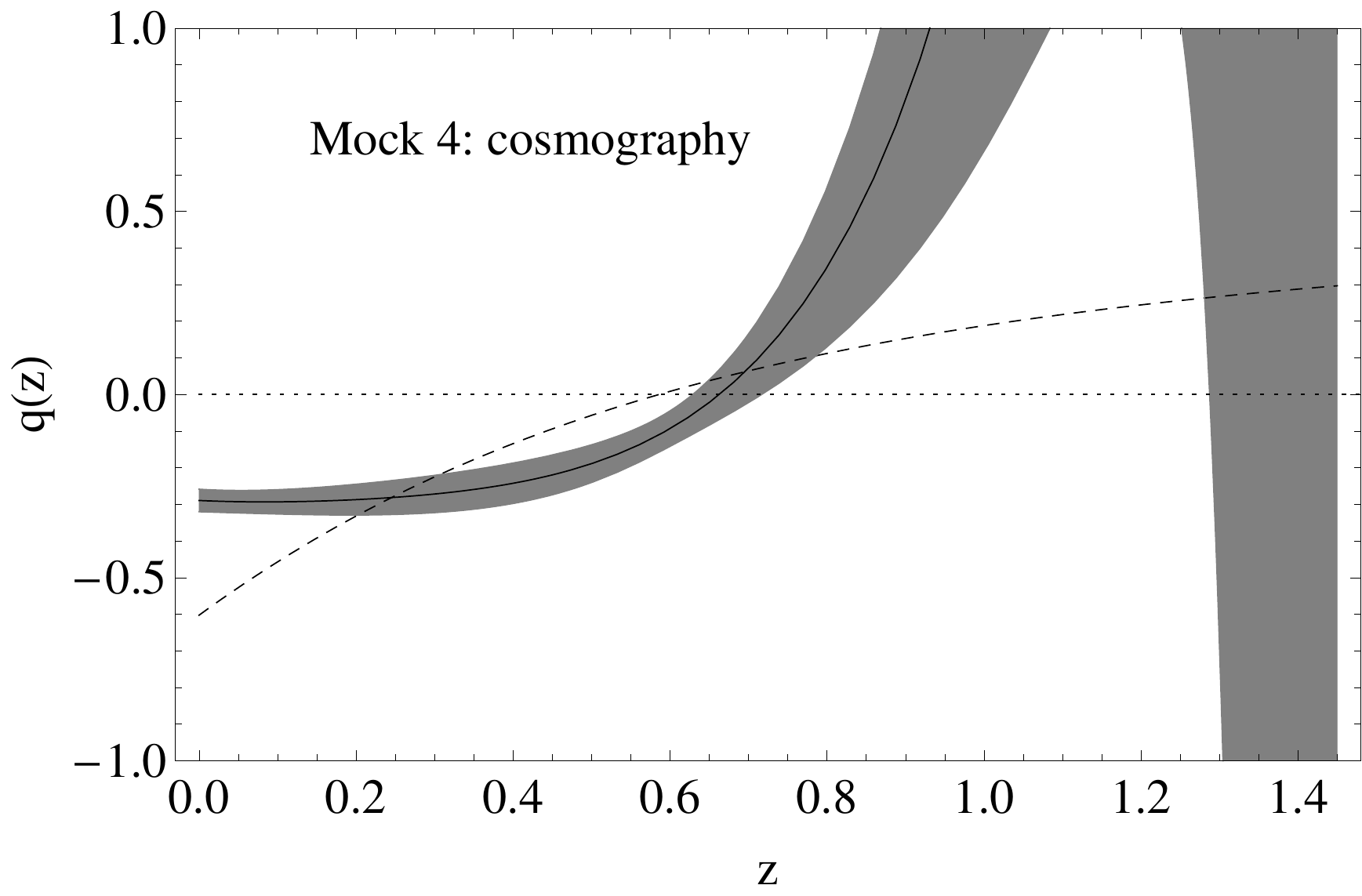}}}
\caption{The deceleration parameter $q(z)$ for all four mocks. The dashed line corresponds to the real model.\label{qzcosmo}}
\end{figure*}

In Figs. \ref{dmucosmo} and \ref{qzcosmo}, we show the residues $\mu_{cosmo}(z)-\mu_{real}(z)$ for all four mocks and the deceleration parameter $q(z)$ for all four mocks, respectively. The dashed line corresponds to the real models. As can be seen in Fig. \ref{qzcosmo}, even though cosmographic models have very small errors, unfortunately there is a big discrepancy between them and the real models at high redshift due to the presence of singularities in the deceleration parameter at $z\sim1$, thus making these models unphysical. So, the problem arises that if we keep fewer terms, say up to second order, then the cosmography models do not fit the data very well, but if we use all the terms, then the model faces the aforementioned problems.

At this point we should note that one can, in principle, continue the expansion of the cosmographic series up to an arbitrary number of terms, but that will not necessarily result in obtaining more information \cite{Vitagliano:2009et}. In order to avoid this problem, one may use, for example, a statistical criterion related to the F-test to decide the right order to truncate the expansion, as this test is specifically built for nested models, as was done in Ref. \cite{Xia:2011iv}. Also, we should stress that the cosmographic expansion may suffer from lack of convergence at $z\gtrsim1$. This, too, is a well known problem in the literature (see for example Ref. \cite{Vitagliano:2009et}) and many different parametrizations have been proposed to solve it, e.g. expanding in terms of $\frac{z}{1+z}$ instead of just $z$, but we will not discuss this further.

\subsection{$w_{0}w_{a}CDM$ models}
\begin{figure*}[t!]
\centering
\vspace{0cm}\rotatebox{0}{\vspace{0cm}\hspace{0cm}\resizebox{0.42\textwidth}{!}{\includegraphics{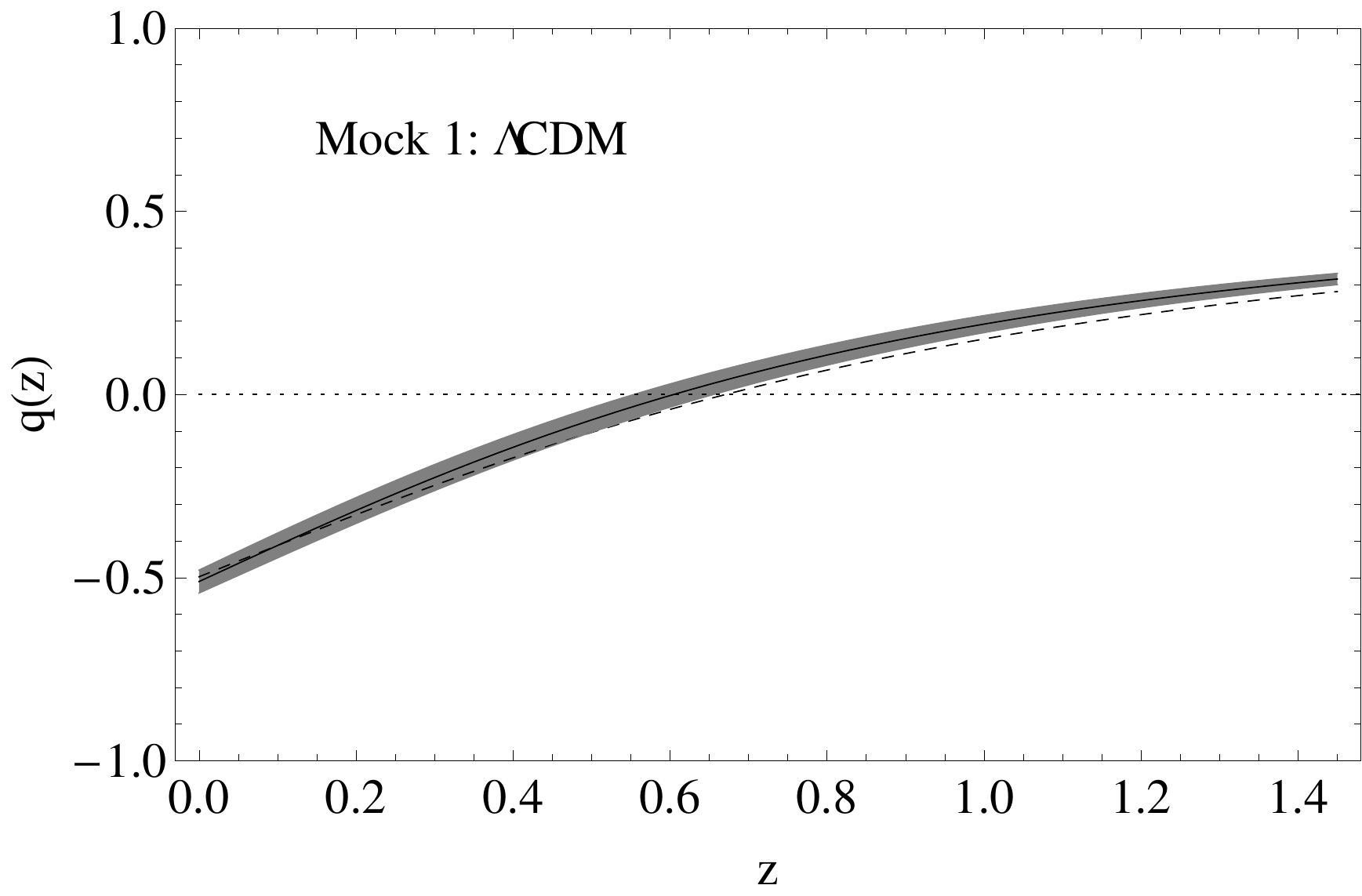}}}
\vspace{0cm}\rotatebox{0}{\vspace{0cm}\hspace{0cm}\resizebox{0.42\textwidth}{!}{\includegraphics{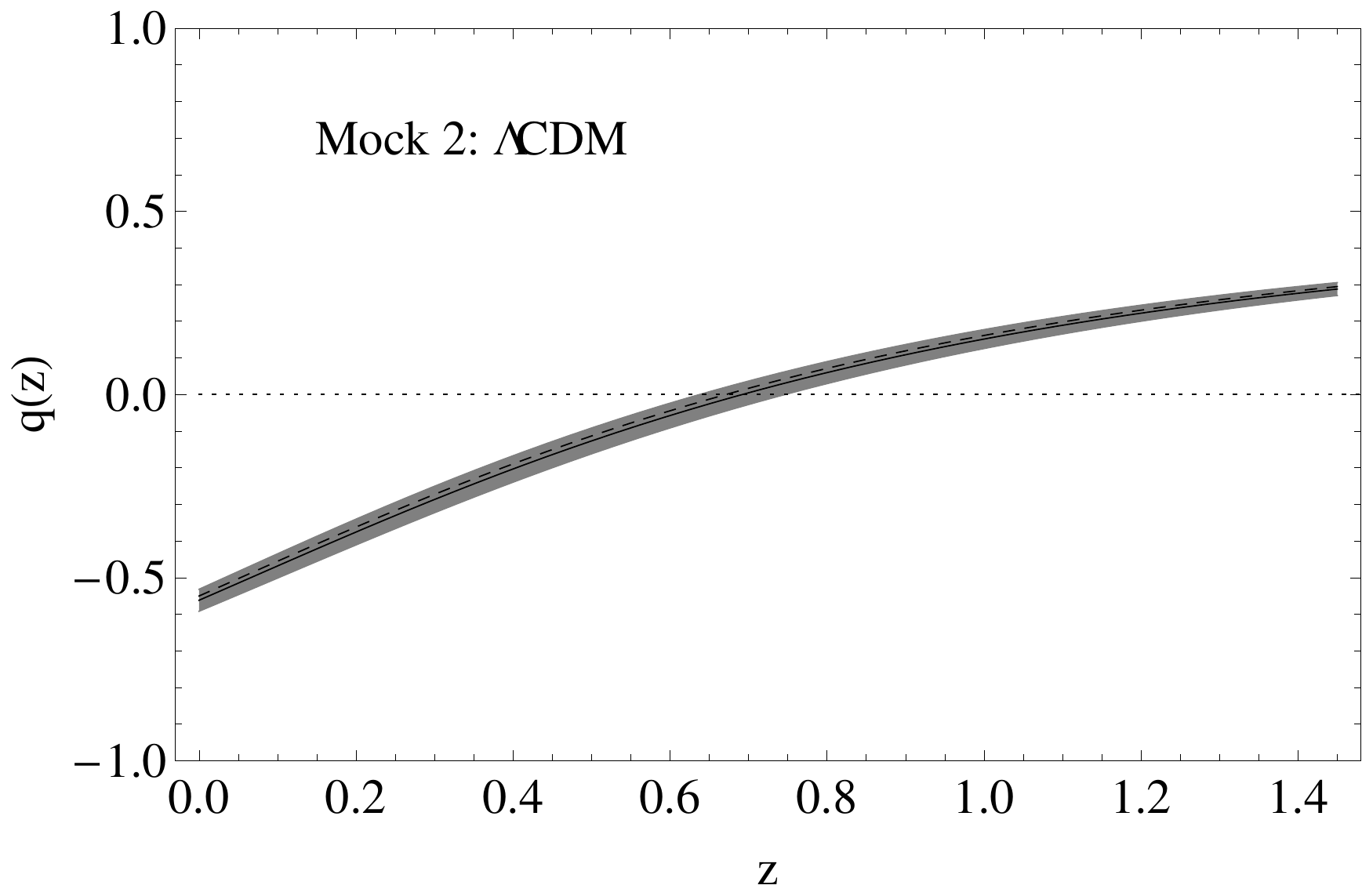}}}
\vspace{0cm}\rotatebox{0}{\vspace{0cm}\hspace{0cm}\resizebox{0.42\textwidth}{!}{\includegraphics{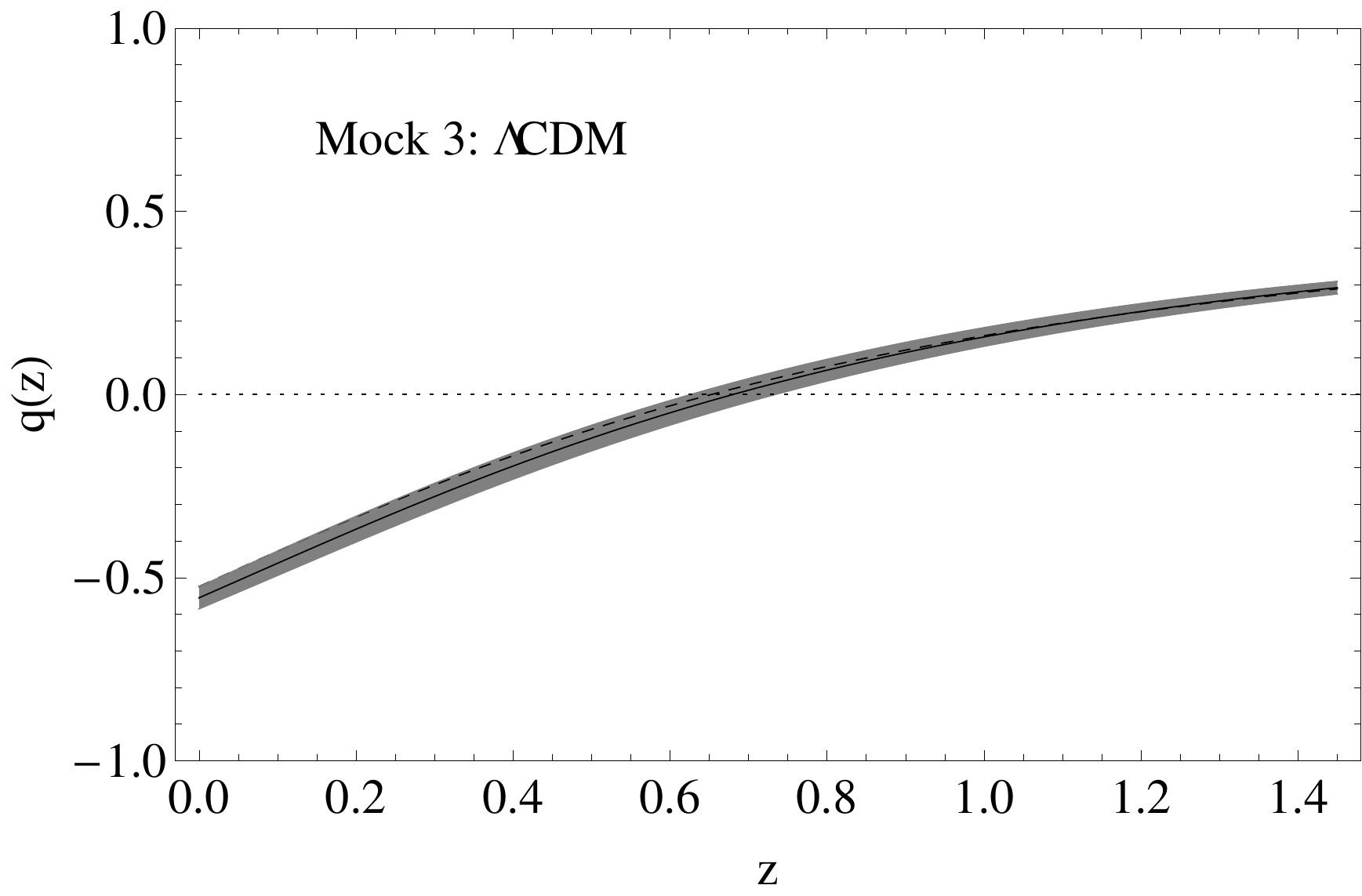}}}
\vspace{0cm}\rotatebox{0}{\vspace{0cm}\hspace{0cm}\resizebox{0.42\textwidth}{!}{\includegraphics{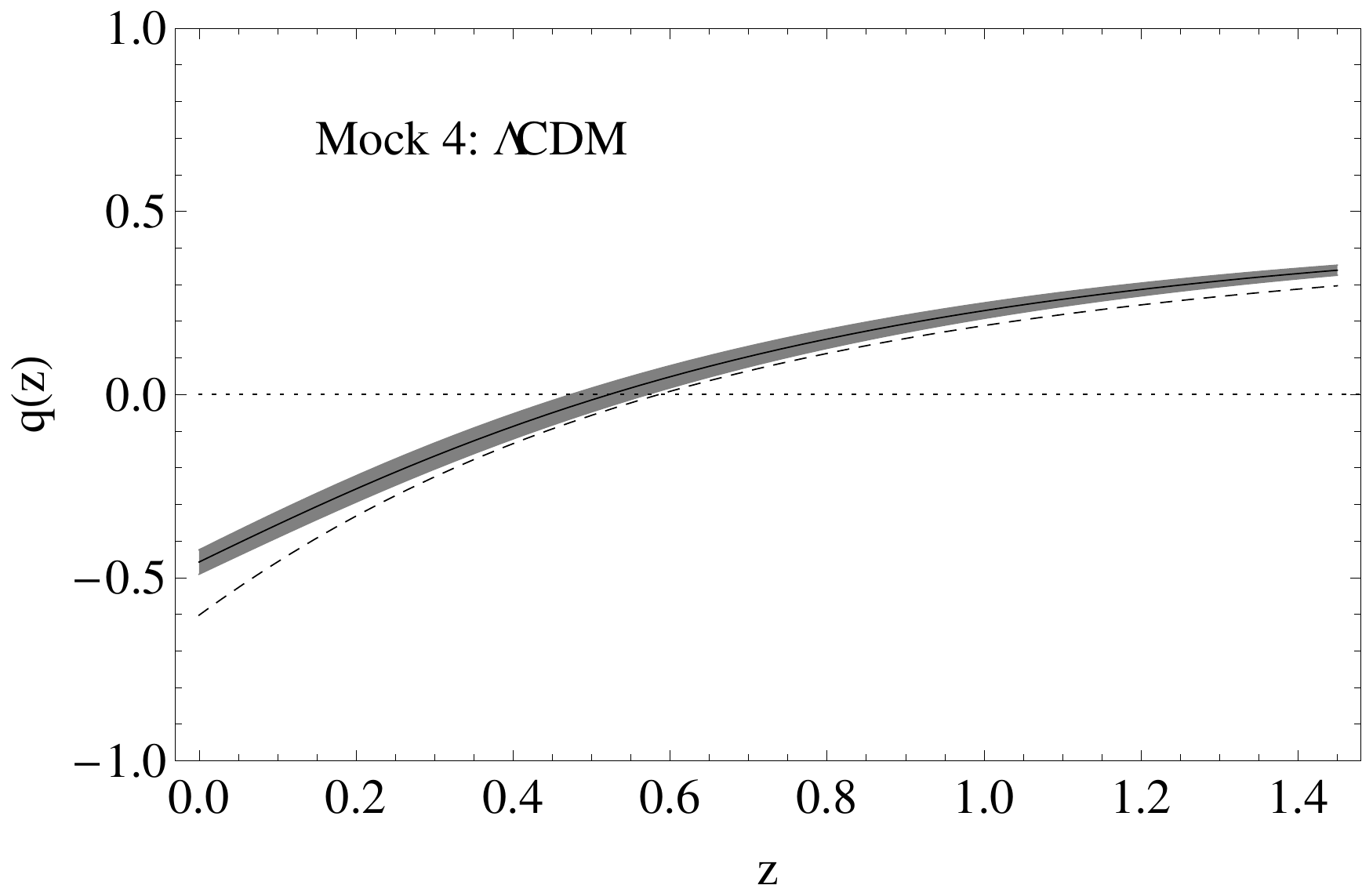}}}
\caption{The deceleration parameter $q(z)$ for all four mocks. The dashed line corresponds to the real model.\label{qzLCDM}}
\vspace{1cm}
\centering
\vspace{0cm}\rotatebox{0}{\vspace{0cm}\hspace{0cm}\resizebox{0.42\textwidth}{!}{\includegraphics{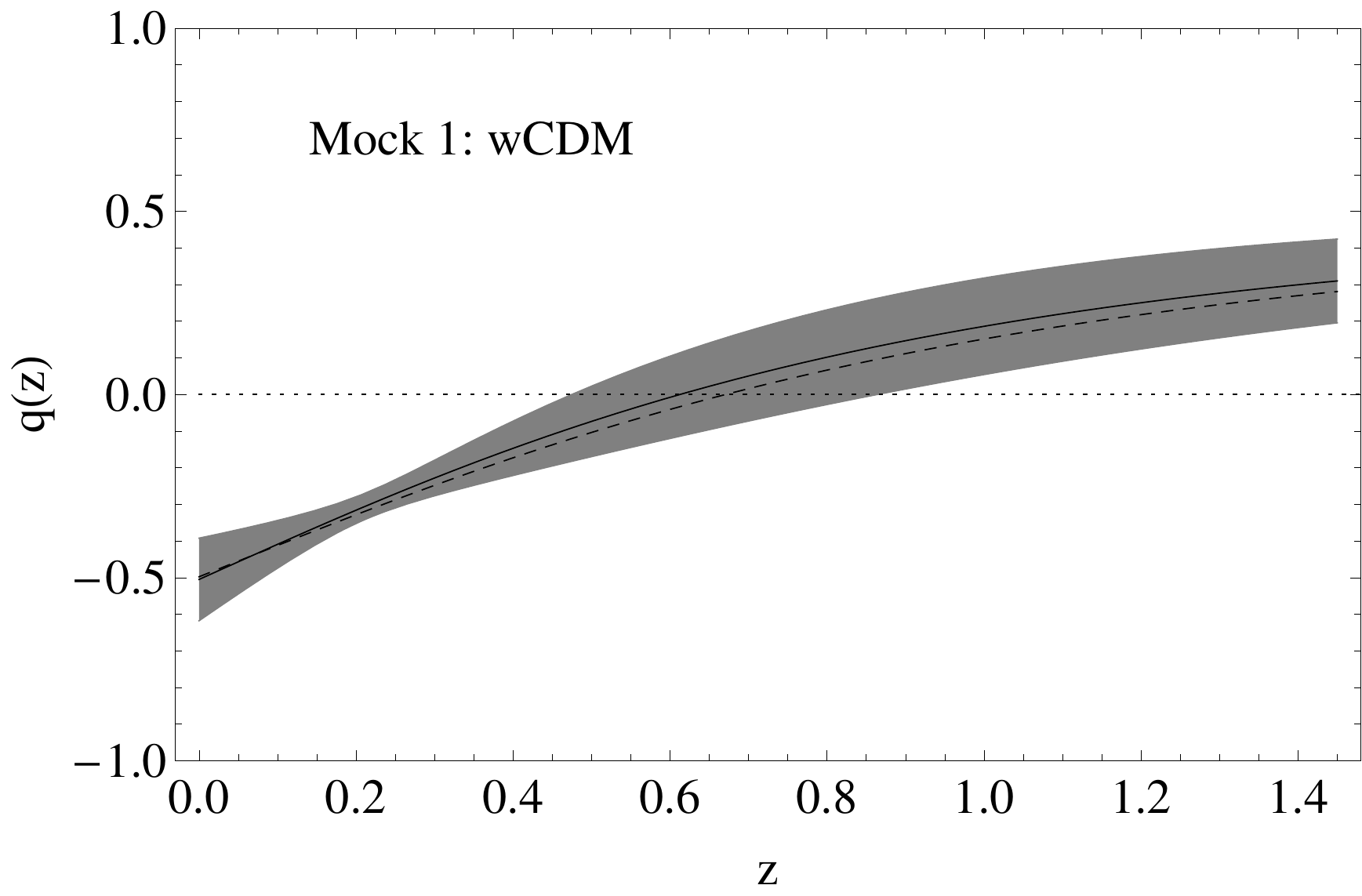}}}
\vspace{0cm}\rotatebox{0}{\vspace{0cm}\hspace{0cm}\resizebox{0.42\textwidth}{!}{\includegraphics{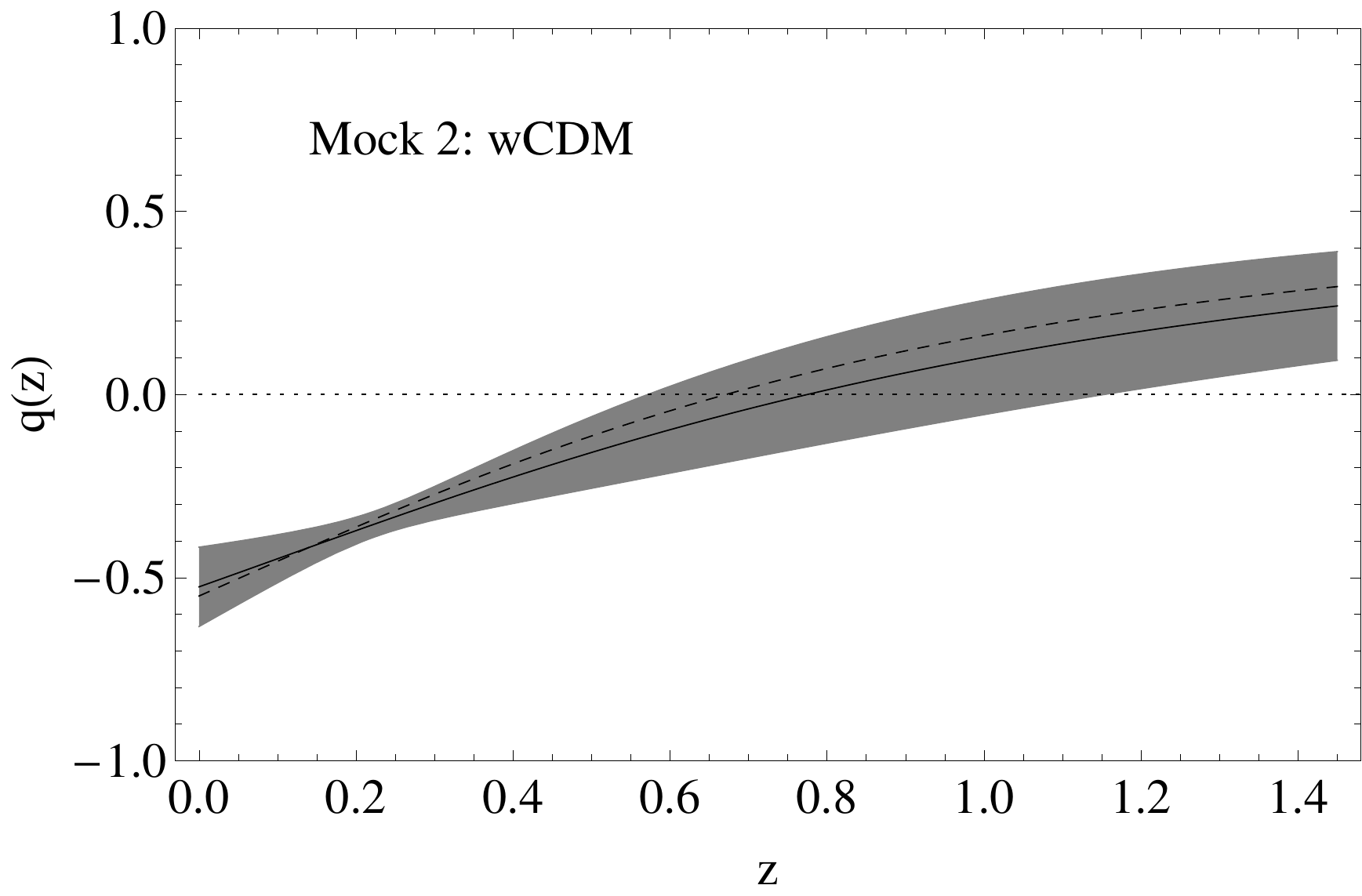}}}
\vspace{0cm}\rotatebox{0}{\vspace{0cm}\hspace{0cm}\resizebox{0.42\textwidth}{!}{\includegraphics{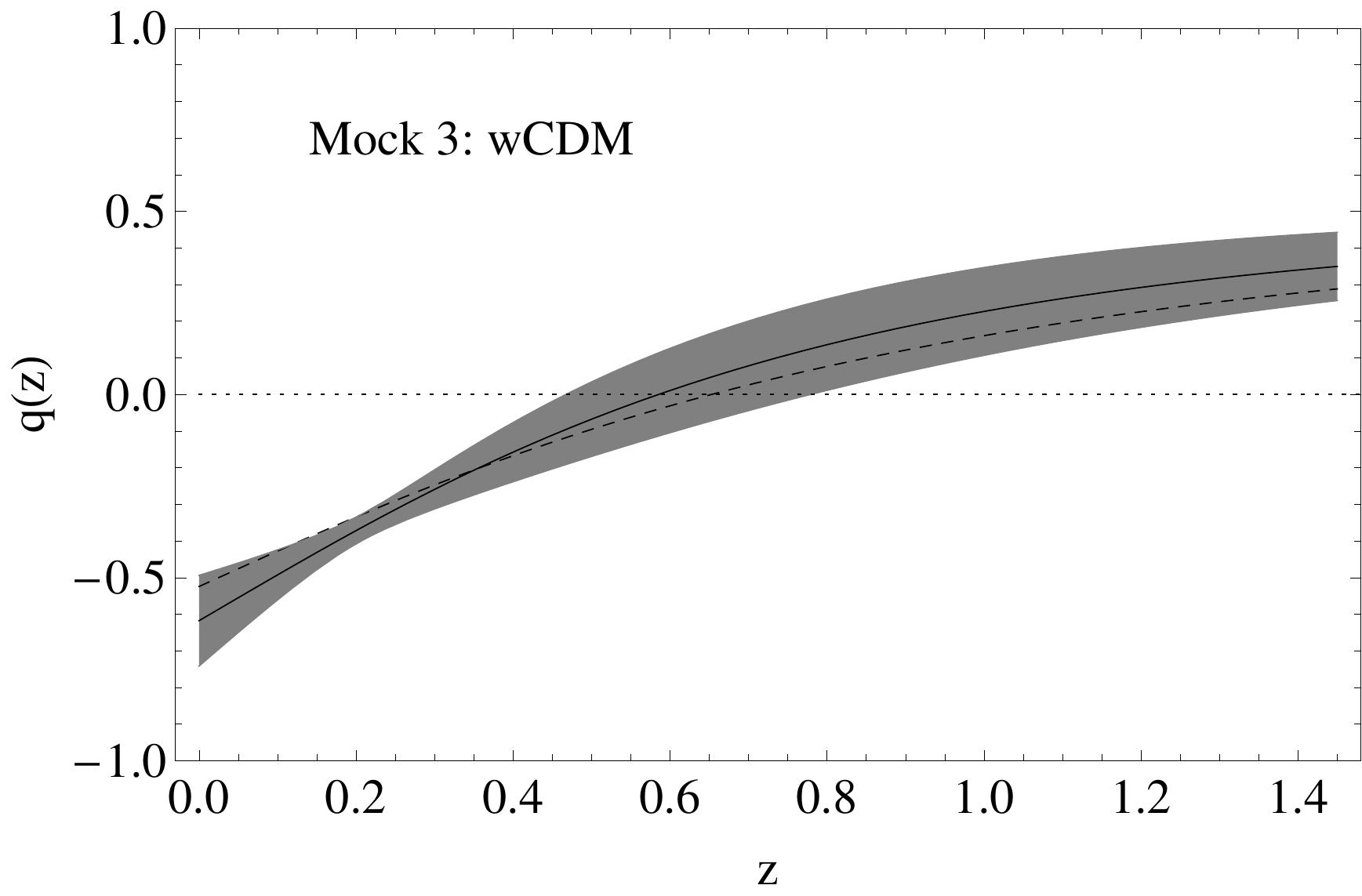}}}
\vspace{0cm}\rotatebox{0}{\vspace{0cm}\hspace{0cm}\resizebox{0.42\textwidth}{!}{\includegraphics{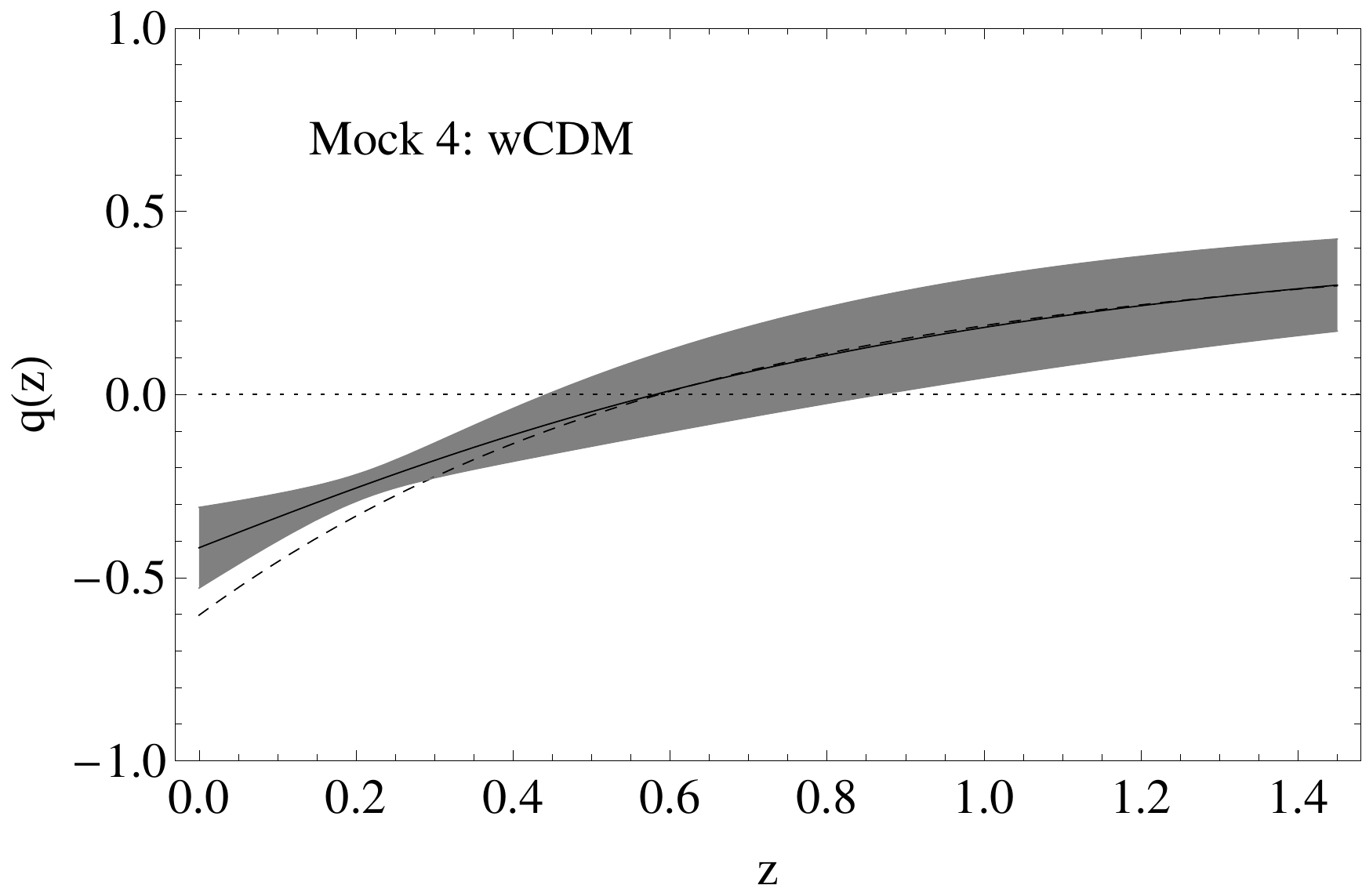}}}
\caption{The deceleration parameter $q(z)$ for all four mocks. The dashed line corresponds to the real model.\label{qzwCDM}}
\end{figure*}

\begin{figure*}[t!]
\centering
\vspace{0cm}\rotatebox{0}{\vspace{0cm}\hspace{0cm}\resizebox{0.42\textwidth}{!}{\includegraphics{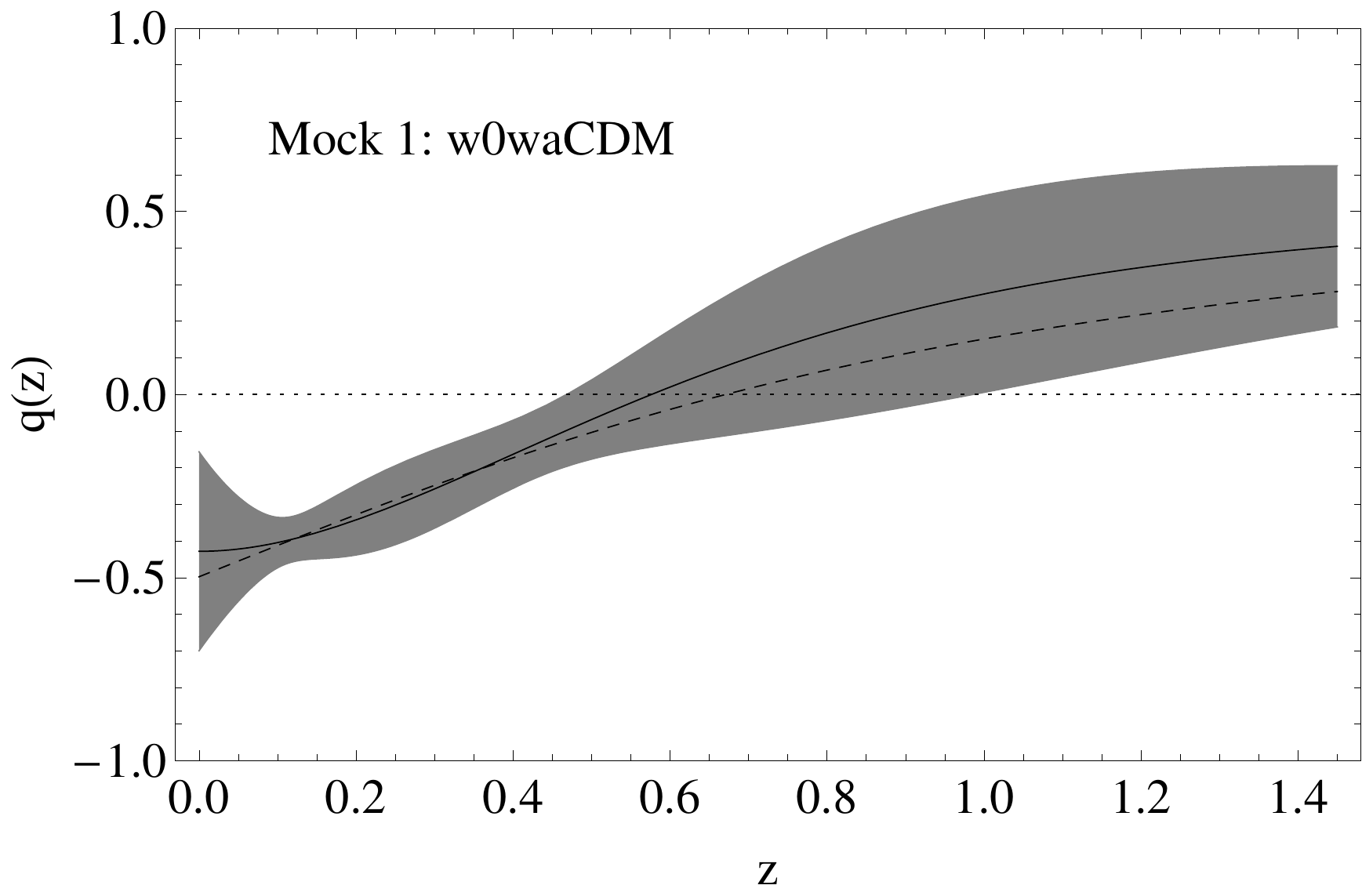}}}
\vspace{0cm}\rotatebox{0}{\vspace{0cm}\hspace{0cm}\resizebox{0.42\textwidth}{!}{\includegraphics{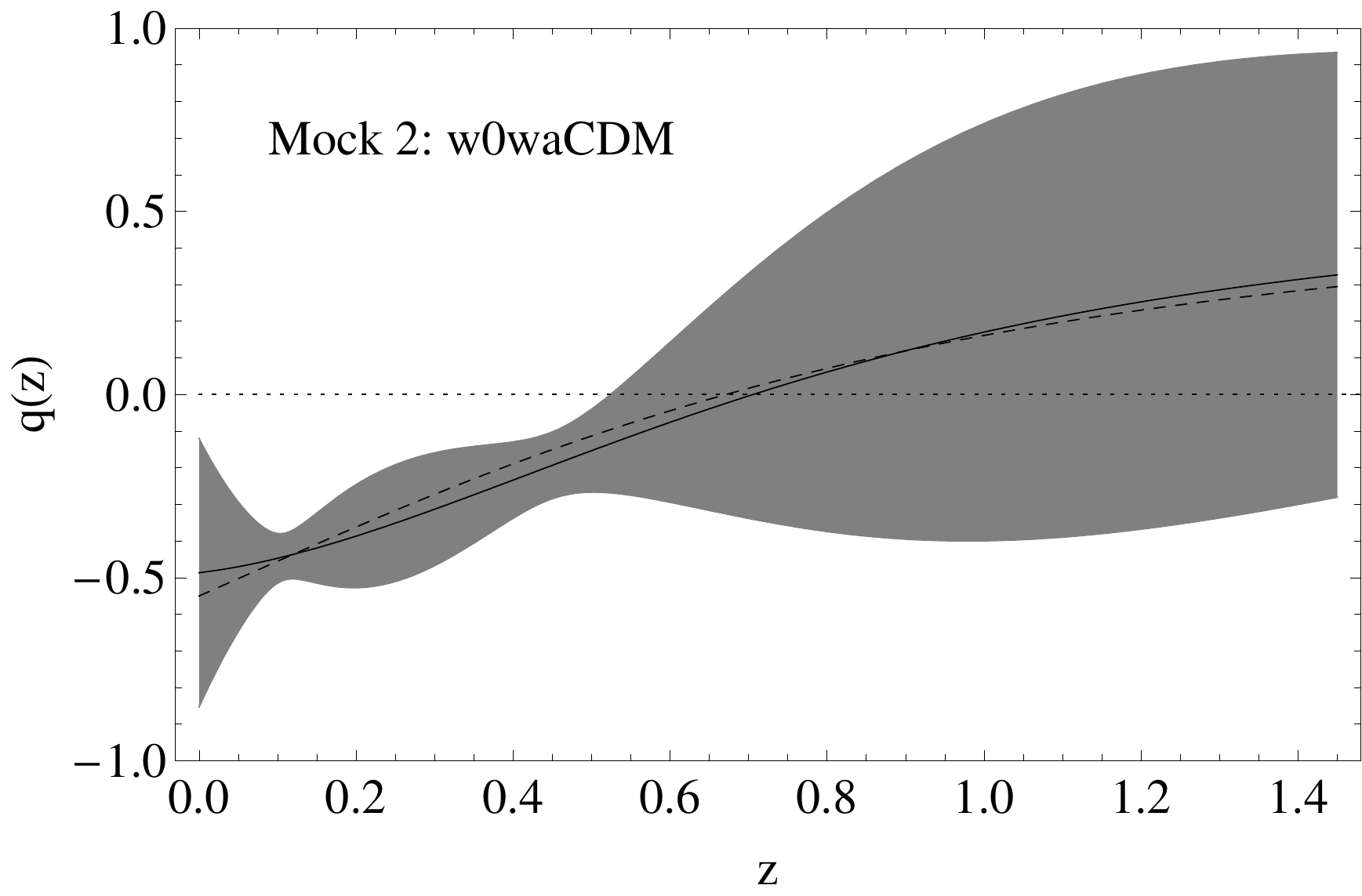}}}
\vspace{0cm}\rotatebox{0}{\vspace{0cm}\hspace{0cm}\resizebox{0.42\textwidth}{!}{\includegraphics{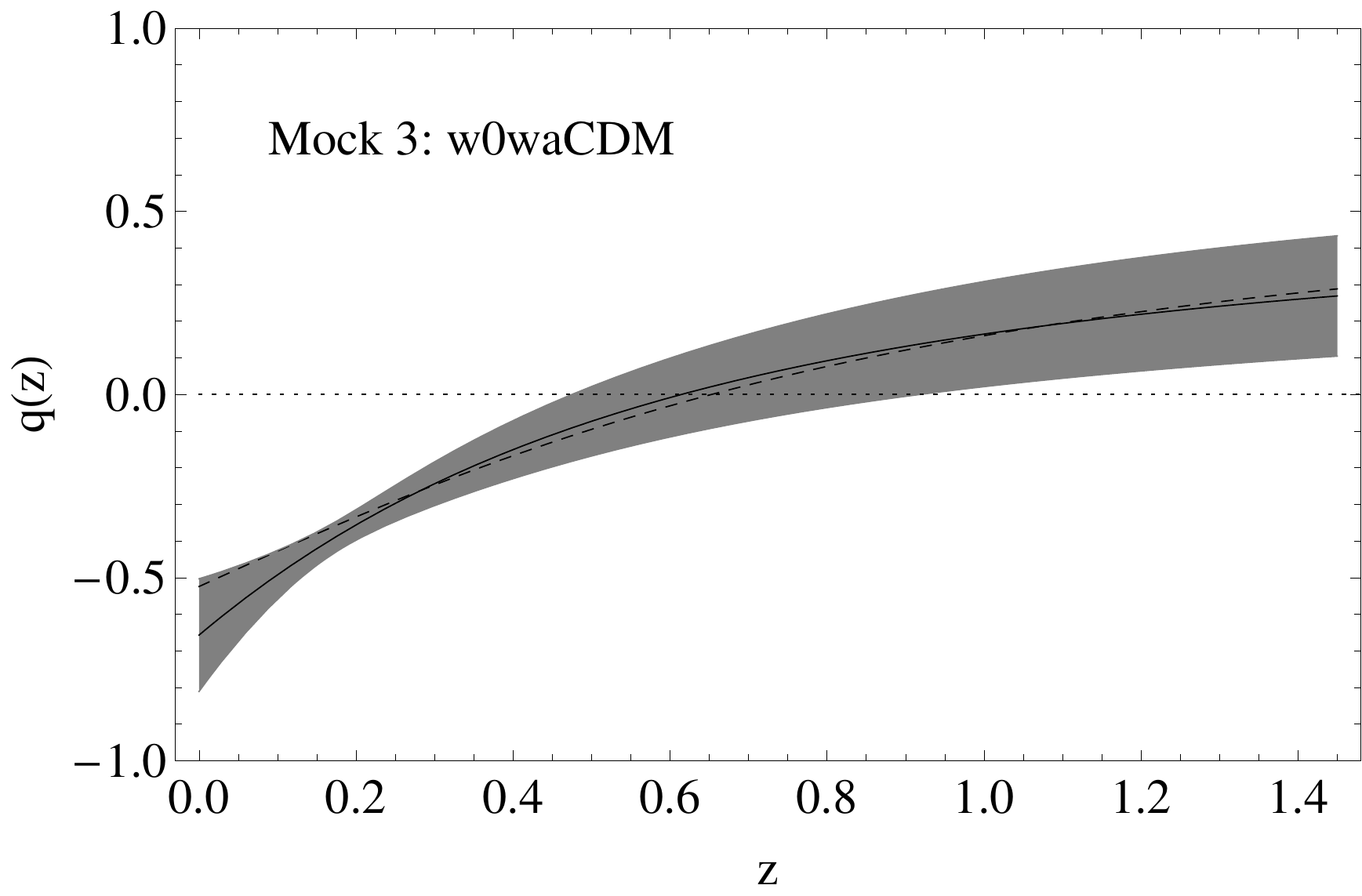}}}
\vspace{0cm}\rotatebox{0}{\vspace{0cm}\hspace{0cm}\resizebox{0.42\textwidth}{!}{\includegraphics{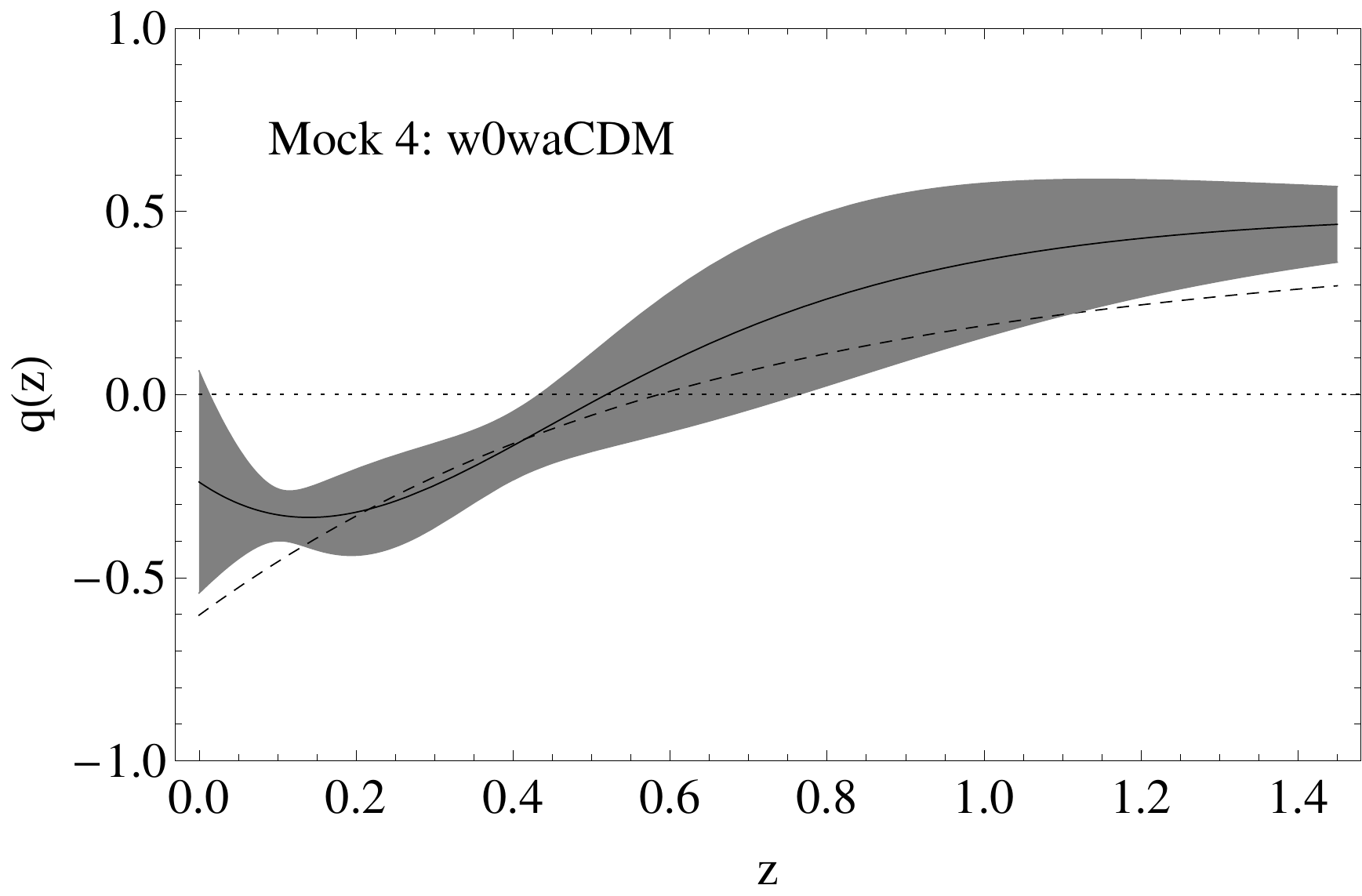}}}
\caption{The deceleration parameter $q(z)$ for all four mocks. The dashed line corresponds to the real model.\label{qzw0waCDM}}
\end{figure*}

For completeness we also fit the mock data with the original DE models, given by Eq.~(\ref{friedman0}). In Figs. \ref{qzLCDM}, \ref{qzwCDM} and \ref{qzw0waCDM} we show the deceleration parameter $q(z)$ for all four mocks for the \lcdm, $w=const$ and $w(a)=w_0+w_a(1-a)$ models. As expected, overall the agreement is quite good, except for some cases. To be more specific, as can be seen, all models fail to fit Mock 4, which is based on a $(w_0,w_a)$ model, with the discrepancy being larger at small redshifts and especially for the \lcdm model.

\section{Comparison\label{compar}}
In this section we will compare the different methods based on how successfully they reconstructed the real models. However, it is quite obvious that comparing all the different model-independent methods to each other is hardly an easy task as the various methods have different intrinsic characteristics; for example, the PCA gives results only on the specific redshift bins, while the GAs provide a smooth and differentiable function at all $z$, but they are nonparametric, while the other methods, based on the approximants and the polynomials, have varying numbers of parameters. This clearly means that the two popular methods mentioned in the Introduction, the $\chi^2/dof$ and the use of the Bayesian evidence, despite all their flaws, cannot be used in this case in order to make a fair and consistent comparison.

However, since we already know the real cosmology, we can make the comparison to zero order by creating a new $\chi^2$ defined as
\be
\chi^2_{comp}=\sum_{i=1}^N\left(\frac{q_{bf_i}-q_{real,i}}{\sigma_{bf,i}}\right)^2,
\label{chi2comp}
\ee
where $(q_{bf_i},\sigma_{bf,i})$ are the predictions of the best-fit models and the corresponding errors, while $q_{real,i}$ is the value of the deceleration parameter for the real model we used to create the mock data at a specific redshift. In order to have a fair comparison with the PCA we decided to test the rest of the models in the same redshift values, i.e. the mean redshift $z$ of the bins, for both six and ten bins.

\begin{table}[t!]
\begin{center}
\caption{The $\chi^2_{comp}$ for various models for all mocks for both six and ten bins. For easy reference, Mock 1 was created with the $w=$const. model $(\om=0.30,w=-0.95)$,  Mock 2 with a \lcdm model $(\om=0.30)$, Mock 3 with the Hu-Sawicki $f(R)$ model $(\om=0.30, b=0.11)$ and Mock 4 with a $w_0w_a$CDM model $(\om=0.30, w_0=-1.05,w_a=0.50)$. For further details and an in-depth analysis of the results, see the text. } \label{comparison}
\begin{tabular}{|c|cccc|cccc|}
  \hline
  \hspace{5pt} \textbf{Method} \hspace{5pt} & \hspace{5pt} \hspace{5pt}& \hspace{5pt} $\chi^2_{comp}$\hspace{5pt}& \hspace{5pt}  for six bins\hspace{5pt}& \hspace{5pt}  \hspace{5pt} & \hspace{5pt}  \hspace{5pt}& \hspace{5pt} $\chi^2_{comp}$\hspace{5pt}& for ten bins\hspace{5pt} \hspace{5pt}& \hspace{5pt}  \hspace{5pt}\\
  \hline
  \hspace{5pt} \hspace{5pt} & \hspace{5pt} Mock 1 \hspace{5pt}& \hspace{5pt} Mock 2 \hspace{5pt}& \hspace{5pt} Mock 3 \hspace{5pt}& \hspace{5pt} Mock 4 \hspace{5pt}& \hspace{5pt} Mock 1 \hspace{5pt}& \hspace{5pt} Mock 2 \hspace{5pt}& \hspace{5pt} Mock 3 \hspace{5pt}& \hspace{5pt} Mock 4 \hspace{5pt}\\
  \hline
  \hspace{5pt} PCA  \hspace{5pt} & \hspace{5pt} 5.319 \hspace{5pt}& \hspace{5pt}2.541 \hspace{5pt}& \hspace{5pt} 1.267\hspace{5pt}& \hspace{5pt} 13.247 \hspace{5pt}& \hspace{5pt} 3.272 \hspace{5pt}& \hspace{5pt} 4.387 \hspace{5pt}& \hspace{5pt} 3.670 \hspace{5pt}& \hspace{5pt} 18.944 \hspace{5pt}\\
  \hline
  \hspace{5pt} GA  \hspace{5pt} & \hspace{5pt} 0.633 \hspace{5pt}& \hspace{5pt}0.736 \hspace{5pt}& \hspace{5pt}6.588 \hspace{5pt}& \hspace{5pt} 8.065 \hspace{5pt}&
  \hspace{5pt} 1.352 \hspace{5pt}& \hspace{5pt} 1.199 \hspace{5pt}& \hspace{5pt} 17.299 \hspace{5pt}& \hspace{5pt} 15.231 \hspace{5pt}\\
  \hline
  \hspace{5pt} Pade $d_L$  \hspace{5pt} & \hspace{5pt} 1.570 \hspace{5pt}& \hspace{5pt}1.047 \hspace{5pt}& \hspace{5pt} 2.572 \hspace{5pt}& \hspace{5pt} 5.000 \hspace{5pt}& \hspace{5pt} 2.063 \hspace{5pt}& \hspace{5pt} 1.908 \hspace{5pt}& \hspace{5pt} 4.005 \hspace{5pt}& \hspace{5pt} 9.756 \hspace{5pt}\\
  \hline
  \hspace{5pt} Pade 2  \hspace{5pt} & \hspace{5pt} 0.930 \hspace{5pt}& \hspace{5pt}0.889 \hspace{5pt}& \hspace{5pt} 1.568 \hspace{5pt}& \hspace{5pt} 5.063 \hspace{5pt}& \hspace{5pt} 1.578 \hspace{5pt}& \hspace{5pt} 1.406 \hspace{5pt}& \hspace{5pt} 2.892 \hspace{5pt}& \hspace{5pt} 10.421 \hspace{5pt}\\
  \hline
  \hspace{5pt} Pade 2a  \hspace{5pt} & \hspace{5pt} 0.912 \hspace{5pt}& \hspace{5pt}2.086 \hspace{5pt}& \hspace{5pt} 2.181 \hspace{5pt}& \hspace{5pt} 4.861 \hspace{5pt}& \hspace{5pt} 1.392 \hspace{5pt}& \hspace{5pt} 2.732 \hspace{5pt}& \hspace{5pt} 2.982 \hspace{5pt}& \hspace{5pt}10.548 \hspace{5pt}\\
  \hline
  \hspace{5pt} \lcdm  \hspace{5pt} & \hspace{5pt} 9.852 \hspace{5pt}& \hspace{5pt}0.883 \hspace{5pt}& \hspace{5pt} 2.503 \hspace{5pt}& \hspace{5pt} 23.170 \hspace{5pt}& \hspace{5pt} 13.586 \hspace{5pt}& \hspace{5pt} 1.476 \hspace{5pt}& \hspace{5pt} 4.860 \hspace{5pt}& \hspace{5pt} 39.920 \hspace{5pt}\\
  \hline
  \hspace{5pt} wCDM  \hspace{5pt} & \hspace{5pt} 0.486 \hspace{5pt}& \hspace{5pt}0.971 \hspace{5pt}& \hspace{5pt} 1.852 \hspace{5pt}& \hspace{5pt} 4.498 \hspace{5pt}& \hspace{5pt} 0.847 \hspace{5pt}& \hspace{5pt} 1.588 \hspace{5pt}& \hspace{5pt} 3.424 \hspace{5pt}& \hspace{5pt} 10.223 \hspace{5pt}\\
  \hline
  \hspace{5pt} $w_0w_a$CDM  \hspace{5pt} & \hspace{5pt} 0.742 \hspace{5pt}& \hspace{5pt}0.214 \hspace{5pt}& \hspace{5pt} 1.038 \hspace{5pt}& \hspace{5pt}5.510 \hspace{5pt}& \hspace{5pt} 1.070 \hspace{5pt}& \hspace{5pt} 0.482 \hspace{5pt}& \hspace{5pt} 1.851 \hspace{5pt}& \hspace{5pt} 5.452 \hspace{5pt}\\
  \hline
\end{tabular}
\end{center}
\end{table}

Finally, as we mentioned in the earlier sections, for some of the models, such as the cosmography, the Chebyshev polynomials for both $q(z)$ and $d_L(z)$ and the Taylor expansions, the best-fit deceleration parameter $q(z)$ has singularities and huge oscillations when the real models do not, thus making it unphysical. As a result, we excluded them from the rest of the comparison.

In Table \ref{comparison} we show  $\chi^2_{comp}$ for various models for all mocks and for both six and ten bins. At this point we should remind the reader that Mock 1 was created with the $w=$const. model $(\om=0.30,w_0=-0.95,w_a=0)$,  Mock 2 with a \lcdm model $(\om=0.30, w_0=-1, w_a=0)$, Mock 3 with the Hu-Sawicki $f(R)$ model $(\om=0.30, b=0.11)$ and Mock 4 with a $w_0w_a$CDM model $(\om=0.30, w_0=-1.05,w_a=0.50)$. For the dark energy models we used Eq.~(\ref{friedman0}), while for the $f(R)$ model we used Eq.~(\ref{Hu1}).

According to the values of the Table we rank the different methods as follows, going from the best (left) to the worst (right).

For six bins:
\begin{itemize}
  \item Mock 1: wCDM, GA, $w_0w_a$CDM, Pade 2a, Pade 2, Pade dL, PCA, \lcdm
  \item Mock 2: $w_0w_a$CDM, GA, \lcdm, Pade 2, wCDM, Pade dL, Pade 2a, PCA
  \item Mock 3: $w_0w_a$CDM, PCA, Pade 2, wCDM, Pade 2a, \lcdm, Pade dL, GA
  \item Mock 4: wCDM, Pade 2a, Pade dL, Pade 2, $w_0w_a$CDM, GA, PCA, \lcdm
\end{itemize}

For ten bins:
\begin{itemize}
  \item Mock 1: wCDM, $w_0w_a$CDM, GA, Pade 2a, Pade 2, Pade dL, PCA, \lcdm
  \item Mock 2: $w_0w_a$CDM, GA, Pade 2, \lcdm, wCDM, Pade dL, Pade 2a, PCA
  \item Mock 3: $w_0w_a$CDM, Pade 2, Pade 2a, wCDM, PCA, Pade dL, \lcdm, GA
  \item Mock 4: $w_0w_a$CDM, Pade dL, wCDM, Pade 2, Pade 2a, GA, PCA, \lcdm
\end{itemize}

If we only consider the model-independent methods, i.e. we exclude the usual DE models, then the ranking is as follows, again going from the best (left) to the worst (right).

For six bins:
\begin{itemize}
  \item Mock 1: GA, Pade 2a, Pade 2, Pade dL, PCA
  \item Mock 2: GA, Pade 2, Pade dL, Pade 2a, PCA
  \item Mock 3: PCA, Pade 2, Pade 2a, Pade dL, GA
  \item Mock 4: Pade 2a, Pade dL, Pade 2, GA, PCA
\end{itemize}

For ten bins:
\begin{itemize}
  \item Mock 1: GA, Pade 2a, Pade 2, Pade dL, PCA
  \item Mock 2: GA, Pade 2, Pade dL, Pade 2a, PCA
  \item Mock 3: Pade 2, Pade 2a, PCA, Pade dL, GA
  \item Mock 4: Pade dL, Pade 2, Pade 2a, GA, PCA
\end{itemize}

It is quite clear that no method out of all the model-independent ones can be the best at fitting all the different models at once. More specifically, certain methods seem to be the best in describing some of the models but do not perform so well at others, e.g. the GAs work very well for the wCDM and \lcdm models but underperform on the more complicated $f(R)$ and $w_0w_a$CDM models.

Also, in general, the PCAs seem not to do very well compared to the other methods, regardless of the model or the number of redshift bins. Regarding the latter, changing the binning at which the comparison is made seems to slightly affect the ranking itself for several of the methods. Finally, not surprisingly the Pad\'e approximants seem to do reasonably well in all cases, thus proving their flexibility in fitting a variety of different models.

\section{Conclusions}
We have entered an era of huge data sets of cosmological probes, thus making it necessary to be able to reconstruct the underlying cosmology as accurately as possible. Contrary to the traditional way of testing only the one or two most popular models and thus running the risk of obtaining biased results, we advocate the choice to use complementary model-independent techniques, in the sense that they assume no underlying theoretical model and have a minimum number of assumptions.

In this vein, we tested several model-independent methods, including the principal components analysis, the genetic algorithms, various Pad\'e approximants, different polynomial expansions and also cosmography, by fitting them to mock SnIa data based on different cosmological models. The inclusion of all these different methods obviously raises the question of how we can compare them since they all have different characteristics, e.g. the best fit of the PCA is only known at certain redshift points, while the GAs are completely nonparametric, thus making the traditional comparison based on Bayesian inference problematic.

The answer, to zero order, put forward in the present analysis was to calculate the $\chi^2_{comp}$ between the reconstructed and real deceleration parameter $q(z)$ and rank the methods accordingly. The main conclusions for following this methodology are as follows. First, it is clear that no one method out of all the model-independent ones can be the best at fitting all the different ``real" cosmologies at once. More specifically, certain methods seem to be the best in describing some of the models but do not perform so well at others, e.g. the GAs work very well for the wCDM and \lcdm models but underperform on the more complicated $f(R)$ and $w_0w_a$CDM models. This is clearly an issue that deserves further investigation as to why it happens and how it can be fixed. On the other hand, the PCA seems to underperform compared to the other methods, on most of the mocks while the Pad\'e approximants do reasonably well on all of the cases.

On the other hand, regarding the usual DE models, it is clear that the $w_0w_a$CDM model, based on $w(a)=w_0+w_a(1-a)$, is the most flexible of the three, but this comes at a high price, as it is the best even in cases where the data originated from a different real cosmology, thus potentially driving us to misleading conclusions about the underlying cosmological model. One possibility to solve this would be the inclusion of different kinds of data, like the BAO and CMB, in order to break the degeneracies, but as we have mentioned this is beyond the scope of the present analysis and is left for a future paper.

Of course, it should be mentioned that the method of comparison itself, by calculating the $\chi^2_{comp}$ of Eq.~(\ref{chi2comp}) and ranking the methods accordingly, could possibly be improved upon, since as was mentioned it is only a zero-order approach to the problem of ranking the very inhomogeneous set of model-independent methods present in the current analysis. However, doing that is not an easy task if one wants to test all of the methods consistently and especially given the two special cases of the PCA and the GA that present the most difficulty among the group of methods.

Finally, perhaps the most important message of the present analysis is that when analyzing the cosmological data, given our ignorance in the dark sector of the cosmological ingredients of the Universe, one should try to use a variety of different methods, both model-independent and  otherwise, in order to extract the maximum amount of information with the least amount of bias, instead of using only one or two specific models something that is becoming more and more important as we move towards an era of huge data sets.

\section*{Acknowledgements}
We would like to thank D. Sapone for useful discussions in the early stages of the work and R. Crittenden, E. Sanchez and V. Vitagliano for fruitful discussions related to the analysis. We acknowledge financial support from the Madrid Regional Government (CAM) under the Program No. HEPHACOS S2009/ESP-1473-02, from MICINN under Grant No. AYA2009-13936-C06-06 and Consolider-Ingenio 2010 PAU (CSD2007-00060), as well as from the European Union Marie Curie Initial Training Network No. UNILHC PITN-GA-2009-237920. S.~N. is supported by CAM through a HEPHACOS Fellowship.


\begin{thebibliography}{99}
\bibitem{Ade:2013zuv}
  P.~A.~R.~Ade {\it et al.}  [Planck Collaboration],
  arXiv:1303.5076 [astro-ph.CO].

\bibitem{Suzuki:2011hu}
  N.~Suzuki, D.~Rubin, C.~Lidman, G.~Aldering, R.~Amanullah, K.~Barbary, L.~F.~Barrientos and J.~Botyanszki {\it et al.},
  Astrophys.\ J.\  {\bf 746}, 85 (2012)
  [arXiv:1105.3470 [astro-ph.CO]].

\bibitem{Copeland:2006wr}
  E.~J.~Copeland, M.~Sami and S.~Tsujikawa,
  Int.\ J.\ Mod.\ Phys.\ D {\bf 15}, 1753 (2006)
  [hep-th/0603057].

\bibitem{Nesseris:2012cq}
  S.~Nesseris and J.~Garcia-Bellido,
  arXiv:1210.7652 [astro-ph.CO].

\bibitem{Nesseris:2004wj}
S.~Nesseris and L.~Perivolaropoulos,
Phys.\ Rev.\ D {\bf 70}, 043531 (2004).

\bibitem{Nesseris:2005ur}
S.~Nesseris and L.~Perivolaropoulos,
Phys.\ Rev.\ D {\bf 72}, 123519 (2005).

\bibitem{Nesseris:2006er}
S.~Nesseris and L.~Perivolaropoulos,
JCAP {\bf 0701}, 018 (2007).

\bibitem {Hu07}
W. Hu and I. Sawicki, Phys. Rev. D., \textbf{76}, 064004 (2007)

\bibitem{Shafieloo:2007cs}
  A.~Shafieloo,
  Mon.\ Not.\ Roy.\ Astron.\ Soc.\  {\bf 380}, 1573 (2007)
  [astro-ph/0703034 [ASTRO-PH]].

\bibitem{Kunz:2007rk}
  M.~Kunz,
  Phys.\ Rev.\ D {\bf 80}, 123001 (2009)
  [astro-ph/0702615].

\bibitem{Basilakos:2013nfa}
  S.~Basilakos, S.~Nesseris and L.~Perivolaropoulos,
  Phys.\ Rev.\ D {\bf 87}, 123529 (2013)
  [arXiv:1302.6051 [astro-ph.CO]].

\bibitem{Shapiro:2005nz}
  C.~Shapiro and M.~S.~Turner,
  Astrophys.\ J.\  {\bf 649}, 563 (2006)
  [astro-ph/0512586].

\bibitem{Huterer:2004ch}
  D.~Huterer and A.~Cooray,
  Phys.\ Rev.\ D {\bf 71}, 023506 (2005)
  [astro-ph/0404062].

\bibitem{Amendola:2011qp}
  L.~Amendola, A.~C.~O.~Leite, C.~J.~A.~P.~Martins, N.~J.~Nunes, P.~O.~J.~Pedrosa and A.~Seganti,
  Phys.\ Rev.\ D {\bf 86}, 063515 (2012)
  [arXiv:1109.6793 [astro-ph.CO]].

\bibitem{Bogdanos:2009ib}
  C.~Bogdanos and S.~Nesseris,
  JCAP {\bf 0905}, 006 (2009)
  [arXiv:0903.2805 [astro-ph.CO]].

\bibitem{Nesseris:2010ep}
  S.~Nesseris and A.~Shafieloo,
  Mon.\ Not.\ Roy.\ Astron.\ Soc.\  {\bf 408}, 1879 (2010)
  [arXiv:1004.0960 [astro-ph.CO]].

\bibitem{Nesseris:2012tt}
  S.~Nesseris and J.~Garcia-Bellido,
  JCAP {\bf 1211}, 033 (2012)
  [arXiv:1205.0364 [astro-ph.CO]].

\bibitem{Tsoulos}
I.~G.~Tsoulos, D.~Gavrilis, E.~Dermatas, Computer Physics
Communications {\bf 177}, 976 (2007).

\bibitem{Huterer:2000mj}
  D.~Huterer and M.~S.~Turner,
  Phys.\ Rev.\ D {\bf 64}, 123527 (2001)
  [astro-ph/0012510].

\bibitem{Gerke:2002sx}
  B.~F.~Gerke and G.~Efstathiou,
  Mon.\ Not.\ Roy.\ Astron.\ Soc.\  {\bf 335}, 33 (2002)
  [astro-ph/0201336].

\bibitem{Alam:2003fg}
  U.~Alam, V.~Sahni, T.~D.~Saini and A.~A.~Starobinsky,
  Mon.\ Not.\ Roy.\ Astron.\ Soc.\  {\bf 354}, 275 (2004)
  [astro-ph/0311364].

\bibitem{press92}W.~H.~Press {\it et. al.}, ``Numerical Recipes'',
Cambridge University Press (1994).

\bibitem{Visser:2004bf}
  M.~Visser,
  Gen.\ Rel.\ Grav.\  {\bf 37}, 1541 (2005)
  [gr-qc/0411131].

\bibitem{Aviles:2012ay}
  A.~Aviles, C.~Gruber, O.~Luongo and H.~Quevedo,
  Phys.\ Rev.\ D {\bf 86}, 123516 (2012)
  [arXiv:1204.2007 [astro-ph.CO]].

\bibitem{Vitagliano:2009et}
  V.~Vitagliano, J.~-Q.~Xia, S.~Liberati and M.~Viel,
  JCAP {\bf 1003}, 005 (2010)
  [arXiv:0911.1249 [astro-ph.CO]].

\bibitem{Xia:2011iv}
  J.~-Q.~Xia, V.~Vitagliano, S.~Liberati and M.~Viel,
  Phys.\ Rev.\ D {\bf 85}, 043520 (2012)
  [arXiv:1103.0378 [astro-ph.CO]].

\end{thebibliography}
\end{document}